\documentclass[twocolumn,prd,floatfix,preprintnumbers,a4paper,nofootinbib,superscriptaddress,groupedaddress]{revtex4-1}
\usepackage[utf8]{inputenc}
\usepackage{amsmath,amssymb}
\usepackage{amsfonts}
\usepackage{bm}
\usepackage{courier}
\usepackage{graphics}
\usepackage{float}
\usepackage{amsmath}
\usepackage{amssymb}
\usepackage[normalem]{ulem} 
\usepackage[mathscr]{euscript}
\usepackage{acronym}
\usepackage{float}
\usepackage[caption=false]{subfig}
\usepackage{lipsum}
\usepackage{mathtools}
\usepackage{multirow}
\usepackage{graphicx}
\usepackage{mathtools}
\usepackage{multirow}
\usepackage{graphicx}
\usepackage[usenames,dvipsnames,table,xcdraw]{xcolor}
\usepackage[colorlinks, pdfborder={0 0 0}]{hyperref} 
\usepackage[nameinlink,capitalise]{cleveref}
\usepackage{tensor}
\usepackage{verbatim}
\usepackage{wrapfig}
\usepackage{soul} 

\graphicspath{{fig}}
\newcommand{\Msun}{\ensuremath{M_\odot}}

\begin{document}
\title{
The landscape of massive black-hole spectroscopy\\ with LISA and  Einstein Telescope } 

	\author{
		Swetha Bhagwat$^{1}$,
		Costantino Pacilio$^{1}$,
		Enrico Barausse$^{2,3}$,
		Paolo Pani$^{1}$.
	}
	\affiliation{$^{1}$Dipartimento di Fisica, ``Sapienza'' Universit\`a di Roma \& Sezione INFN Roma1, P.A. Moro 5, 00185, Roma, Italy}
	\affiliation{${^2}$SISSA, Via Bonomea 265, 34136 Trieste, Italy and INFN Sezione di Trieste}
	\affiliation{${^3}$IFPU - Institute for Fundamental Physics of the Universe, Via Beirut 2, 34014 Trieste, Italy}

\begin{abstract}
Measuring the quasi-normal mode~(QNM) spectrum emitted by a perturbed black-hole~(BH) --~also known as BH spectroscopy~-- provides an excellent opportunity to test the predictions of general relativity in the strong-gravity regime. We investigate the prospects and precision of BH spectroscopy in massive binary black hole ringdowns, one of the primary science objectives of the future Laser Interferometric Space Antenna~(LISA) mission. 
We simulate various massive binary BH population models, featuring competing prescriptions for the Delays between  galaxy and BH mergers, for the impact of supernova feedback on massive BH growth, and for the initial population of high redshift BH seeds (light versus heavy seeds).
For each of these scenarios, we compute the average number of expected events for precision BH spectroscopy using a Fisher-matrix analysis. 
We find that, for any heavy seed scenario, LISA will measure the dominant mode frequency within ${\cal O}(0.1) \%$ relative uncertainty and will estimate at least 3 independent QNM parameters within $1 \%$ error. The most optimistic heavy seed scenarios produce $\mathcal{O}(100)$ events with $1 \%$ measurability for 3 or more QNM quantities during LISA’s operational time.
On the other hand,  light seed scenarios produce lighter merger remnants, which ring at frequencies higher than LISA's sensitivity. 
Interestingly, the light seed models give rise to a fraction of mergers in the band of Einstein Telescope, allowing for the measurement of 3 QNM parameters with $\sim 10 \%$ relative errors in approximately a few to ten events/yr.
More precise BH spectroscopy in the light seed scenarios would require instruments operating in the deciHertz band.
\end{abstract}

\preprint{ET-0465A-21}

\maketitle

\section{Introduction}

When a binary black hole~(BH) merges, it forms a distorted BH which then settles down by emitting gravitational waves~(GWs) at characteristic complex frequencies~\cite{QNM-Chandrasekhar,Vishveshwara}. This part of the signal is called the ringdown and the characteristic frequency spectrum is called the quasi-normal modes~(QNMs)~\cite{Kokkotas-review,Berti:2009kk}. Kerr BHs in the general theory of relativity~(GR) satisfy the no-hair theorem, which mandates that its (infinite) QNM spectrum be uniquely and fully determined by just 2 observable parameters — namely the mass $M_f$ and the dimensionless spin $\chi_f$ of the remnant BH~\cite{No-hair-original}.
Verifying whether the observed QNM spectrum satisfies the no-hair theorem allows us to perform clean and robust null-hypothesis tests of the Kerr metric and of GR in the strong-field regime~\cite{QNM-review-first-tgr-proposal,Berti:2018vdi,PhysRevD.85.082003,Maselli:2019mjd,Volkel:2020daa}. BHs in most modified gravity theories~\cite{Berti:2015itd,Yagi:2016jml} either have a different dynamics or are not described by a Kerr metric~\cite{Barausse:2008xv,Pani:2009wy};  therefore we expect their QNM spectrum to differ from that of a Kerr BH in GR. Furthermore, the QNM spectra also allow us to probe the nature of the remnant, e.g., if it is a GR BH or an exotic compact object~\cite{Barausse:2014tra,Cardoso:2019rvt,Cardoso:2016oxy,2021arXiv210506410M,Maggio:2020jml,2021PhRvD.103b4041B,2019LRR....22....4C}.

Information from the ringdown can also be combined with the inspiral-merger part of the signal to perform consistency tests, wherein one checks if the mass and spin of the remnant inferred by the pre-ringdown signal is consistent with that measured using the ringdown. This test could be done even when only a single QNM parameter is measured. Currently, the best measurement of the dominant mode frequency agrees with the GR predictions within $ \sim 16 \%$ measurement uncertainty for GW150914~\cite{TGR-gw150914, Ghosh:2021mrv, Ghosh:2016qgn,LIGOScientific:2019fpa,LIGOScientific:2020tif,LIGOScientific:2021sio}.

Performing model-independent tests of the no-hair theorem using \emph{BH spectroscopy} alone is a more ambitious program, which requires measuring at least 3 independent QNM parameters~\cite{Berti:2005ys,gossan-et-al,RD-TGR-first-proposal}. The first two QNM parameters are inverted to find the mass and spin of the remnant, assuming the latter is a Kerr BH. This fixes the whole QNM spectrum. Then, the measurement of the third (and possibly more) QNM parameter(s) can be used to check for consistency with GR’s predictions.

Currently, with the LIGO-Virgo data, the statistical uncertainty in the measurement of the QNM parameters limits the precision of ringdown-based tests. BH spectroscopy was attempted for GW150914 using overtones~\cite{Isi:2019aib} and for GW190814 and GW190412 using the secondary angular QNM~\cite{2020PhRvD.102l4070C}. BH spectroscopy with overtones has potential limitations related to resolvability of the QNMs, the number of overtones to be included, and is sensitive to the choice of start time of the ringdown~\cite{Bhagwat:2019dtm,Ota:2019bzl,2020PhRvD.101d4033B,JimenezForteza:2020cve}. On the other hand, the analysis of~\cite{2020PhRvD.102l4070C} has a low signal-to-noise ratio~(SNR) in the secondary angular mode, limiting the constraining power of the test.

Given the current state of affairs, it is relevant to forecast the prospects for BH spectroscopy with the future GW detectors. In particular, the coalescence of massive BH~(MBH) binaries could provide an ideal setting for BH spectroscopy~\cite{Berti:2005ys,Berti:2016lat,Barack:2018yly}, as the ringdown SNR scales as $\sim M_f^{3/2}$, where $M_f$ is the remnant mass~\cite{Berti:2016lat}. 

In this work, we quantify the landscape of BH spectroscopy for astrophysically-motivated population models of MBH binaries. We estimate the measurement errors expected for multiple-QNM parameters in the ringdowns produced by these populations.
Our main focus is on MBH binaries detectable by the forthcoming Laser Interferometric Space Antenna~(LISA)~\cite{LISA:2017pwj}. However, for reasons that will be explained in a moment, we shall also investigate the potential of 
next-generation ground-based detectors~\cite{Kalogera:2021bya}, such as Cosmic Explorer~(CE)~\cite{Evans:2016mbw,Essick:2017wyl} and Einstein Telescope~(ET)~\cite{Hild:2010id,Maggiore:2019uih}, to perform ringdown-based tests using (light) MBH binaries. In particular, we shall focus on ET since it has a better sensitivity at low frequency and is better suited to detect MBH ringdowns.

The motivation for our study is twofold. First, LISA is expected to have an unprecedented sensitivity towards MBH ringdowns~\cite{Berti:2005ys,2016PhRvD..93b4003K}; for instance, an equal-mass binary BH with a total mass of $10^6 \Msun$ at $2\,{\rm Gpc}$ will have an optimal ringdown SNR $\rho_{\rm rd}\sim 2000$ in the LISA data. Testing the nature of massive compact objects and the underlying theory of gravity in the strong-field regime using BH spectroscopy is one of the main science objectives of the LISA mission~\cite{Barausse:2020rsu,Cardoso:2019rvt} and of its possible extensions~\cite{Sesana:2019vho,Baibhav:2019rsa}. However, to get a more realistic picture of the science objectives achievable by LISA, these studies need to fold-in knowledge on the expected binary MBH population in the universe. In order to assess this issue, we study 8 MBH binary populations~\cite{Barausse:2020mdt,Barausse:2020gbp} that should bracket the expected astrophysical modelling uncertainties. These models produce different distributions of the coalescence parameters, such as mass ratio and spins of the binary BH system, its redshift, as well as the remnant mass and spin; these determine the expected event rates seen by LISA (see Sec.~\ref{sec:MBH-pop} for a brief summary of the population models). A major ingredient for these population models in the context of this work is the seed mass function — which broadly classifies our models into heavy seeds~(HS) and light seeds~(LS). The seed mass plays a dominant role in determining the remnant BH’s mass and, in turn, the scale of the QNM frequencies in the ringdown. While the LS models produce lighter remnants whose ringdown is dominated by unfavourably high frequencies with respect to the LISA’s sensitivity curve, the HS scenarios produce ringdowns that lie in the sweet spot of LISA’s power spectral density. Therefore, we foresee the best ringdown tests with LISA within the HS scenario.

Secondly, most studies assessing LISA’s potential for ringdown tests are based either on the \emph{detectability} of secondary modes (i.e., requiring that SNR in the secondary mode be above a given detectability threshold, see e.g., \cite{Berti:2016lat,2019PhRvD..99b4005B, 2020PhRvD.101h4053B,2020PhRvD.102b4023B}) or on their \emph{resolvability} from the fundamental mode (i.e., requiring that the frequencies and damping times of the two QNMs are sufficiently different to be measured unambiguously at a given SNR~\cite{Berti:2005ys,Berti:2007zu,2016PhRvD..94h4024B,Ota:2021ypb}). 
While useful, the information carried by these analyses is limited because it does not quantify the \emph{measurability} of different modes (i.e., the precision at which QNM measurements can be done and hence the quality of a BH spectroscopy test achievable)~\cite{2020PhRvD.101d4033B}. Our work focuses on measurability of QNM parameters.

For this study, we first simulate 100 years of data containing analytical ringdown signals for each of the MBH population models. We then estimate the uncertainty in the parameter estimation for the QNM spectra using a numerical Fisher matrix~(FM) formalism for up to 5 QNM parameters. We also identify the combinations of the QNM parameters that produce the smallest measurement errors for each of the MBH binary population. The relative measurement uncertainty of the QNM parameters, which we refer to as `measurability’ (following the definition in~\cite{2020PhRvD.101d4033B, JimenezForteza:2020cve}) decides how well we can constrain a putative modified theory or alternative remnant model with the observations, and how constraining the ringdown-based null tests of GR will be using these events. We present the details of our analysis framework in Sec.~\ref{sec:Framework}.

This paper is organized as follows -- In Sec.~\ref{sec:MBH-pop}, we present a brief discussion on the binary MBH population models used in this study and their implications for BH spectroscopy. Next, in Sec.~\ref{sec:Framework}, we outline our framework and define the notion of detectability, resolvability, and measurability as used in this work. Our results for the prospects of BH spectroscopy with LISA and ET are presented in Sec.~\ref{sec:results-lisa} and Sec.~\ref{sec:results-et}, respectively. Finally, we conclude with a discussion and future directions in Sec.~\ref{sec:conc}. 

\section{Exploring the MBH population in the context of ringdown} \label{sec:MBH-pop}
\subsection{MBH population models}
\label{sec:MBH-pop2}
We describe the population of MBH binaries targeted by LISA using the semi-analytic model of Ref.~\cite{EB2012}, with
successive improvements described in Refs.~\cite{sesana2014,antonini1,antonini2,BonettiII,bonetti2019,Barausse:2020mdt}. The model tracks the evolution of MBHs in their galactic hosts
as a function of cosmic time. Galaxies are modelled as
dark matter halos accreting chemically pristine gas from the intergalactic medium. This gas can either accrete to the centre of the halo along cold filaments (at high redshift or in low-mass systems)~\cite{Dekel2006,Cattaneo2006,Dekel2009}, or it can get shock-heated to the halo's virial temperature, then cooling down to the center of the halo. The cold gas accumulating in the center can then give rise to disk structures (because of conservation of angular momentum)~\cite{1998MNRAS.295..319M}, where star formation can take place and contribute to the gas chemical evolution. Galactic disks (gaseous or stellar) can also become unstable to bar instabilities, or be disrupted by galaxy mergers, thus forming gaseous and stellar spheroids (which can also undergo star formation, typically in bursts). On smaller scales, the model also includes additional components, such as nuclear star clusters~\cite{antonini1,antonini2}, a nuclear gas ``reservoir'' from which MBHs can accrete~\cite{Granato:2003ch}, as well as MBHs (for which mass and spin are consistently evolved under accretion and mergers). Feedback on the growth of structures is accounted for in the form of both supernova~(SN) explosions (which tend to quench star formation in low-mass galaxies)~\cite{fb1,fb2,fb3} and jets/disk winds from
active galactic nuclei~(AGNs), whose effect is dominant in large galaxies~\cite{Croton2006,2008ApJS..175..390H,2006MNRAS.370..645B}.
Besides suppressing star formation, both SN and AGN feedback also
eject gas from the nuclear region from which MBHs accrete. In particular, SN explosions may quench MBH accretion in systems with escape velocities $\lesssim 270$ km/s~\cite{habouzit}, thus hindering the growth of MBHs in shallow potential wells
and the hardening of MBH binaries (by suppressing gas-driven migration).

The halo merger history is followed by using an extended Press-Schechter formalism~\cite{PS}, modified to reproduce results from N-body simulations~\cite{Parkinson2008}.
Mergers of galaxies track those of halos, but with Delays accounting for: \textit{(i)}
the survival of the smaller halo within the larger one as a subhalo,
which is slowly dragged to the center of the system by dynamical friction
while undergoing tidal disruption and evaporation~\cite{boylankolchin,taffoni}; and for \textit{(ii)}
 the dynamical friction (including again tidal effects leading to disruption and evaporation)
 between the baryonic components~\cite{binneytremaine}.
 On smaller scales, the mergers of MBHs (when the latter are present) track those of galaxies, but again 
 with potentially significant Delays. These account
 for the evolution of MBH pairs at separations ranging 
 from kpc down to the binary's influence radius (including the effect
 of dynamical friction~\cite{2017ApJ...840...31D} and incorporating the results of hydrodynamic simulations~\cite{changa}). Furthermore, at the smaller separations where a bound binary forms, the Delays account for
 gas-driven migration (if a gaseous disk is present)~\cite{MacFadyen2008,Cuadra2009,Lodato2009,Roedig2011,Nixon2011,Duffel2019,Munoz2019},
 three-body interactions with stars (stellar hardening)~\cite{Quinlan1996,Sesana2015}
 and Kozai-Lidov and/or chaotic interactions with another MBH/MBH binary (when the latter are present as a result of an earlier galaxy merger)~\cite{Hoffman2007,BonettiI,BonettiII,BonettiIII,bonetti2019}. When a MBH binary finally reaches sufficiently small separations (depending on the mass, but typically $\sim 10^{-2}$--$10^{-3}$ pc), it is driven to coalescence by GW emission alone in less than a Hubble time. 
 When the merger happens, the MBH mass and spin are evolved using semi-analytic prescription reproducing the results of numerical-relativity simulations~\cite{Barausse:2012qz,Hofmann:2016yih}. GW-induced kicks are also accounted for~\cite{kick}, and can result in the ejection of the merger remnant from the host galaxy.
 
A crucial ingredient of the model is also given by the initial conditions for the MBH population at high redshift. In this work, we follow, e.g.,~Refs.~\cite{2016PhRvD..93b4003K,bonetti2019,Barausse:2020mdt,Barausse:2020gbp} and adopt two possible scenarios. In the \emph{LS scenario}, we assume that the MBH population grows from seeds of a few hundred $M_\odot$, produced as remnants of Pop~III stars~\cite{Madau2001}. In more detail,
we populate large halos collapsing from the $3.5\sigma$ peaks of the primordial density field at $z\gtrsim 15$ with seed BHs, whose mass we assume to be about  $2/3$ of the
initial Pop~III stellar mass (to account for SN winds). The star's initial mass
is drawn from a log-normal distribution peaking at $300 M_\odot$ and with root-mean-square of 0.2
dex (with an exclusion region between 140 and $260\,M_\odot$ to account for pair instability SNe). 

We also consider a \emph{HS scenario} where MBHs form already with mass $\sim 10^5 M_\odot$. For concreteness, we use the model of Ref.~\cite{Volonteri2008}, where seeds form from the bar-instability driven collapse of protogalactic disks
at high redshift ($z\gtrsim 15$) and in halos with spin parameter and virial temperature below critical threshold values.
In more detail,  these thresholds are provided by Eq. (4) -- with $Q_c = 3$ -- and Eq. (5) of Ref.~\cite{Volonteri2008}, while the seed mass is set by Eq. (3) of the same work. In reality, it is of course possible (if not likely) that both HSs and LSs form in nature, effectively leading to a  ``mixture'' of the two scenarios, which LISA will shed light on~\cite{Toubiana:2021iuw}.

In this paper, we consider several possible versions of both seed models. Besides the default \emph{SN-Delays} models in which
all the aforementioned physical ingredients are considered, in the
\emph{noSN} models we switch off the effect of SN feedback
 on the nuclear gas reservoir, while in the  \emph{shortDelays}
models we neglect the Delays occurring as MBHs move from  kpc to pc separations. 
In more detail, in the  \emph{shortDelays}
models we switch off the dynamical friction on the satellite galaxy and/or on its MBH, 
but we maintain the  Delays due to dynamical friction between the halos, and the Delays due to stellar hardening, gas-driven migration, and triple/quadruple interactions between MBHs.\footnote{The \textit{shortDelays} models correspond to those with the same name in Refs.~\cite{Barausse:2020gbp,Toubiana:2021iuw}, which are also the same as the \textit{noDelays}  models of \cite{Barausse:2020mdt}.}

\begin{figure*}[th]
\hspace*{-1cm}
\includegraphics[width=0.50\textwidth]{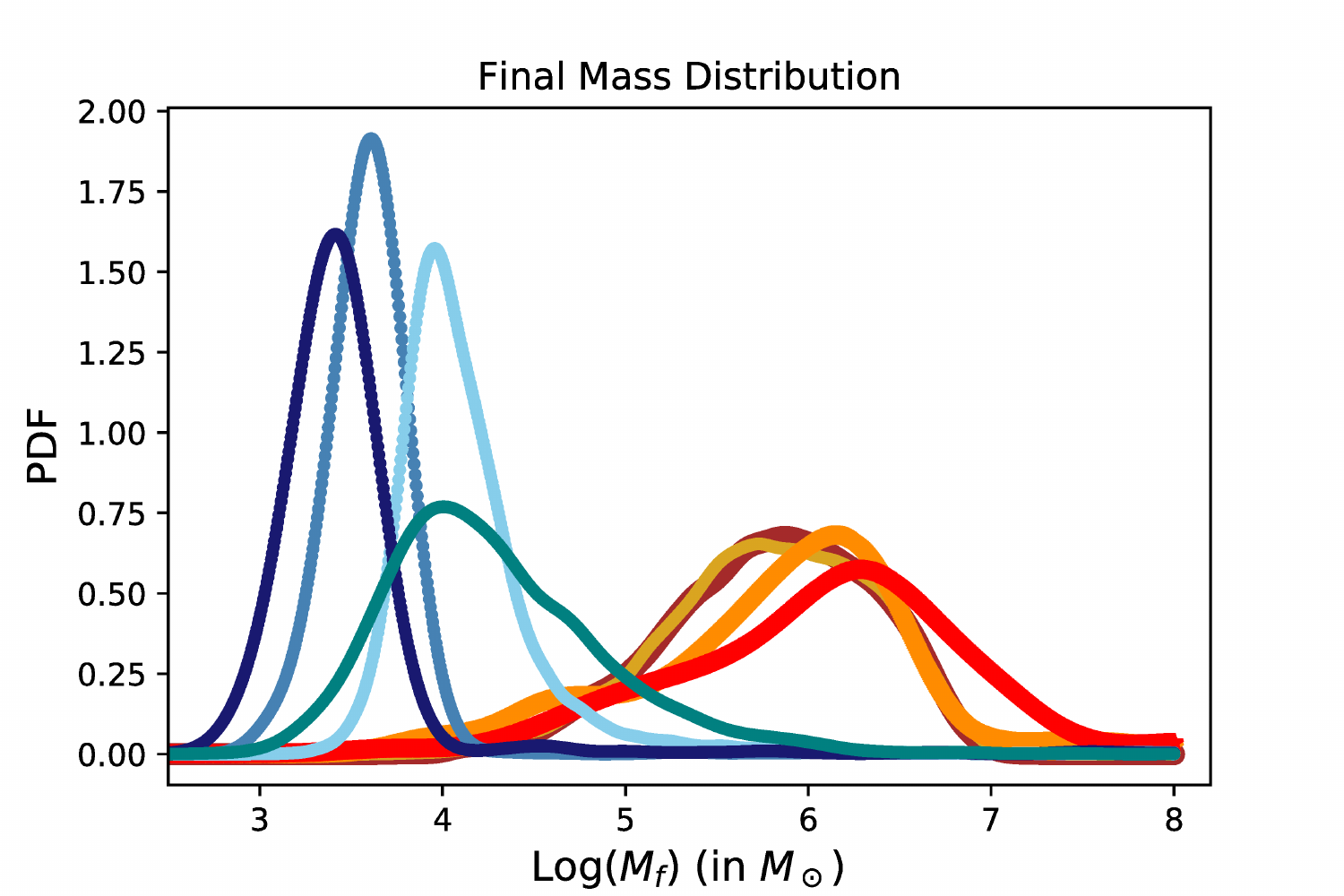} 
\hspace*{-1cm}
\includegraphics[width=0.50\textwidth]{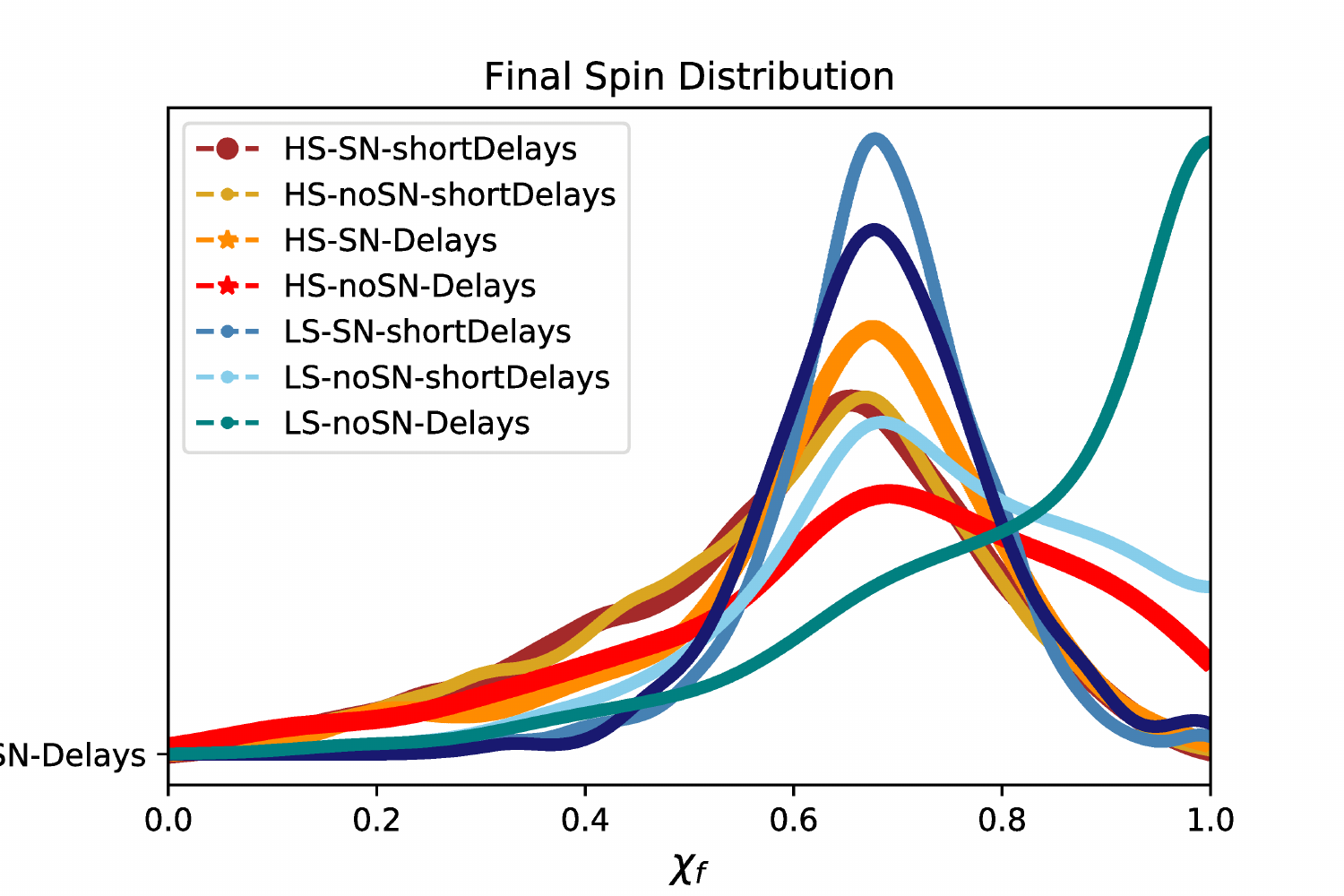} \\
\hspace*{-1cm}
\includegraphics[width=0.50\textwidth]{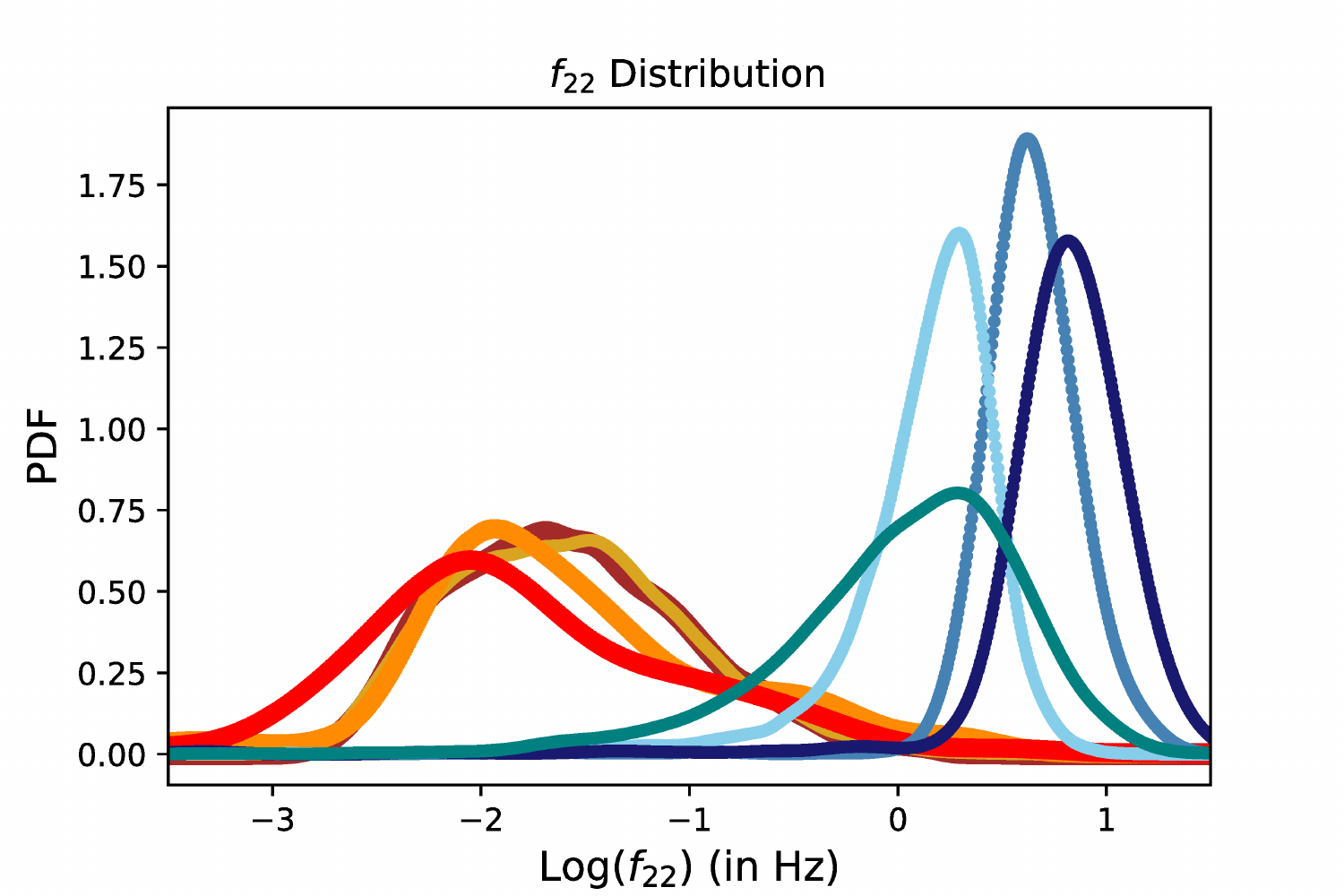} 
\hspace*{-1cm}
\includegraphics[width=0.50\textwidth]{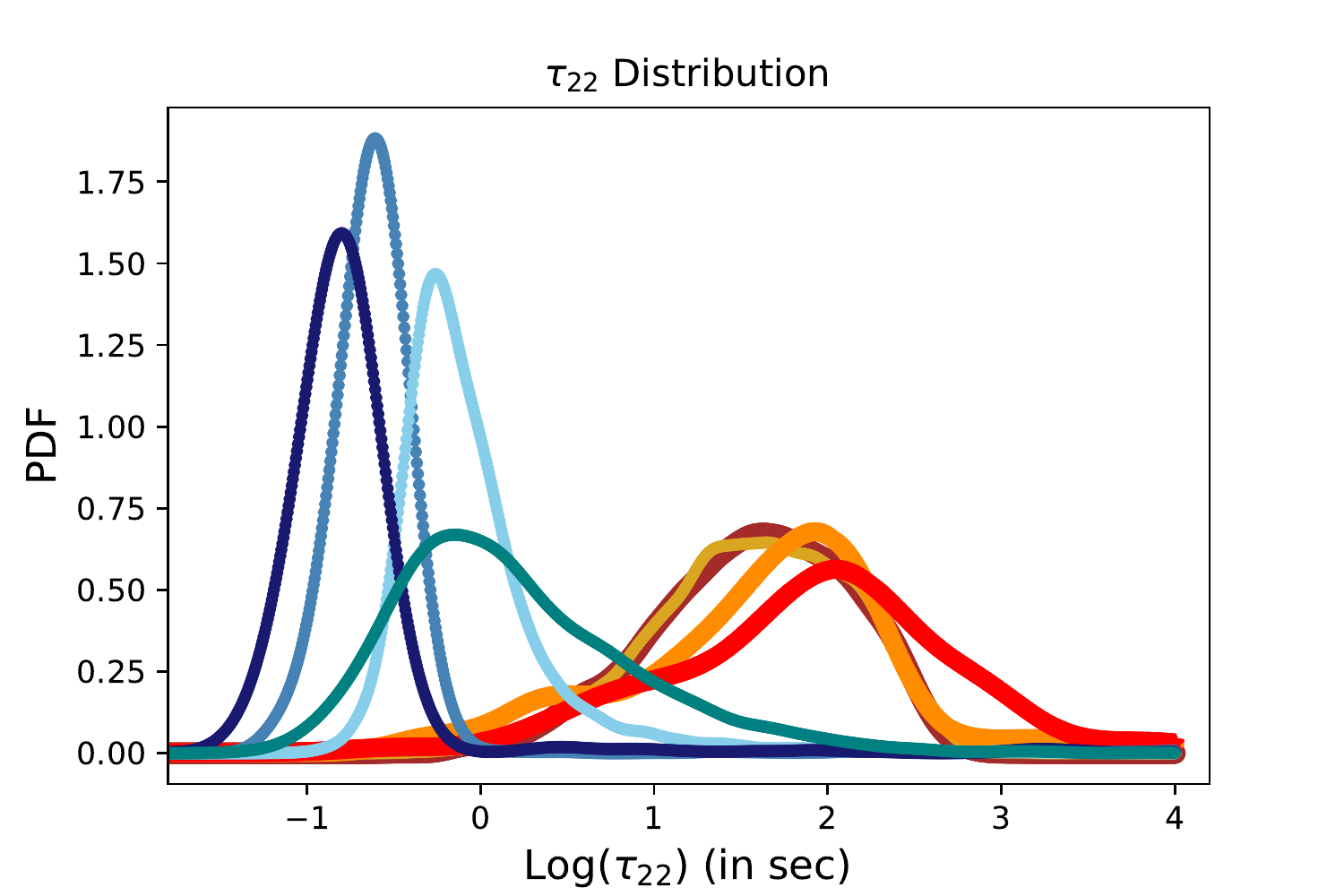}
\caption{Top panels: distributions of remnant detector-frame mass (left panel) and spin (right panel) for different MBH population models. Bottom panels: the corresponding distribution of the dominant QNM frequency (left panel) and damping time (right panel). 
} 
\label{fig:Freq-tau-22}
\end{figure*}

\subsection{MBH population in the context of BH spectroscopy}
In this section, we consider the implications of the different MBH populations on the ringdown analysis. The QNM spectrum depends dominantly on $M_{f}$ and subdominantly on $\chi_{f}$. The top panels of Fig.~\ref{fig:Freq-tau-22} show the distributions of $M_{f}$ and $\chi_{f}$ for the  binary MBH populations used in our study. Notice the clear segregation of the $M_{f}$ distributions for the LS (warm colored histograms) from the HS models (cool colored histograms). The $M_{f}$ distributions for all the HS models peak around $10^6 \Msun$ with a substantial support in $M_{f} \in [10^5 -10^7] \Msun$, whereas for LS models $M_{f} \in [10^3 - 10^6] \Msun$. 
While the seed mass dominantly decides the mass of the remnant, among the LS models we see that the models without SN feedback allow for a slightly higher mass of the remnant BHs, making them more promising LS models for BH spectroscopy with LISA. In the 
LS models accounting for SN feedback, the latter prevents efficient accretion and thus the growth of MBHs, giving rise to lighter remnants, with a $M_{f}$ distribution peaking around $\sim 10^3 \Msun$. Overall, depending on the underlying population model, the support for the remnant mass in a MBH binary merger can vary between $10^2 - 10^8 \Msun$.

All models except LS \emph{noSN-Delays} have  remnant spin distributions
that peak at $\chi_{f}\sim 0.68$, but have a broad support across $\chi_{f} \in [0.3,0.9]$; note that $\chi_{f} \sim 0.68$ corresponds to the spin of remnant BH for a nonspinning equal mass binary BHs, and that the $\chi_{f}$ distributions are consistent with a clustering of events close to $q \sim 1$ (see Fig.~\ref{fig:assym} below). LS \emph{noSN-Delays} has $\chi_{f}$ distribution railing against extremal spin, as the absence of SN feedback and the long time Delays allow accretion to spin the progenitor BHs up. We observe in our simulated populations that the systems that have nearly extremal spinning remnant also have the heavier progenitor BH with spin $\sim 1$. Nonetheless, we have checked that the results of our analysis are robust when we cap the BH spin to $0.9$.

Next, the QNM frequencies and damping-times scale inversely and linearly with $M_{f}$, respectively; we see how the features in the final mass distribution translate to the QNM frequencies and damping times in the bottom panels of Fig.~\ref{fig:Freq-tau-22}. In particular, the distribution of $f_{22}$ determines whether the QNMs fall within the sensitivity range of a given instrument. For the HS models, the distribution of $f_{22}$ peaks at $\sim 10\,{\rm mHz}$ and has a support approximately in the range $[1\,{\rm mHz},1\,{\rm Hz}]$. Note that the LISA power spectral density~\cite{Babak:2021mhe, Robson:2018ifk} has a sweet spot at $\sim 10$ mHz and is therefore perfectly suited for a ringdown analysis in these scenarios.\footnote{
Interestingly, overall the distributions have small support at $f_{22}<1\,{\rm mHz}$,
as expected from the exponential suppression of the high end of the MBH mass function. This suggests that the details of the LISA power spectral density at low frequency are not relevant for this analysis.}
LS models on the other hand have $f_{22}$ peaking at $\sim 5$ Hz. with support in $f_{22} \in [10^{-1}, 50]\,{\rm Hz}$, making them suboptimal for BH spectroscopy with LISA.  For example, a remnant BH with $M_f=10^{3} \Msun$ and $\chi_f=0.68$ will produce a ringdown at $\sim 17\,{\rm Hz}$, outside the LISA sensitivity band. Note also that the HS models have $\tau_{22} \in [1, 10^{3}]\,$s with a peak around $100\,$s, while the LS models have $\tau_{22} \in [0.05,5]\,$s. This means that typically the ringdown will last significantly longer than the ringdowns currently detected in the LIGO-Virgo band, and it is more likely to be contaminated by other simultaneous signals from other sources.

\begin{figure*}
\includegraphics[width=0.47\textwidth]{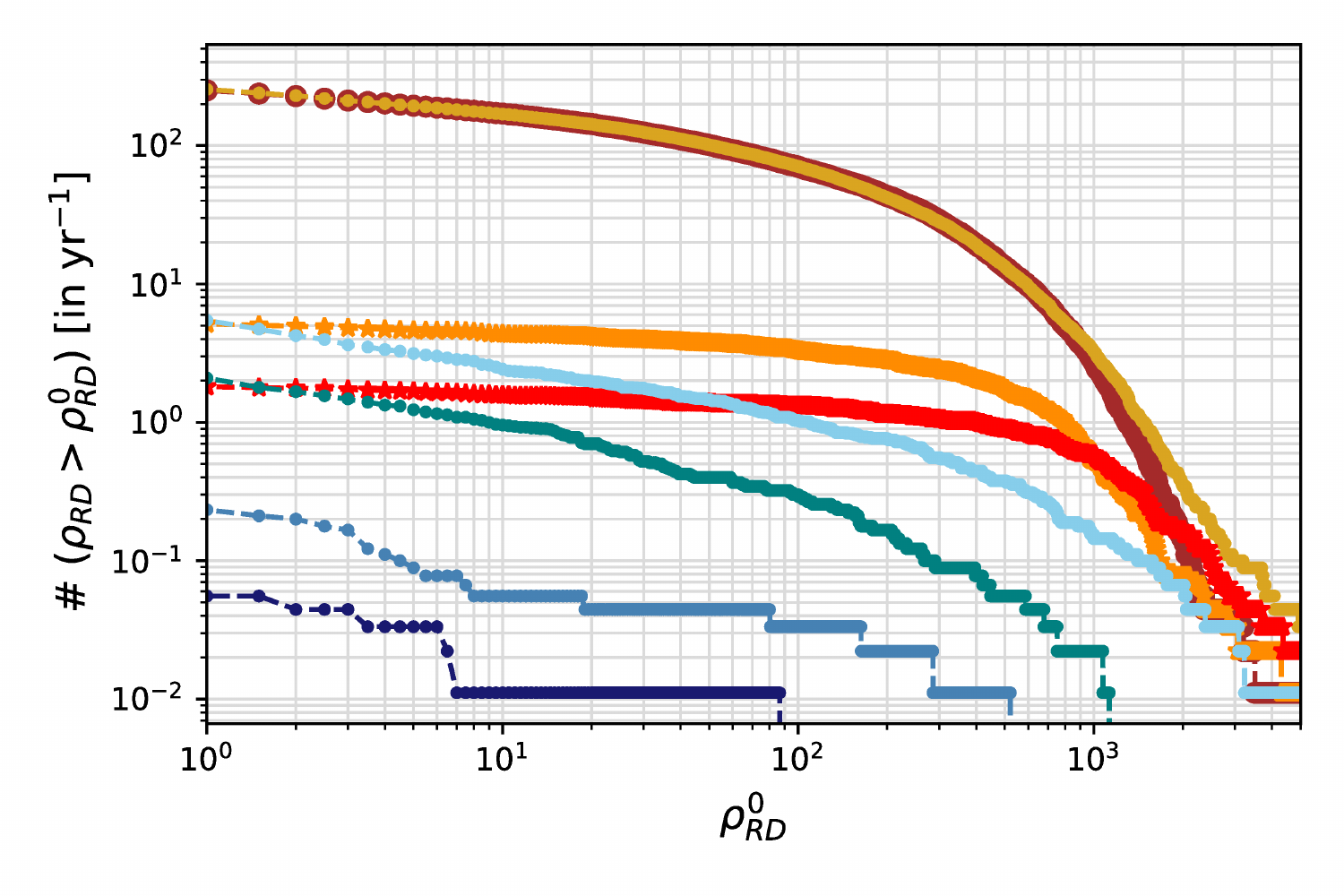}
\includegraphics[width=0.47\textwidth]{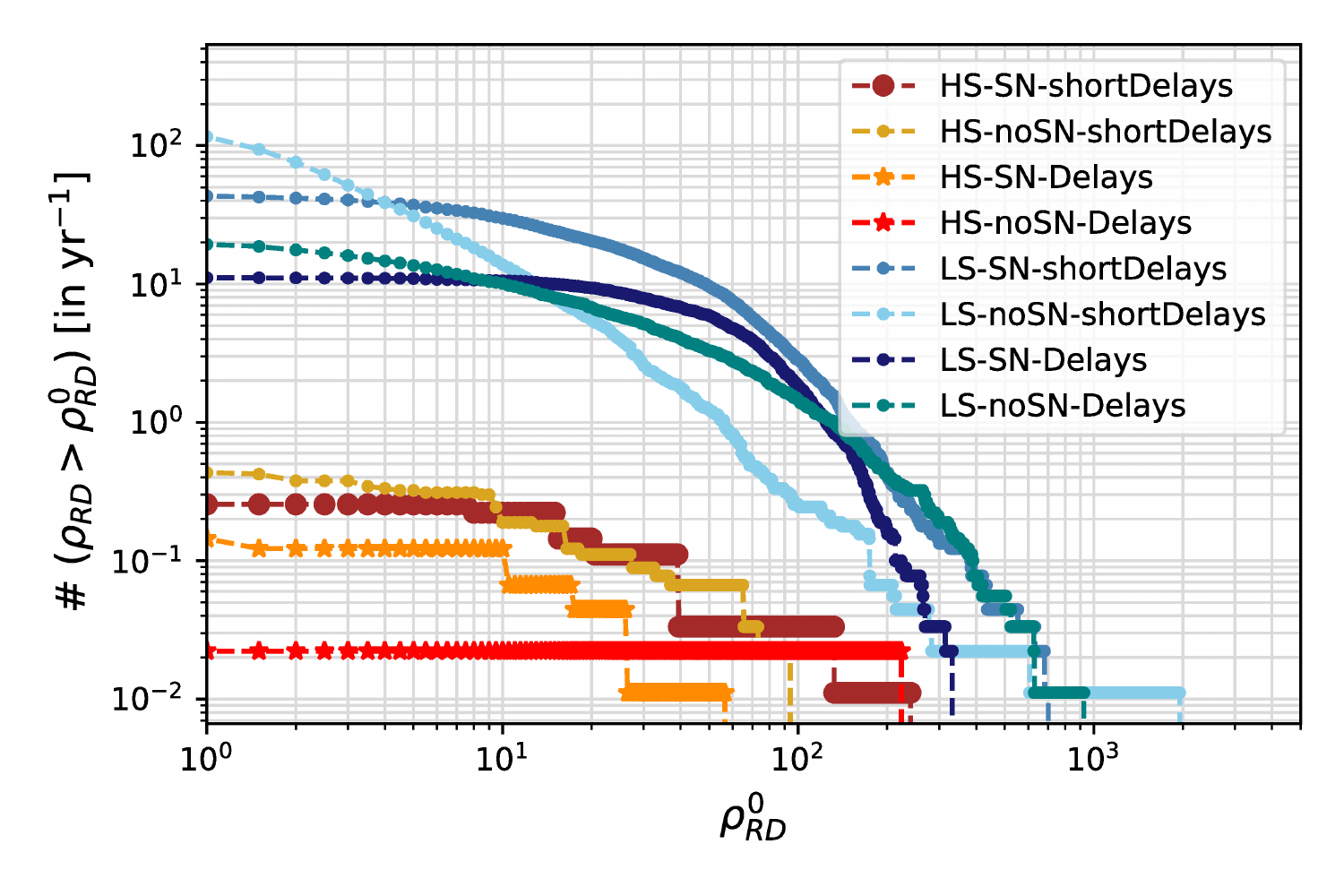}
\caption{Cumulative number of events/yr with $\rho_{\rm rd}$ greater than or equal to a threshold SNR indicated on the $x$-axis of the plot. We plot the optimal ringdown SNR. The left and right panels refer to the case of a LISA and ET detection, respectively. The warm (cold) color scheme indicates the HS (LS) models.  We see that LISA and ET are complementary: LISA will be sensitive to ringdown signals mostly for the HS scenarios, whereas it could detect only the high-mass tail of the merger remnants in the LS scenarios. On the other hand, ET will be insensitive to the HS populations but will detect the low-mass tail of the remnants in the LS scenario.}
\label{fig:snr-cumu}
\end{figure*}

In Fig.~\ref{fig:snr-cumu}, we show the number of events/yr with $\rho_{\rm rd} > \rho^{0}_{\rm rd}$ for LISA (left panel) and ET (right panel). (Details of the SNR calculation and noise curves are given in the next section.) The number of events above a certain SNR in a detector will depend prominently on the mass-spin distribution of the remnant BH, distance of the system, and the intrinsic event rate predicted in a population model.
HS models are optimal for spectroscopy using LISA, with the \emph{shortDelays} models predicting $\sim 2$ events/yr and \emph{Delays} models predicting $\sim 1$ event/yr with $\rho_{\rm rd}$ as large as 1000. Furthermore, HS \emph{Delays} models have a low event rate and LISA will see $\sim 5$ events/yr with $\rho_{\rm rd} \in [1,1000]$, while for the HS \emph{shortDelays} models LISA will see $\sim 100-200$ events/yr with $\rho_{\rm rd} \in [100, 1000]$. Refer to Table~\ref{tab:events-per-year} for the intrinsic merger rates predicted by the various population models.

\begin{table}[]
\resizebox{0.45\textwidth}{!}{%
\begin{tabular}{|c|c|c|}
\hline
\rowcolor[HTML]{C0C0C0} 
\textbf{Seed} & \textbf{Population Models } & \textbf{Total number of BBH events/yr} \\ \hline
\textbf{}     & SN-shortDelays               & $\sim 317/{\rm yr}$                      \\ \hline
\textbf{HS}   & SN-Delays                       & $\sim 6/{\rm yr}$                        \\ \hline
              & noSN-shortDelays            & $\sim 322/{\rm yr}$                      \\ \hline
\rowcolor[HTML]{FFFFFF} 
              & noSN-Delays                    & $\sim 2.5/{\rm yr}$                      \\ \hline \hline
              & SN-shortDelays               & $ \sim 45/{\rm yr}$                       \\ \hline
\textbf{LS}   & SN-Delays                        & $\sim 12/{\rm yr}$                       \\ \hline
              & noSN-shortDelays             & $\sim 290/{\rm yr}$                       \\ \hline
              & noSN-Delays                    & $\sim 45/{\rm yr}$                      \\ \hline
\end{tabular}%
}
\caption{Total intrinsic event rates for massive binary mergers in the catalogs, obtained by averaging 100 realizations of the population predicted by each of the 8 SMBH population models considered in this work, and ignoring SNR thresholds.}
\label{tab:events-per-year}
\end{table}

However, the HS \emph{Delays} models have events with $\rho_{\rm rd}$ distribution sharply peaked around $\rho_{\rm rd} \sim 10^{3}$ with a long tail towards  lower SNR, while the HS \emph{shortDelays} models have a $\rho_{\rm rd}$ distribution with a broad support in $\rho_{\rm rd} \in [1,500]$. We also see from our simulated populations that the HS \emph{Delays} models merge at closer luminosity distance distribution ($\sim 1-50$~Gpc) compared to the \emph{shortDelays} ($\sim 25-125$)~Gpc ones. Therefore, although the HS \emph{Delays} models predict lower event rates, the events they produce are likely to have a very loud ringdown in LISA and serve as golden events for spectroscopy.

As for the LS scenario, LISA might see just a few events for LS \emph{noSN} models -- a few events with $\rho_{\rm rd} \in [10,100]$ for \emph{noSN-shortDelays}, and $\sim 1$ or 2 events/yr with $\rho_{\rm rd} \sim 10$ for \emph{noSN-Delays}. From the distribution in Fig.~\ref{fig:Freq-tau-22}, we see that both the \emph{noSN} models have $M_{f}$ peaking close to $10^{4} \Msun$ but the \emph{noSN-Delays} model has a tail that extends further to higher mass compared to \emph{noSN-shortDelays}. Furthermore, the \emph{noSN-Delays} model has a closer redshift distribution compared to \emph{noSN-shortDelays} (the former has a considerable support for $z \in [1,10]$ and the latter for $z \in [1,20]$). Naively, it seems puzzling that in Fig.~\ref{fig:snr-cumu} the \emph{noSN-Delays} model shows lower rates than the \emph{noSN-shortDelays} model. This can be attributed to the intrinsic event rates in each of the models (see Table~\ref{tab:events-per-year}) — while the \emph{noSN-Delays} model predicts 1.5 mergers/yr, the \emph{noSN-shortDelays} model predicts $\sim 290$ events/yr. Subleading to this is the fact that $\chi_{f}$ of the remnant BH formed in \emph{noSN-Delays} population is higher, shifting $f_{22}$ to a slightly higher frequency. 
Furthermore, LISA will be insensitive to ringdowns produced by the LS models that incorporate  SN feedback, as those scenarios predict light remnants. Note also that the worst-performing HS model (i.e., \emph{noSN-Delays})
yields rates comparable to the best case LS model (i.e., \emph{noSN-shortDelays}).

As shown in Fig.\,\ref{fig:snr-cumu},
LISA and ET will be complementary with respect to the HS versus LS scenarios. Indeed, the right panel of Fig.\,\ref{fig:snr-cumu} shows that ET is insensitive to the ringdown produced by all the HS models, but will have interesting rates in the LS scenarios.
In particular, the LS \emph{noSN-shortDelays} model has significantly smaller rates ($\leq 10$ events/yr with $\rho_{\rm rd}\geq10$) compared to the other LS models. However, the latter have comparable $\rho_{\rm rd}$ -- i.e., $\sim 1-3$ events/yr with $\rho_{\rm rd}\geq100$ and 10-20 events/yr with $\rho_{\rm rd}\geq10$. 

Finally, in order to perform BH spectroscopy, the subdominant modes have to be sufficiently excited. The amplitude of the QNM excitation depends on the perturbation conditions setup before the ringdown, which in turn depend on the asymmetry of the binary BH system. Generally, to excite the odd angular modes, an \emph{asymmetric binary} (i.e., with either a mass ratio $q$ sufficiently different from unit\footnote{Note, however, that in the large-$q$ limit, the total energy output of the ringdown decreases for a fixed total binary mass~\cite{Berti:2007fi}. For instance, the EMOP energy in the subdominant modes peaks at $q={\cal O}({\rm a~few})$ depending on the mode and on individual spins~\cite{Baibhav:2017jhs}.} or sufficiently large component spins\footnote{As a direct measure of spin-induced asymmetry of the binary, we shall use the effective spin $\chi_{\rm eff}\in[-1,1]$, which is the mass-weighted
average of the individual dimensionless component spins $\chi_1$ and
$\chi_2$, projected along the unit vector parallel to the
binary's orbital angular momentum~\cite{Ajith:2009bn}.}) is needed.  In Fig.~\ref{fig:assym}, we plot the asymmetry of the MBH binary on the $q-\chi_{\rm eff}$ plane and find that all models predict fairly asymmetric binary systems, with a considerable spread in $q$ and/or in $\chi_{\rm eff}$. 
Furthermore, we note that while all the HS models produce binaries with a large range of $\chi_{\rm eff}$, the \emph{shortDelays} HS models also produce more binaries BH events with large $q$, making them more promising for BH spectroscopy. (Note that even though the probability density functions of $q$ for the HS models are similar -- c.f. Fig.~11 of Ref.~\cite{Barausse:2020gbp} -- the \emph{shortDelays}  models predict a higher event rate, thus sampling the large-$q$ tails of the distribution more efficiently.)

\begin{figure*}
\includegraphics[width=0.45\textwidth]{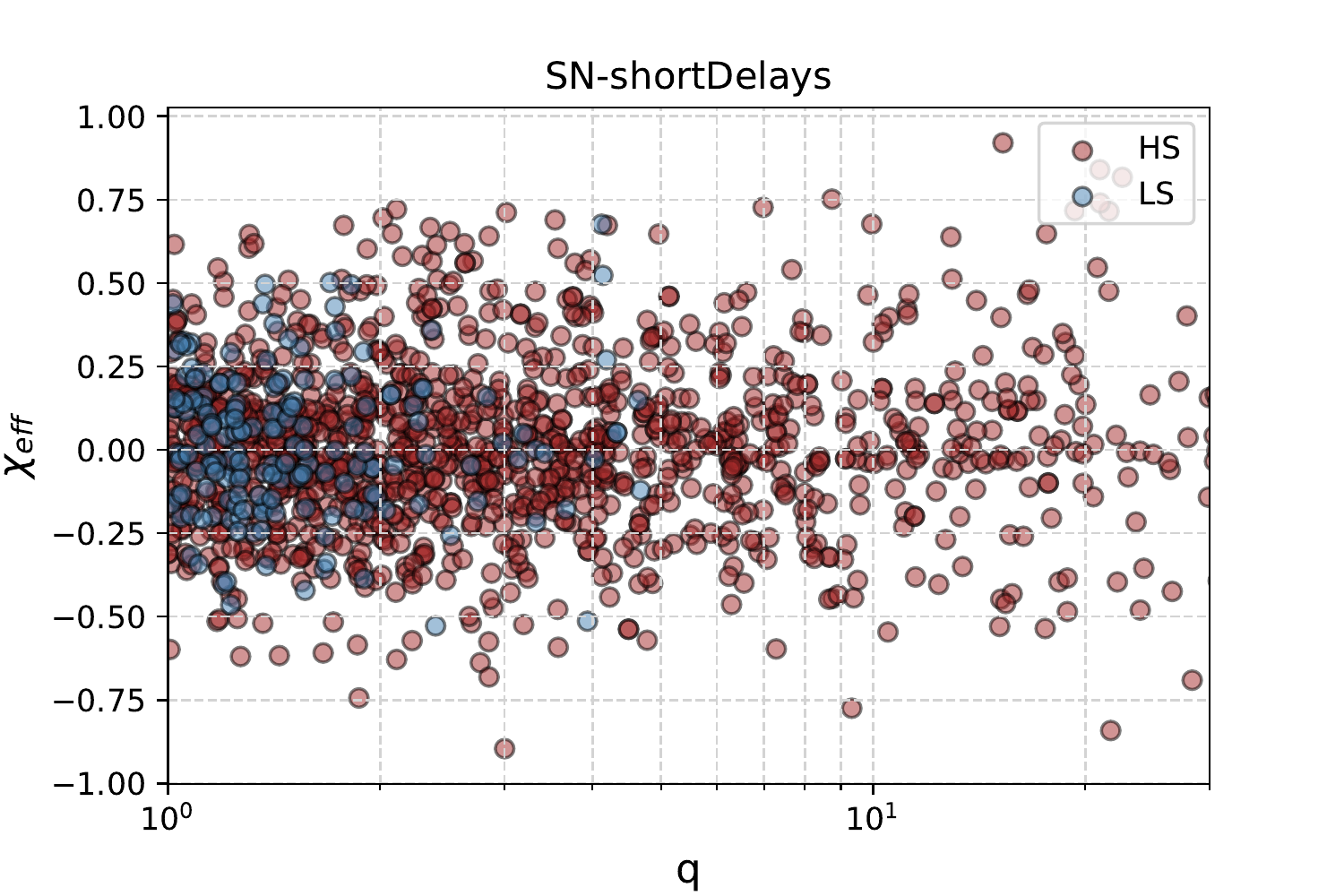}
\includegraphics[width=0.45\textwidth]{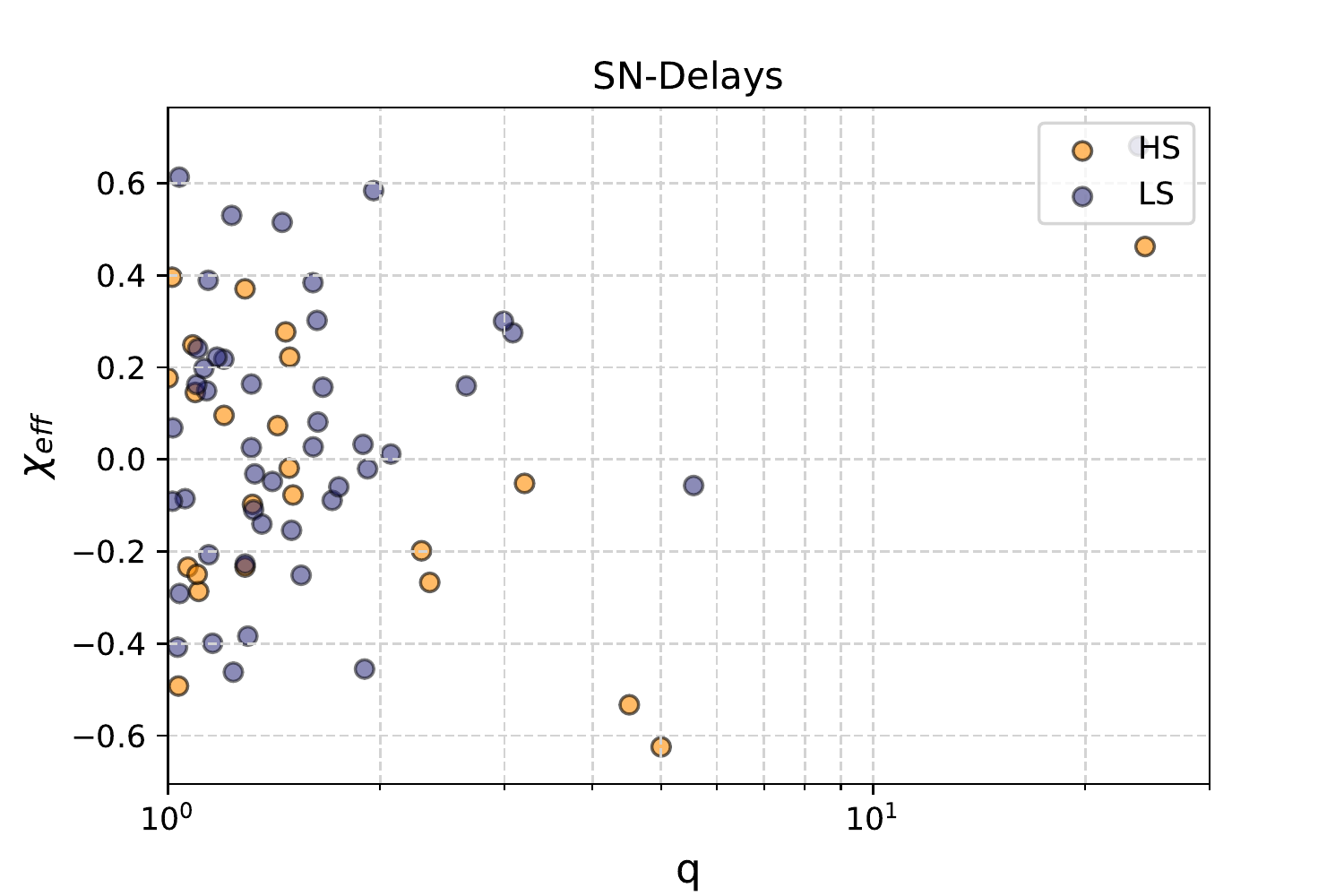} \\ 
\includegraphics[width=0.45\textwidth]{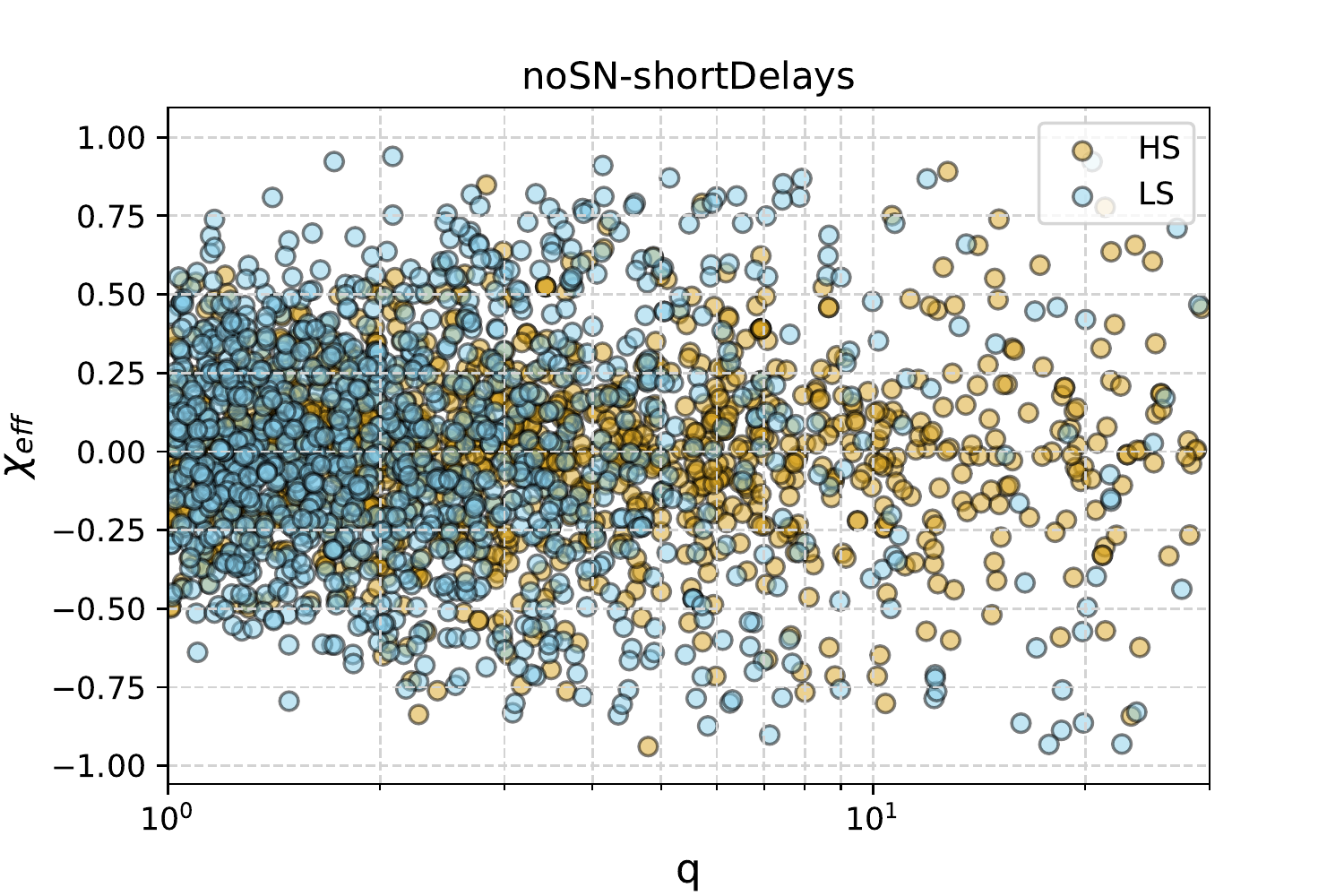}
\includegraphics[width=0.45\textwidth]{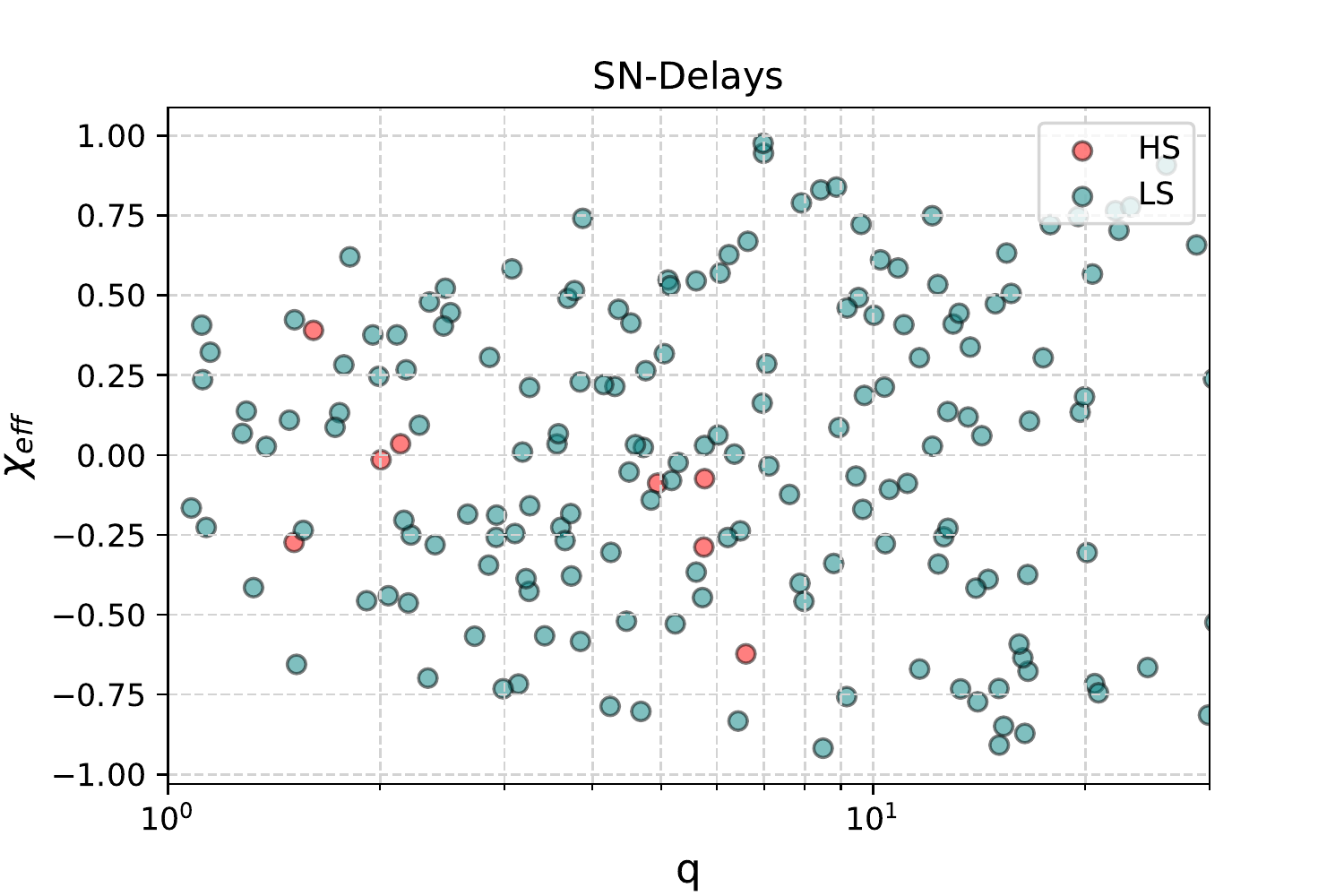}
\caption{Realizations of the binary mass ratio $q$ and effective spin $\chi_{\rm eff}$ for the HS and LS models. Here we plot a scatter of events (intrinsic) corresponding to a 4-year data realization. The color scheme in each panel follows that of Figs. \ref{fig:Freq-tau-22} and \ref{fig:snr-cumu}.}
\label{fig:assym}
\end{figure*}

\section{Framework}\label{sec:Framework}

\subsection{Computation of measurement uncertainties}
\label{sec:Fisher}

We model the ringdown waveform as a linear superposition of damped sinusoids with individual mode amplitudes ${\cal A}_{lmn}$, phases $\phi_{lmn}$, and the QNM frequencies and damping times $\{f_{lmn},\tau_{lmn}\}$ as independent parameters:
\begin{equation}
\label{eq:rdmodel}
\begin{split}
    &h(t)=h_+(t)+ih_\times(t)\,,\\
    &h_{+}(t)=\sum_{lmn}{\cal A}_{lmn}\cos\left(2\pi f_{lmn}t+\phi_{lmn}\right)e^{-t/\tau_{lmn}}\,\mathcal{Y}^{lm}_{+}(\iota)\,,
    \\
    &h_\times(t)=\sum_{lmn}{\cal A}_{lmn}\sin\left(2\pi f_{lmn}t+\phi_{lmn}\right)e^{-t/\tau_{lmn}}\,\mathcal{Y}^{lm}_{\times}(\iota)\,,\\
\end{split}
\end{equation}
where $\iota$ is the inclination angle of the remnant spin and the $\{+,\times\}$ harmonics are defined in terms of spin-weighted spherical harmonics as \cite{q-from-rd,Berti:2007zu}
\begin{equation}
    \mathcal{Y}_{+, \times}(\iota) = \tensor[_{-2}]{\mathcal{Y}}{^{lm}}(\iota,0) \pm (-1)^l\tensor[_{-2}]{\mathcal{Y}}{^{l-m}}(\iota,0)\,.
\end{equation}

Here, the integers $\{l,m,n\}$ correspond to the multipolar, azimuthal, and overtone indices, respectively. For each $(l,m)$, we focus only on the fundamental ($n=0$) tone, neglecting the overtones. This is motivated by two considerations: a)~resolving overtones from the fundamental mode and measuring them is challenging~\cite{Bhagwat:2019dtm,JimenezForteza:2020cve}; b)~For MBH binary populations, the initial binary BH systems are typically asymmetric (either $q>1$ or $\chi_{\rm eff}\neq0$, see Fig.~\ref{fig:assym}) and the subdominant angular modes (unlike with equal mass systems) are sufficiently excited~\cite{Bhagwat:2019dtm,Ota:2019bzl,JimenezForteza:2020cve}. Thus, for MBH binaries, we expect BH spectroscopy with multiple angular modes to be more practical. Henceforth, we drop the index $n$ for simplicity.

We estimate the statistical uncertainties in the measurements of the QNM parameters, i.e., frequency and damping time, for each event in a population of MBHs. Let $\Theta = \bigcup\limits_{lm} \{f_{lm}, \tau_{lm},{ {\cal A}}_{lm},\phi_{lm}\}$ be the set of independent parameters of the ringdown model in Eq.~\eqref{eq:rdmodel}.
In this agnostic form, the ringdown waveform depends on $4$ real parameters (frequency, damping time, amplitude and phase) for each angular mode included in its modelling. In our analysis, we include up to 4 of the loudest modes, i.e., $(l,m)=(2,2),(3,3), (2,1), (4,4)$.

To compute the uncertainties, we use an FM formalism~\cite{TheFisher}, which provides an accurate estimate of the statistical errors in the high-SNR limit with Gaussian noise and in the absence of errors from waveform systematics~\cite{Use-and-abuse-of-FM}. 

For a given waveform $h(t)$ defined in terms of $\Theta=\{\theta_i\}$ parameters, the FM is defined as
\begin{equation}
 \Gamma_{ij}=\Big\langle\frac{\partial h}{\partial \theta_i}\Big\vert\frac{\partial h}{\partial \theta_j}\Big\rangle_{\theta=\hat{\theta}}\,,\label{eq:FM}
\end{equation}
where $\hat{\theta}$ are the true (injected) values, and the scalar product is defined as
\begin{equation}
    \langle h_1\vert h_2\rangle=4\Re\int_{f_{\rm{min}}}^{f_{\rm max}}\frac{\tilde{h}_1(f)\tilde{h}^*_2(f)}{S_n(f)}df \,,
\end{equation}
where the detector spectral density is denoted by $S_n$ and $\tilde h$ is the Fourier transform of the signal. The statistical error on the $i$-th parameter is then given by $\sigma_{\theta_{i}}=\Sigma^{1/2}_{ii}$, where ${\bf \Sigma} = {\bf \Gamma}^{-1}$. 

We evaluate Eq.~\eqref{eq:FM} numerically using directly Eq.~\eqref{eq:rdmodel} and a Python implementation based on the \texttt{sympy} package~\cite{SYMPY}. We also assume that different angular modes are independent and thereby ignore the mixing.~\footnote{Note that this assumption would break for different overtones of the same angular mode~\cite{Berti:2005ys,Bhagwat:2019dtm,JimenezForteza:2020cve}.} 
We adopt the LISA SciRD\footnote{We neglect the white-dwarf confusion noise, which anyway affects signals with frequency smaller than $\approx 1\,{\rm mHz}$. As shown in Fig.~\ref{fig:Freq-tau-22}, the ringdown in the catalogs is always at higher frequencies due to the lack of very massive remnants. Therefore, including the white-dwarf confusion noise would not change our results.} power-spectral density~\cite{Babak:2021mhe} and the ET-D configuration~\cite{Hild:2010id}, respectively.
In our implementation of the FM, we perform an average over the sky-position and polarization angle. For LISA, this average is incorporated in the power spectral density~\cite{Robson:2018ifk,Babak:2021mhe}, while for ET we explicitly set $ \langle F_{+}^{2} \rangle =  \langle  F_{\times}^{2}  \rangle = 1/5 $, where $F_{+, \times}$ are the sky-position dependent detector response function. For further details refer to the formalism in~\cite{Berti:2005ys}.

Next, the ringdown SNR is defined as $\rho_{\rm rd}=\langle h\vert h\rangle^{1/2}$. We use the expression of $\rho_{\rm rd}$ given by Eq.~(17) of \cite{2019PhRvD..99b4005B} and assume that the remnant is optimally inclined at $\iota=0$, while still averaging over sky position and polarization angle.

We interpolate the tabulated data provided in~\cite{Berti:data} to estimate the Kerr QNMs,  $\{f_{lm}, \tau_{lm}\}$. 
As shown in Fig.~\ref{fig:assym}, all the models predict a significant spread in the distribution of $\chi_{\rm eff} \in \{-0.75,0.75\}$.
Therefore, accounting for the effect of \emph{both} $q$ and the binary component spins
in the excitation amplitude ${\cal A}_{lm}$ is crucial for assessing the prospects for BH spectroscopy. For example, even an equal-mass BH binary with nonzero $\chi_{\rm eff} $ could excite measurable subdominant QNMs.

We compute the amplitude ratios $\mathcal{A}_{lm}^R = {\cal A}_{lm}/A_{22}$ 
for each mode using the ringdown energy $\mathscr{E}_{lm}$ (also known as EMOP energy) fits presented in~\cite{Baibhav:2017jhs}.\footnote{Note that, while most QNM amplitude fits in literature (e.g.,~\cite{London:2014cma,JimenezForteza:2020cve}) generally assume that all the QNMs start simultaneously, in~\cite{Baibhav:2017jhs} the start time for each mode is separately obtained by maximizing the EMOP energy. While this is a desirable feature, the amplitude calculated from the EMOP energy does not contain information about the relative phase difference between the modes. However, as long as we are looking at different angular modes and not overtones, the modes are fairly independent of each other and the errors $\sigma_{\theta_{i}}$ on the QNM parameters are almost independent of the relative phase difference between modes. We refer to~\cite{Bhagwat:2019dtm,JimenezForteza:2020cve} for more details on this effect.} To compute the amplitude we note that
\begin{align}
    \mathscr{E}_{lm} (q,\chi_{1},\chi_{2}) \propto \int \frac{\partial h_{lm}}{\partial t} \frac{\partial h^{*}_{lm}}{\partial t} dt \,,
\end{align} 
and that the ringdown energy is proportional to the square of amplitudes, namely
\begin{align}
\label{eq:amplitude-ratio}
{\cal A}_{lm}  \propto \sqrt{\mathscr{E}_{lm} \frac{2\tau_{lm}}{1+4\pi^2 f_{lm}^2\tau_{lm}^{2}}}\,.
\end{align}

We use the analytical fitting formula provided in~\cite{gossan-et-al} for the dominant mode amplitude,
\begin{align}
\label{eq:A22_form}
    {\cal A}_{22}=0.864 \frac{q}{(1+q)^{2}}\,.
\end{align}
The expression for ${\cal A}_{22}$ neglects the dependence on $\chi_{\rm eff}$, whose inclusion affects the ringdown SNR by only a small overall factor and is not crucial for a population-based statistical analysis (for instance, see the variation of  ${\cal A}_{22}$ in Fig.~1 of  \cite{Kamaretsos:2012bs}). Finally, without loss of generality, we set $\phi_{22}=0$. 

Although in Fig.~\ref{fig:assym} we show the $q$-axis up to $q=15$ for the sake of visualization, there is a small fraction of events with very large mass ratio (some as large as $q = 10^3$, see Fig.~11 in Ref.~\cite{Barausse:2020gbp}).
The EMOP energy fits in \cite{Baibhav:2017jhs} that we use to calculate the amplitudes are calibrated against numerical relativity waveforms for reasonably low $q$, as well as against extreme mass-ratio waveforms obtained by perturbation theory. Furthermore, the angular modes excitations approximately reach the test particle limit ($q \to\infty$) already at $q\sim 10$ (see Fig.~6 of \cite{JimenezForteza:2020cve}). In other words, the dependence of $\mathcal{A}_{lm}$ at large $q$ is not significant, and we do not expect any fitting errors to influence our results.

\subsection{Detectability, resolvability, and measurability}

The simplest quantifier one could use to assess the impact of secondary QNMs is to check for their \emph{detectability} by setting some SNR threshold for the secondary mode. However, detectability alone is not informative about the ability to \emph{resolve} multipole modes or \emph{measure} their frequencies and damping times, which is at the root of performing BH spectroscopy.

We use two criteria to quantitatively assess the prospects of BH spectroscopy~\cite{JimenezForteza:2020cve}: a)~\emph{resolvability}, and b)~\emph{measurability} of the QNM parameters $\{f_{lm}, \tau_{lm} \}$. The resolvability criterion ensures that the measured QNM parameters can be resolved from each other unambiguously by requiring that the estimated posterior distributions for the QNM parameters satisfy a minimum separation demanded by the Rayleigh criterion~\cite{Berti:2007zu}. The QNM parameters $\{ \theta_{i}, \theta_{j} \} \in \Theta$ are resolvable if the standard deviations of their posterior ($\sigma_{\theta_{i}},\sigma_{\theta_{i}}$) satisfy the following condition,
\begin{align}
         \max[\sigma_{\theta_{i}}, \sigma_{\theta_{j}}]&<|\hat{\theta}_{i} - \hat{\theta}_{j}|,
\end{align}\label{eq:res}
where $\hat{\theta}$ is the most likely estimate (or the true  value, in the case of FM) of the parameter $\theta$.

However, resolvability alone is not adequate to gauge the prospects of BH spectroscopy; we need to quantify our ability to measure the QNM parameters. For instance, two QNM parameters that are sufficiently far apart could be resolved even in a weak signal, but their measurement can have large uncertainties, and it can be poorly informative for a no-hair theorem test. Thus, following~\cite{JimenezForteza:2020cve}, we define measurability $\Delta \theta_{i}$ as the relative statistical uncertainty in the measurement of the parameters $\theta_i$. An event has $x \%$ measurability if it satisfies the following criterion:
\begin{equation}
    \label{eq:percent:def}
    {\rm max}_{_{i}}\left[\Delta \theta_{i}\right] < \frac{x}{100}, \qquad \forall \{{\theta_{i}}\} \in \Theta\,.
\end{equation}
where $ \Theta$ are the set of parameters we are interested in measuring and $\Delta \theta_{i} = \sigma_{{\theta}_{i}}/\rm {\hat{\theta}}_{i}$ . In our study, we shall hierarchically check for resolvability and measurability for each event and for different combinations of QNM quantities. For a signal with $x \%$ measurability, \emph{all} the QNM parameters under consideration are measured with \emph{at least} $x\%$ relative precision. We stress again that it is the measurability of the QNM parameters that determine the precision of BH spectroscopy, and the extent to which one can constrain modified-gravity theories or validate the Kerr metric through null tests.

\begin{figure}
\includegraphics[width=0.49\textwidth]{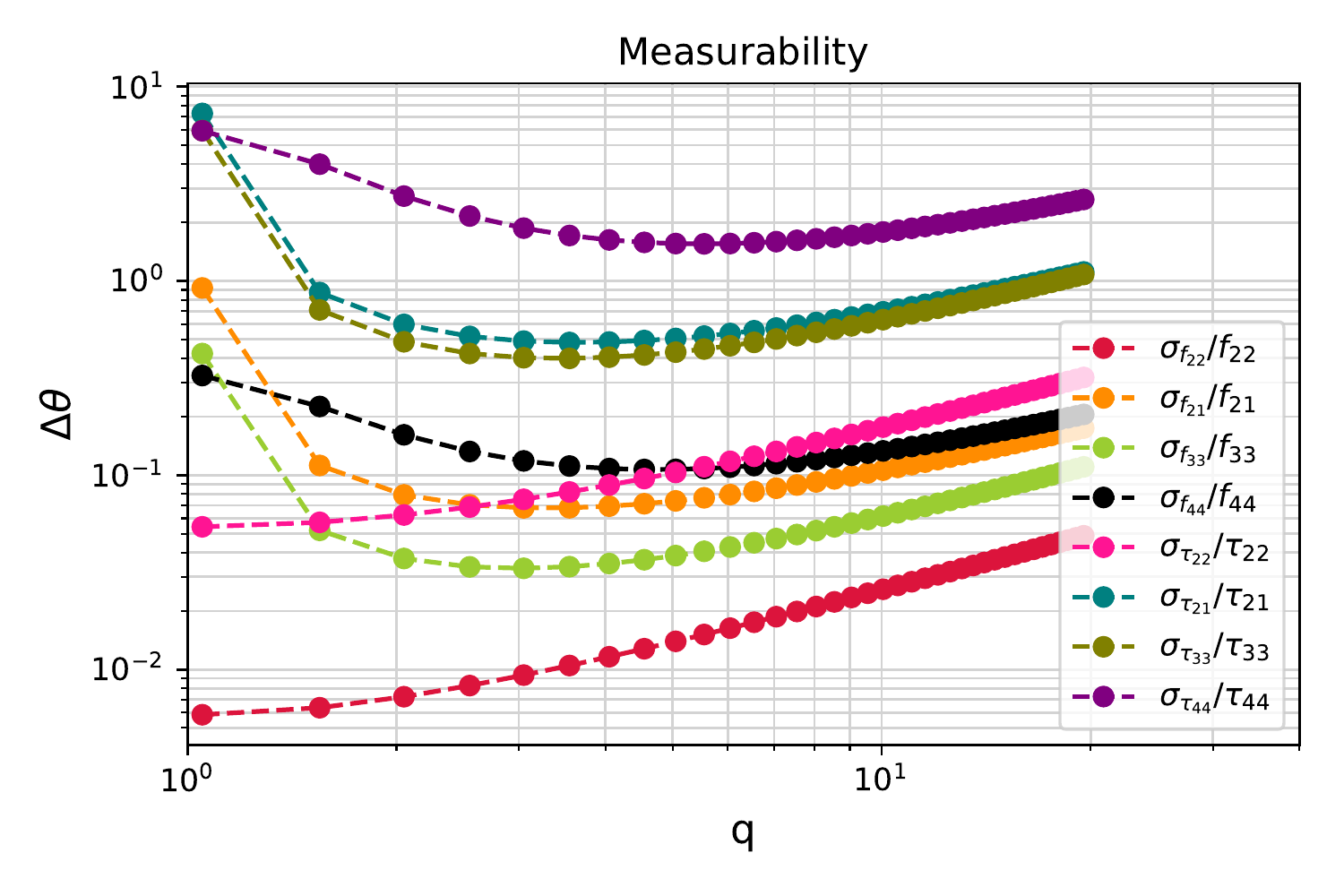}
\caption{Measurability of various QNM parameters for a MBH at $z=10$ detected with LISA. The progenitor system is a nonspinning binary BH with $M_{f} = 10^{6} \Msun$ and $a_{f} = a_{f}(q)$. Note that nonzero $\chi_{\rm eff}$ can alter this plot at a subleading level. }
\label{fig:sigma}
\end{figure}

In Fig.~\ref{fig:sigma} we investigate the variation of measurability of various QNM parameters with respect to $q$ for a nonspinning binary~\cite{Hofmann:2016yih,Pan:2011gk,Bhagwat:2021kfa,Barausse:2009uz}.~\footnote{For a fixed mass, given a binary mass ratio $q$ all the parameters of the ringdown are fixed for a nonspinning binary BH system. Note that the individual BH spins have a subleading effect on the measurability compared to the mass ratio.} Figure~\ref{fig:sigma} displays the relative measurement error i.e.,  $\Delta {\theta}_{i}$  of QNM parameters as a function of $q$, for a fiducial mass of $M_{f} = 10^{6} \Msun$ and for $z=10$. Different choices of $M_f$ and $z$ will only scale $\sigma_{{\theta}_{i}}$ by an overall factor. 

From Fig.~\ref{fig:sigma}, we note the hierarchy of measurability at high $q$ ($q \geq 5$) goes as follows:
$\Delta f_{22} < \Delta f_{33} < \Delta f_{21} < \Delta f_{44}< \Delta \tau_{22}\ll \Delta \tau_{33} \sim \Delta \tau_{21}\ll\Delta \tau_{44}$. 
At low $q$, however, the excitation of subdominant modes is suppressed and the hierarchy of measurability goes as: $\Delta_{f_{22}}\ll \Delta_{\tau_{22}}\ll\Delta_{f_{33}}<\Delta_{f_{21}}<\Delta_{f_{44}} \ll \Delta_{\tau_{33}} \sim \Delta_{\tau_{21}}<\Delta_{\tau_{44}}$. We further highlight two interesting crossing points around $q \sim 1.5$: for $q >1.5$ $\Delta_{f_{33}}<\Delta_{\tau_{22}}$ and $\Delta_{f_{44}}<\Delta_{f_{21}}$ .

\section{Ringdown tests with LISA}
\label{sec:results-lisa}

In this section, we present the results of our ringdown analysis for LISA. We first briefly summarize our  analysis setup.

Our injected waveforms include the four angular modes $(l,m) \in \{(2,2),(3,3),(2,1),(4,4)\}$. For each population model, we simulate 100 realizations of one year of data. The figures and numbers presented throughout the next two sections are the mean expectation obtained by averaging these realizations. Moreover, a Poisson error bars or counting error, i.e., $\sqrt{N}$ (where $N$ is the number of events satisfying a given criterion) is plotted in shaded colors in all the result figures. 

We summarize the prospects of performing the no-hair theorem test with LISA in Sec. \ref{sec:rate-lisa}. This is followed by a detailed discussion on the prospects of measuring 3 or more QNM parameters, including a study on different combination of QNM parameters in Sec. \ref{sec:3-plus-param-lisa}. Finally, in Sec.~\ref{sec:3-minus-param-lisa} we consider the cases in which the events do not necessarily allow for 3 independent QNM parameters to be extracted; this could still be used to perform powerful consistency tests when combined with the information about binary masses and spins obtained from the inspiral part of the signal. Since we wish to quantify the landscape of BH spectroscopy with LISA in its entire effective operational time, we present results for 4 years of LISA data in Sec.~ \ref{sec:3-plus-param-lisa} and \ref{sec:3-minus-param-lisa}.

The same analysis setup and structure is used for discussing the results for ET in Sec.~\ref{sec:results-et}.

\begin{figure*}
\includegraphics[width=0.49\textwidth]{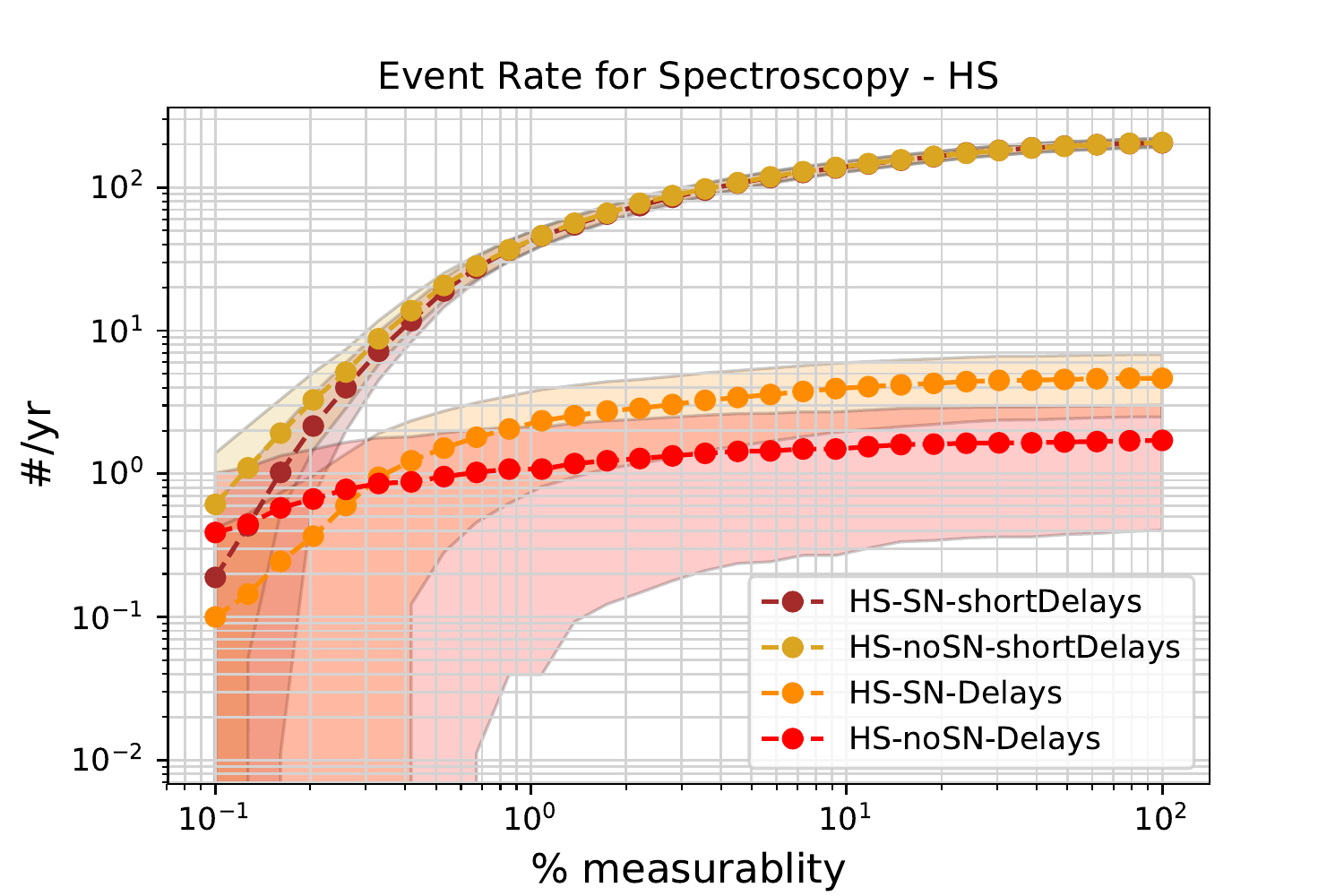}
\includegraphics[width=0.49\textwidth]{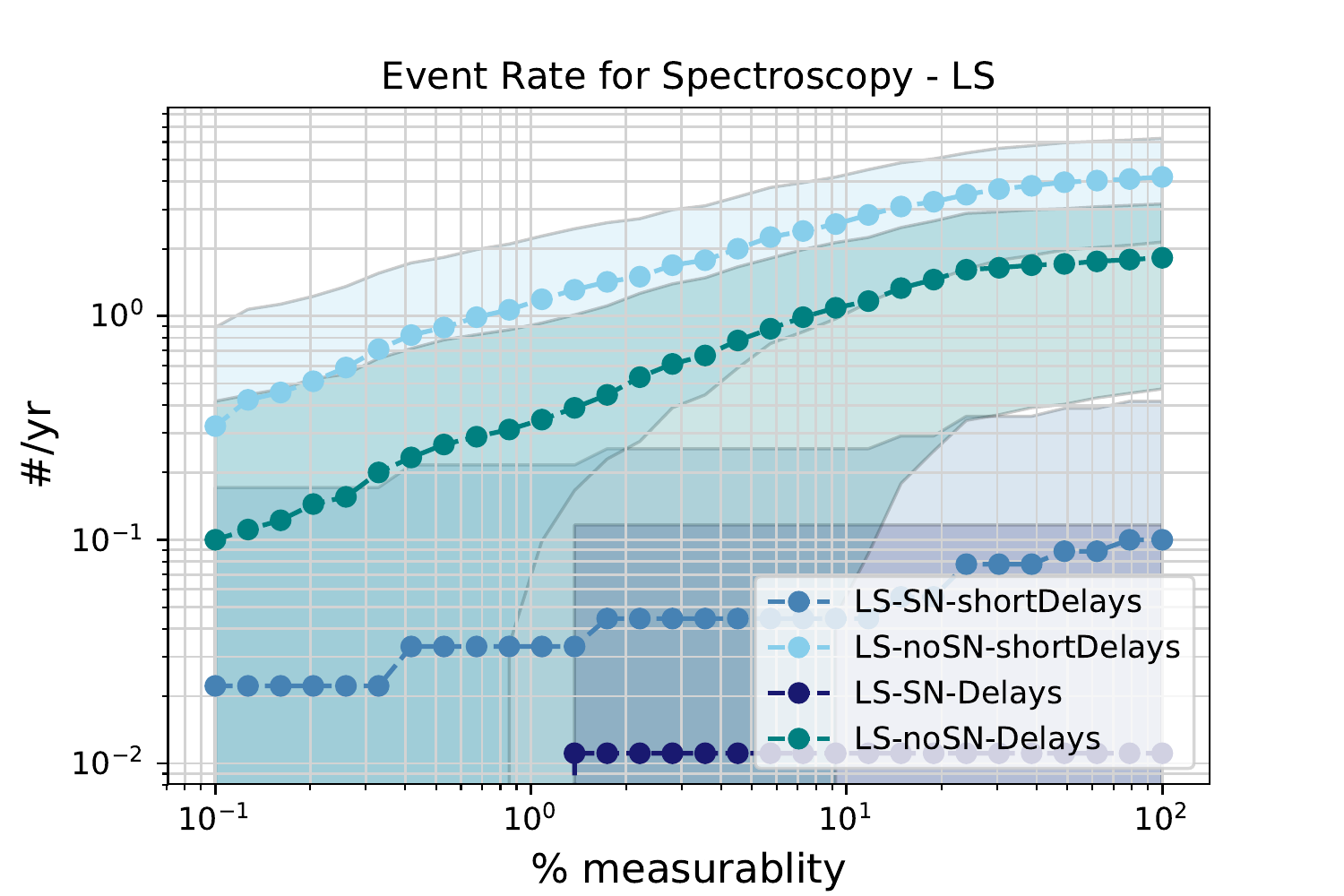}
\caption{
Rates of events for which \emph{any} 3 independent ringdown parameters can be resolved and measured by LISA with a given precision for the HS (left panel) and LS (right panel) scenarios considered in this work.
A Poisson error bars or counting error, i.e., $\sqrt{N}$ (where $N$ is the number of events satisfying a given criterion) is plotted in shaded colors. The solid curves indicate the average expectation obtained using 100 realization of 1-year data. The most favorable MBH population models belong to the HS scenario with shortDelays (green and pink curves in the left panel).}
\label{fig:No-hair-rate}
\end{figure*}

\begin{table*}
\begin{tabular}{|c|c|c|c|c|}
\hline
\rowcolor[HTML]{C0C0C0} 
\textbf{Seed} & \textbf{Population Models} & \textbf{Rate ($0.1\%$ measurable)} & \textbf{Rate ($1\%$ measurable)} & \textbf{Rate ($10\%$ measurable)} \\ \hline \hline \hline
\textbf{}   & SN-shortDelays    & 0/yr & 33.8/yr & 132.8/yr \\ \cline{2-5} 
\textbf{HS} & noSN-shortDelays & 0.2/yr & 34.5/yr  & 132.7/yr   \\ \cline{2-5} 
            & SN-Delays & 0.04/yr & 2.2/yr & 3.9/yr \\ \cline{2-5} 
            & noSN-Delays       & 0.1/yr & 1.0/yr  & 1.5/yr   \\ \hline \hline \hline
            & SN-shortDelays    & 0/yr    & 0/yr    & 0/yr  \\ \cline{2-5} 
\textbf{LS} & noSN-shortDelays   & 0/yr    & 0.7/yr    & 2.1/yr     \\ \cline{2-5} 
            & SN-Delays & 0.21/yr  & 0/yr & 0/yr  \\ \cline{2-5} 
            & noSN-Delays       & 0/yr & 0.2/yr & 0.7/yr   \\ \hline
\end{tabular}%

\caption{Rates of events for which \emph{any} 3 independent ringdown parameters can be resolved and measured by LISA with $0.1,1,10\%$ precision for the 8 different MBH population scenarios considered in this work. }

\label{tab:no-hair-rate}
\end{table*}

\subsection{Summary on prospects of no-hair theorem tests with LISA}
\label{sec:rate-lisa}

Figure~\ref{fig:No-hair-rate} and Table~\ref{tab:no-hair-rate} summarize the overall prospects for performing the no-hair theorem tests with MBHs using LISA.  Here we marginalized over all combinations of 3 QNM parameters and present the overall rate of events that allow for the no-hair theorem test at a given level of measurability. A more detailed result for each combination is presented in detail in Sec.~\ref{sec:3-plus-param-lisa}.

\subsubsection{Heavy seed models}
Firstly, note that the intrinsic rates of binary BH events predicted for the HS \emph{shortDelays} models ($\sim 300-320$ /yr) are about two order of magnitude larger than that for the \emph{Delays} models ($1.5 - 6$ events/yr), see Table~\ref{tab:events-per-year}. However, the distribution of $\rho_{\rm rd}$ for the binary BH events expected for the \emph{Delays} models peaks sharply at $\rho_{\rm rd} \sim 10^{3}$, higher compared to models with \emph{shortDelays}, for which $\rho_{\rm rd}$ is more broadly spread in $\rho_{\rm rd} \in [1, 10^3])$.

From Fig.~\ref{fig:No-hair-rate} (left), we note that the two \emph{shortDelays} models have similar rates and are the most promising cases for BH spectroscopy. They predict $\sim 30$ events/yr with $\leq 1 \%$ measurability (with at least one event a year that would allow for $ \sim 0.5 \%$ measurability). Next, we observe that the \emph{Delays} models predict about $\sim 1-2 $ events/yr with $\sim 1 \%$ measurability. This is not surprising because of the intrinsically low event rates predicted by these models. Thus, if the underlying population of MBH binaries have HSs, LISA will realistically perform BH spectroscopy with a precision of at least $1 \%$. 

\subsubsection{Light seed models}
While, for the HS models, \emph{Delays} versus \emph{shortDelays} is the dominant factor for BH spectroscopy, for the LS models the effect of SN feedback is more pronounced. As discussed in \cite{Barausse:2020mdt}, this is due to the combination of two factors: by halting their growth, SN feedback prevents LS from forming BHs in the frequency band to which LISA is most sensitive; and it also makes gas driven migration of BH pairs (which is particularly important for low-mass systems) inefficient by ejecting gas in the nuclear region of the galaxy, thus reducing the merger rate. We verify this in Fig.~\ref{fig:No-hair-rate} (right) by noting that the two \emph{SN} LS models seem unlikely to allow for spectroscopy with LISA; 
although both \emph{SN-shortDelays} and \emph{SN-Delays} models have an intrinsic rate of ${\cal O}(10)$~event/yr, most of their ringdowns have frequencies that are higher than the LISA sensitivity and are undetectable by LISA.
The \emph{noSN-shortDelays} model seems to be the most promising case among the LS scenarios, with $\sim 2$ events/yr allowing for measurability at $10 \%$ level. Finally, the \emph{noSN-Delays} model predicts less than 1 event/yr with $10 \%$ measurability. Note that this level of measurability for BH spectroscopy is anyway better than what can be achieved with the current data from the LIGO-Virgo detectors. Furthermore, it refers to sources at a different curvature scale, so it would be in principle relevant to constrain different classes of modified theories of gravity~\cite{Barausse:2014tra,Maselli:2019mjd}.

\begin{figure*}[ht]
\hspace*{-1cm}
\includegraphics[width=0.45\textwidth]{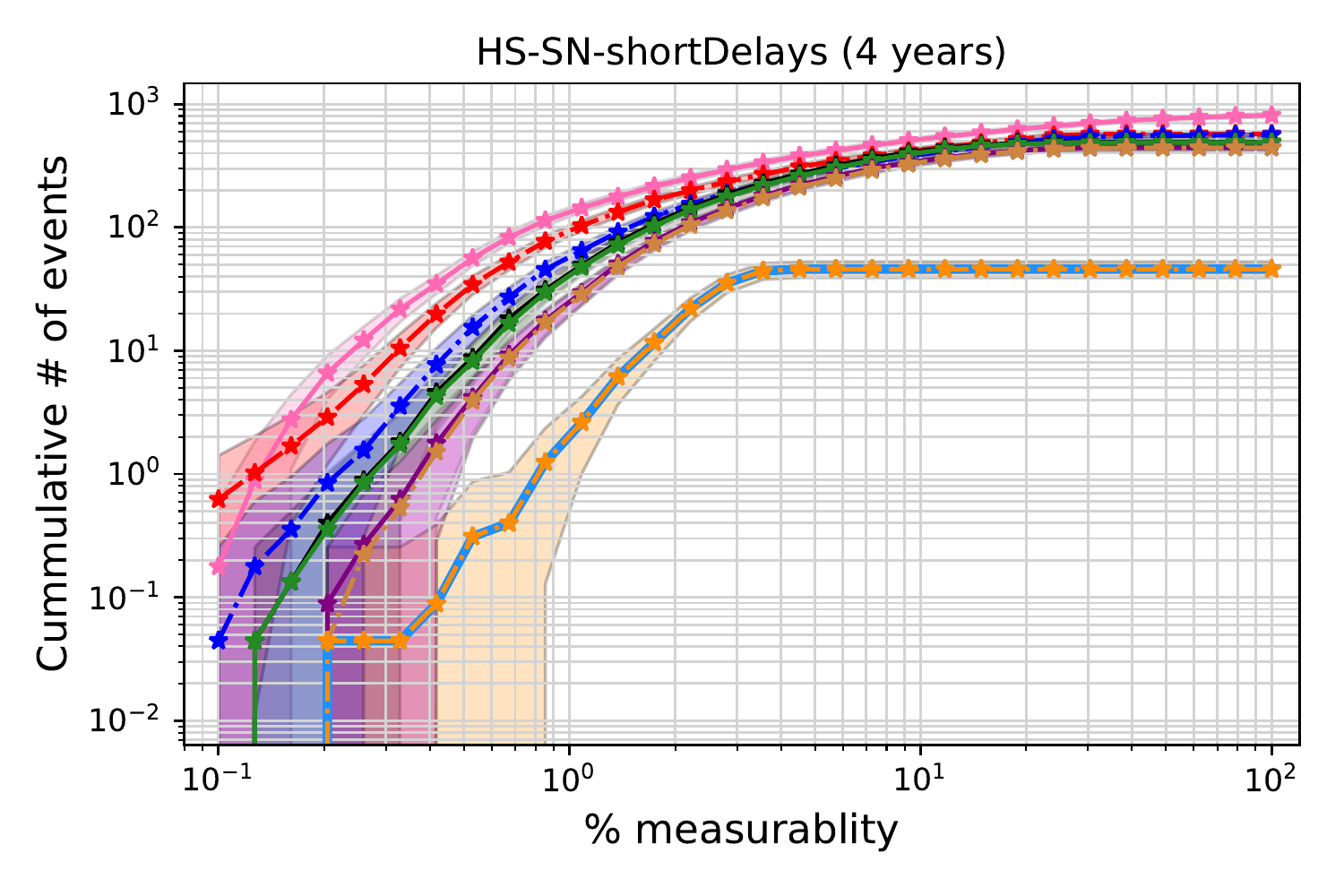}
\includegraphics[width=0.45\textwidth]{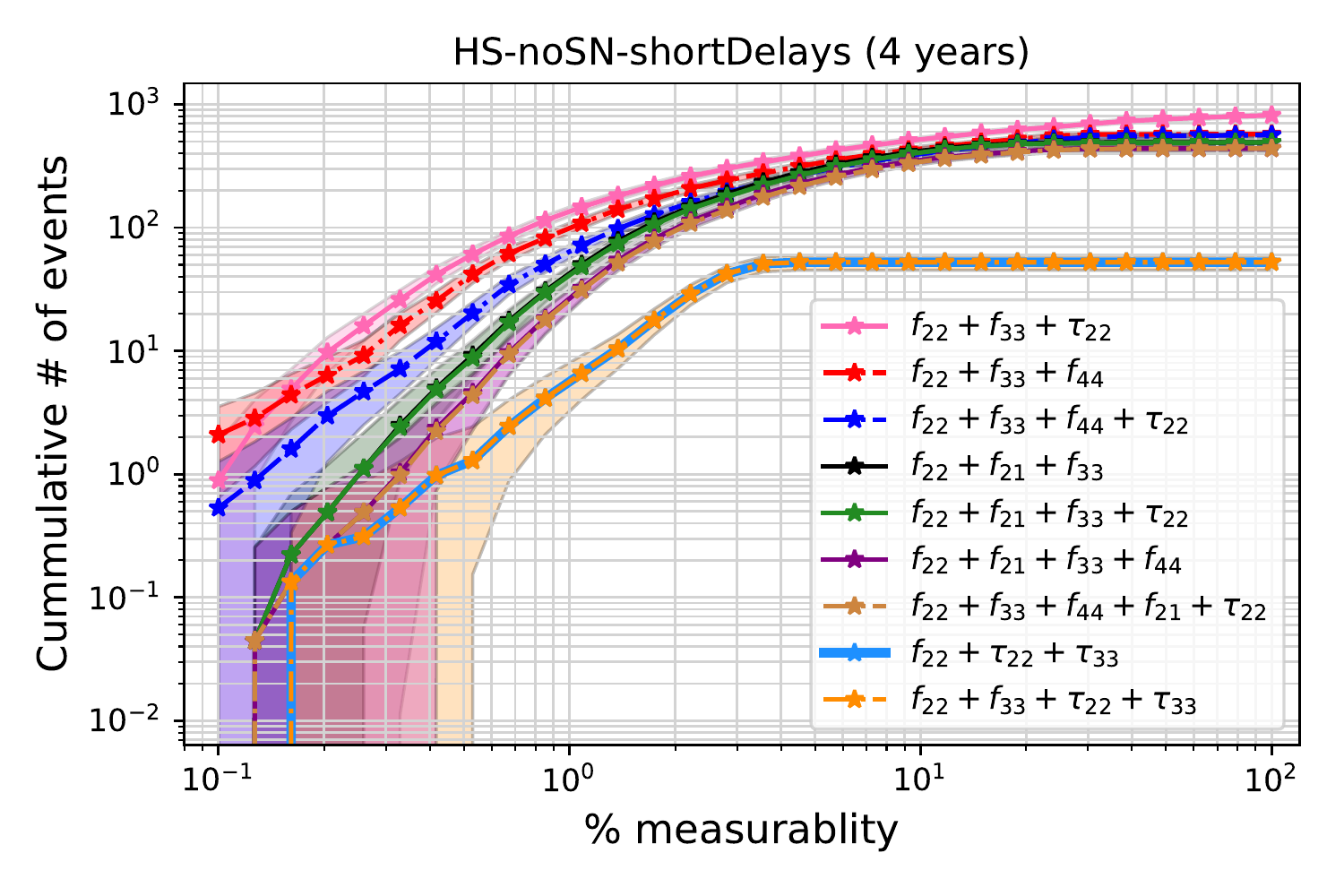} \\ 
\hspace*{-1cm}
\includegraphics[width=0.45\textwidth]{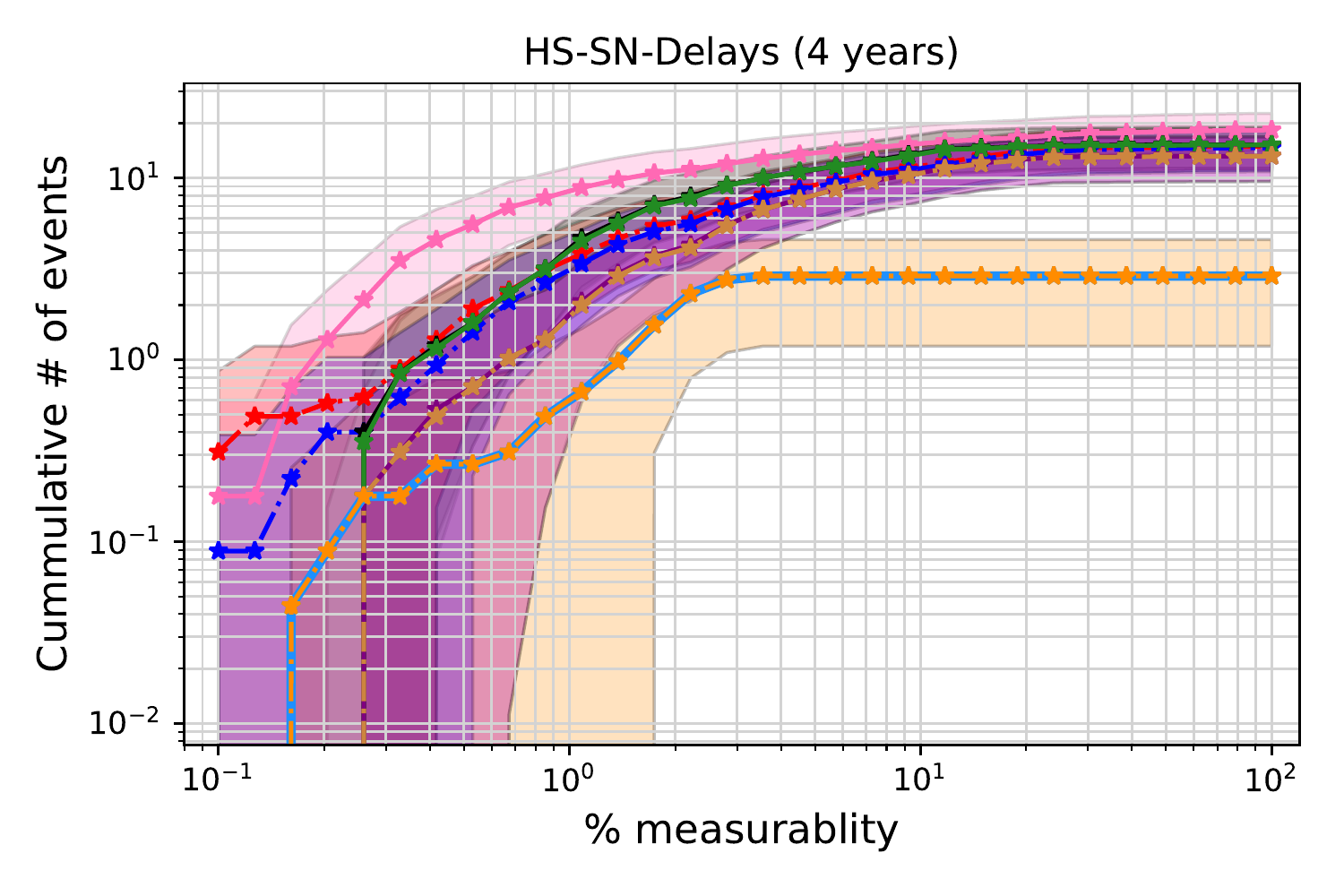}
\includegraphics[width=0.45\textwidth]{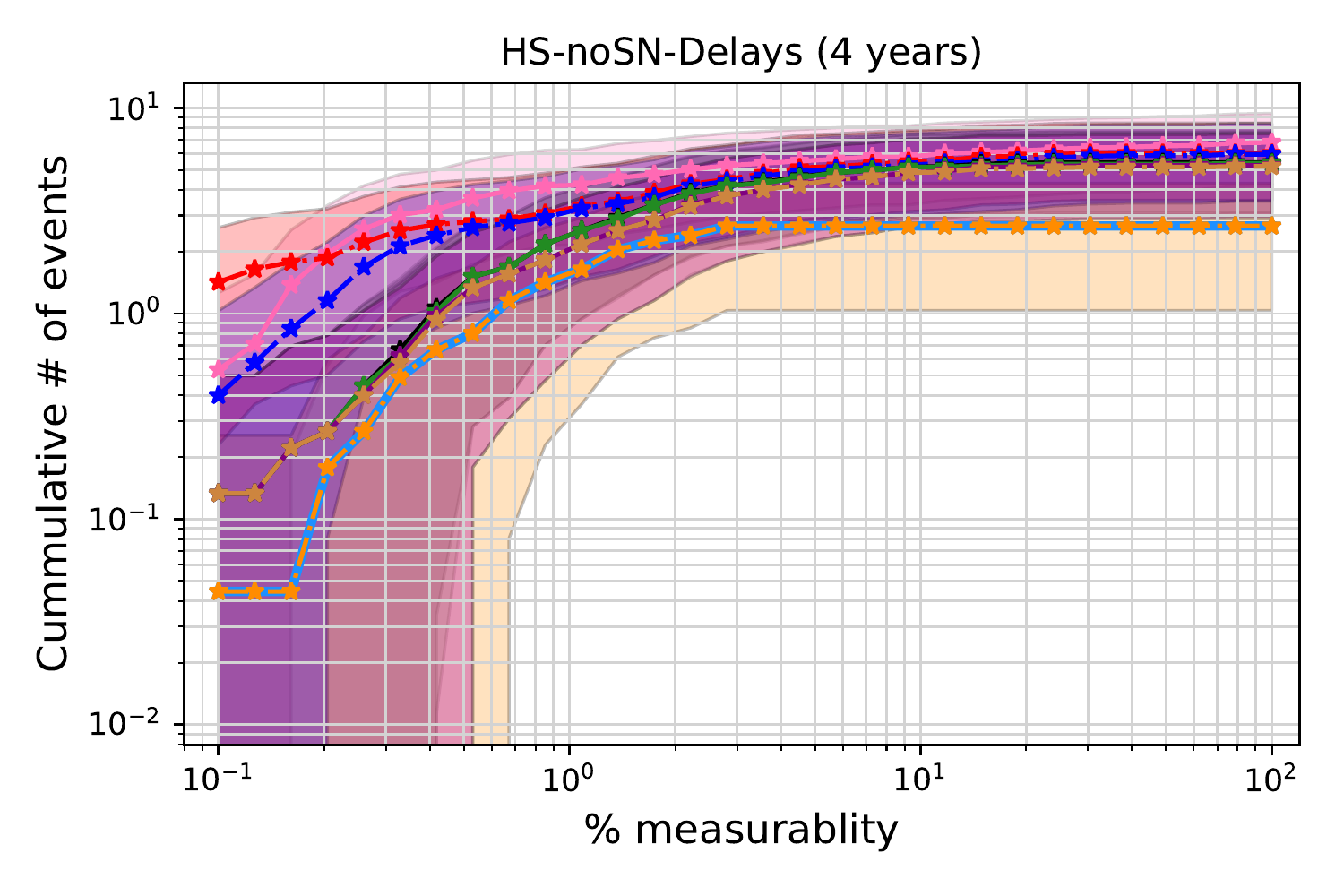}
\caption{Cumulative number of events in 4 year of data that allows for measurability of 3 or more QNM parameters with LISA in the HS scenarios. Note that the black and the green curves, and the brown and the purple curves, overlap.}
\label{fig:3-plus-param-lisa}
\end{figure*}

\subsection{BH spectroscopy with LISA by measuring 3 or more QNM parameters}
\label{sec:3-plus-param-lisa}
We explore the measurability for different combinations of the QNM parameters for the 4 HS models in Fig.~\ref{fig:3-plus-param-lisa}. Firstly, across all the MBH population models we have considered, we find that $f_{22}$ and $f_{33}$ are the easiest QNM parameters to be measured. For initial binary BHs with moderate to large mass ratios, the third best measured parameter is $f_{44}$, while for small or equal mass ratios $\tau_{22}$ performs better. Next, note that while in most cases the measurability of a QNM parameter is limited by the relative uncertainty in its measurement (i.e., $\sigma_{\theta_i} / \theta_i$), certain combinations of QNM frequencies can pose a challenge due to resolvability. The most likely combinations based on the frequency spacing are -- a) $f_{21}$ with $f_{22}$ and b) $f_{44}$ with $f_{33}$. In the population we study, with the former poses a considerable bottleneck.
For instance, in the HS \emph{shortDelays} models, $\sim 125$ events/yr have $f_{21}$ resolvable from $f_{22}$ while $\sim 175$ events/yr have $f_{44}$ that is resolvable from $f_{22}$; therefore, despite having typically lower amplitude of excitation the (4,4) mode performs better in figure \ref{fig:3-plus-param-lisa}. Also note that $\sim 145$ events/yr have $f_{44}$ resolvable from $f_{33}$. We shall see that resolvability criteria substantially reduces the number of signals that allow for measurability of QNM combination involving the $f_{21}-f_{22}$. 

We divide the subsequent discussion into \emph{shortDelays} and \emph{Delays} models based on the similarity in the qualitative behaviors for BH spectroscopy. First, we focus on the shortDelays models, which are the best case scenario for BH spectroscopy with LISA. From Fig.~\ref{fig:3-plus-param-lisa}, we see that the number of events that allow for certain percent measurability in a population depends on the chosen combination of the QNM parameters. Statistically, for the \emph{shortDelays} models, the hierarchy is as follows:

\begin{equation*}
\begin{aligned}
     & \#(\{f_{22}, f_{33}, \tau_{22} \} )>\#(\{f_{22}, f_{33}, f_{44} \} ) > \\
     & \#(\{f_{22}, f_{21}, f_{33} \} ) \sim \#(\{f_{22}, f_{21}, \tau_{33} \} )\,.
\end{aligned}
\end{equation*}
Interestingly, $f_{44}$ performs statistically better than $f_{21}$ even though (2,1) mode is typically excited more than (4,4) (except for systems with $q \sim 1$).

From Fig.~\ref{fig:3-plus-param-lisa}, we see that $\sim 100$ events will allow for $\sim 1\%$ measurability for both $\{ f_{22},f_{33},\tau_{22}\}$ and $\{ f_{22},f_{33},f_{44} \}$.  Note that the set of signals that allow for the former combination and the latter combination can be different, since in Fig.~\ref{fig:3-plus-param-lisa} we only check for the number of events. The difference in the number and set of signals is more pronounced for higher-precision measurability thresholds -- for instance, for a $0.2 \%$ measurability, we find $\sim 8$ events for $\{f_{22},f_{33},\tau_{22} \}$, $\sim 3$ events  for $\{ f_{22},f_{33}, f_{44} \}$ and $\sim 1$ event for $\{f_{22}, f_{33}, f_{44}, \tau_{22} \}$.  However, at $\sim 10 \%$ measurability $300-400$ events will allow for each of the above QNM parameter combinations. Inspired by this, we further investigate the prospects of BH spectroscopy beyond 3 QNM parameters. Interesting, we find that, in 4 years of LISA data, $\sim 10$ events will allow subpercent measurability and $\sim 20$ events will allow $1 \%$ measurability for as much as 5 QNM parameters, namely $\{ f_{22}, f_{33}, f_{21}, f_{44}, \tau_{22} \}$. Extracting more than 4 independent QNM parameters might help improve the precision of the no-hair theorem test, as the spectrum in the ringdown needs to satisfy additional constraints that come from the measurement of the additional parameters.

Next, unlike for the case of \emph{shortDelays} models, in the \emph{Delays} models we see that  the QNM combinations do not have a clear hierarchy in the number of signals allowing for given percent measurability. Recall that these models predict a small event rate and that the error bars on the number of events satisfying a certain criterion goes as $\sqrt{N}$. Therefore, a small event rate would induce a large relative error in the estimation of a number of spectroscopically valuable events, and it is not meaningful to read the hierarchy from Fig.~\ref{fig:3-plus-param-lisa}. However, broadly, the combination $\{f_{22}, f_{33}, \tau_{22} \}$ seems to perform better than the other combinations. 

For the \emph{SN-Delays} model, $ \sim 2$ events will allow for $1 \%$ measurability for 5 QNM parameters. Moreover, $ \sim 5 $ events will allow for either of the combinations $\{ f_{22}, f_{33}, f_{44} \}$ or $ \{f_{22}, f_{21}, f_{33} \}$, and $\sim 10$ events for $\{ f_{22}, f_{33}, \tau_{22}\}$. Similarly, for  the \emph{noSN-Delays} models we will be able to perform 5-parameter QNM spectroscopy with $1 \%$ measurability for $\sim 1-2$ events, whereas 3-4 events will allow for $1 \%$ measurability of either of the following combinations of QNM parameters: $\{ f_{22}, f_{33}, \tau_{22} \}$ or $\{ f_{22}, f_{33}, f_{44} \}$ or $ \{f_{22}, f_{21}, f_{33} \}$. 

\begin{figure*}[ht]
\includegraphics[width=0.45\textwidth]{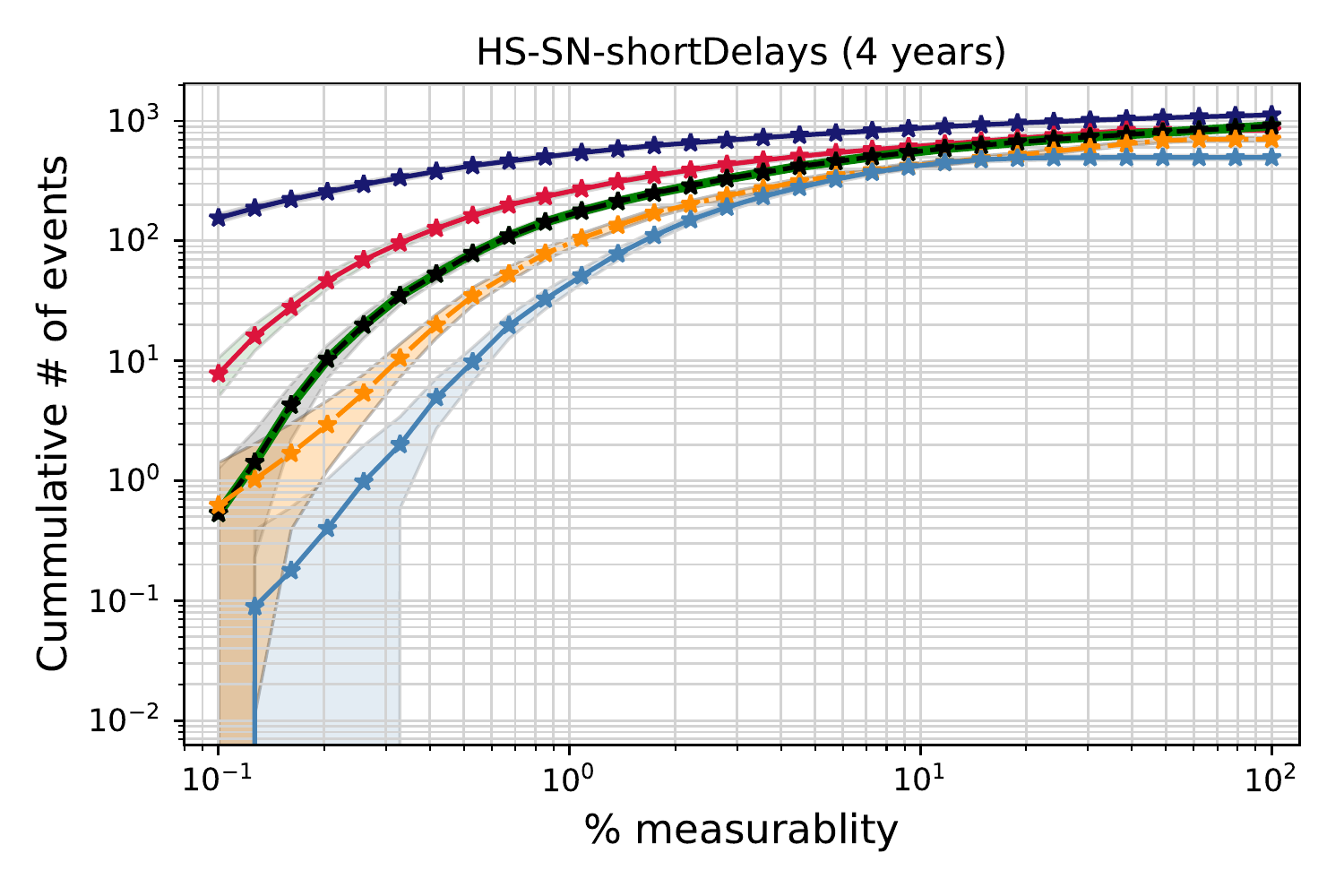}
\includegraphics[width=0.45\textwidth]{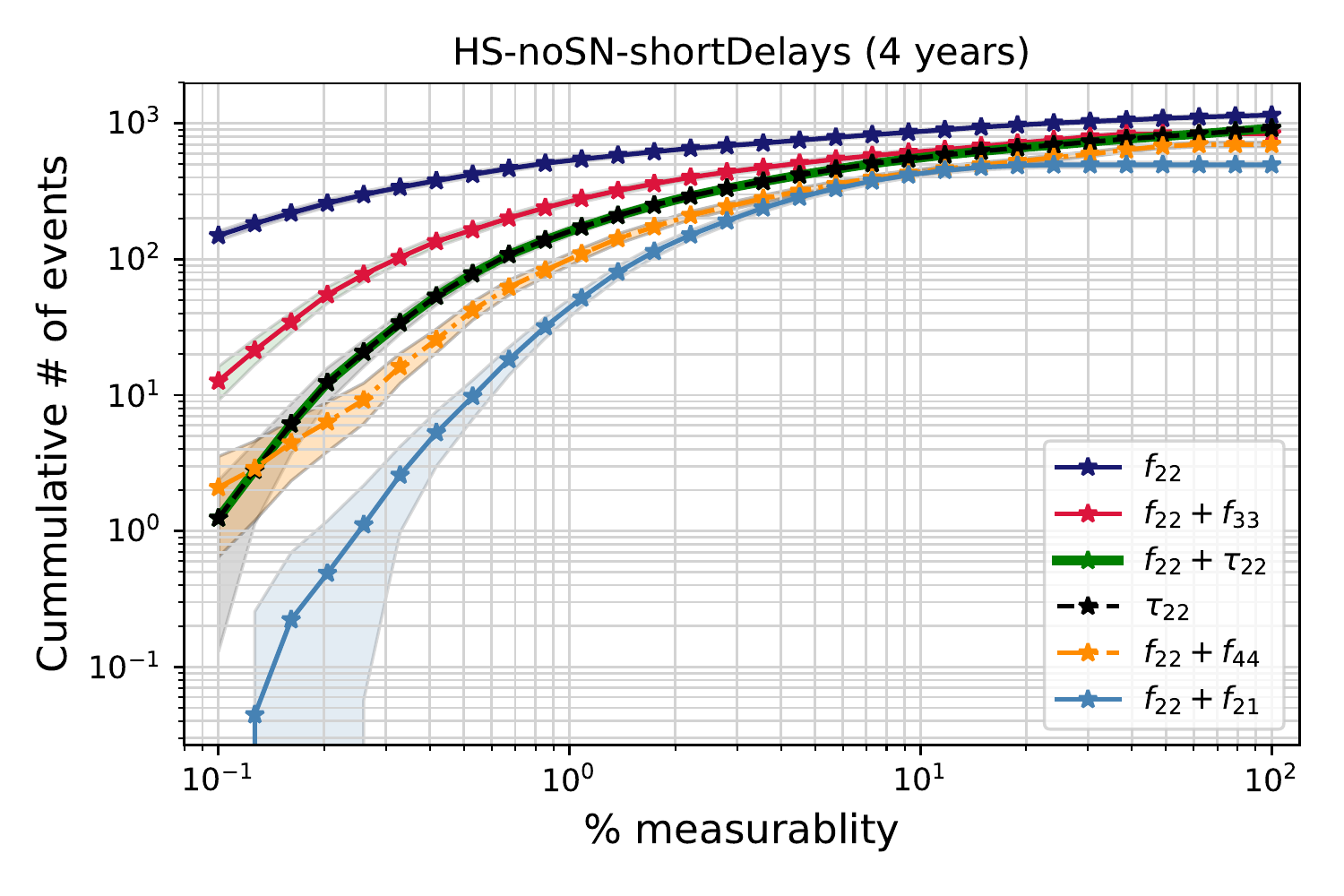} \\ 
\includegraphics[width=0.45\textwidth]{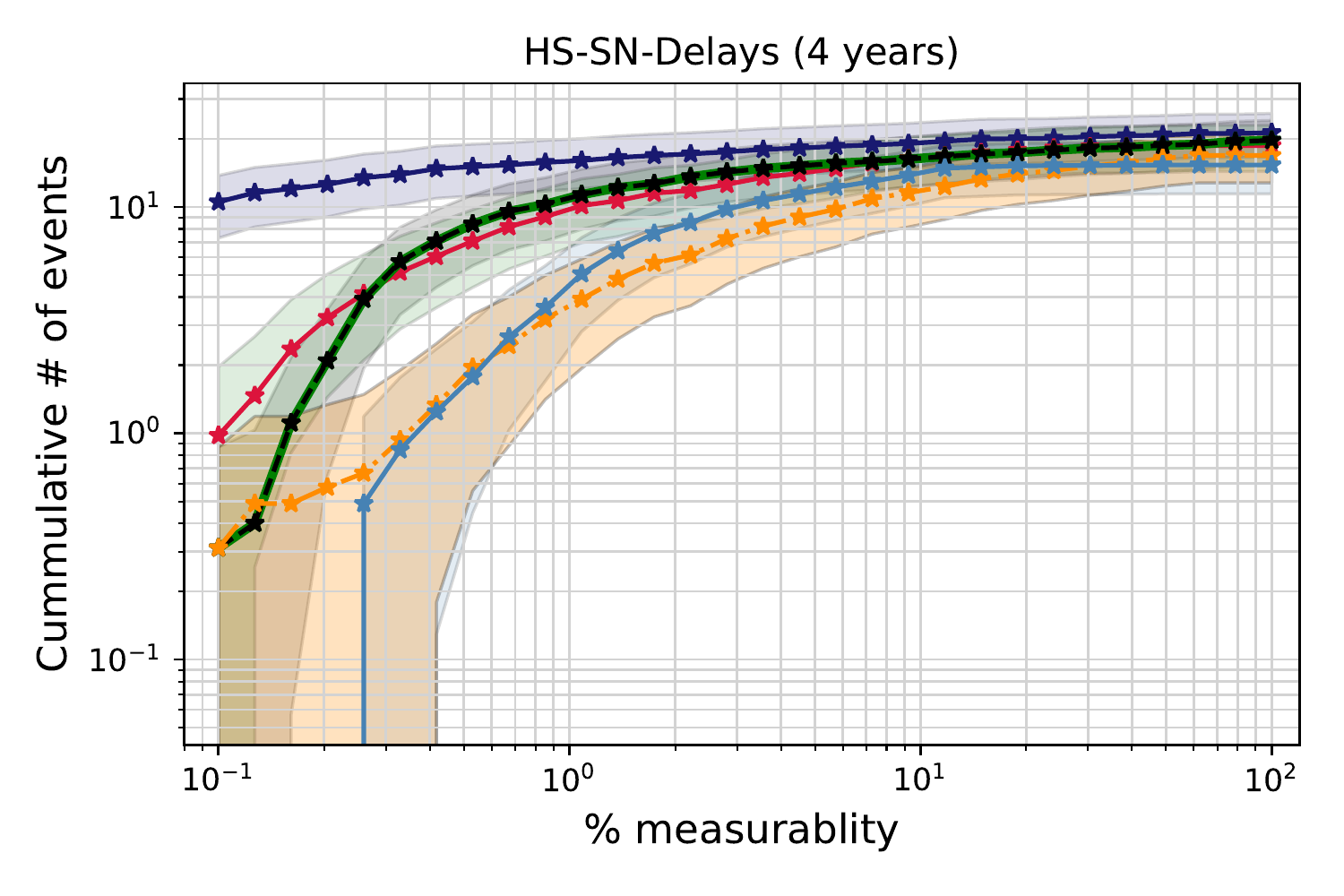}
\includegraphics[width=0.45\textwidth]{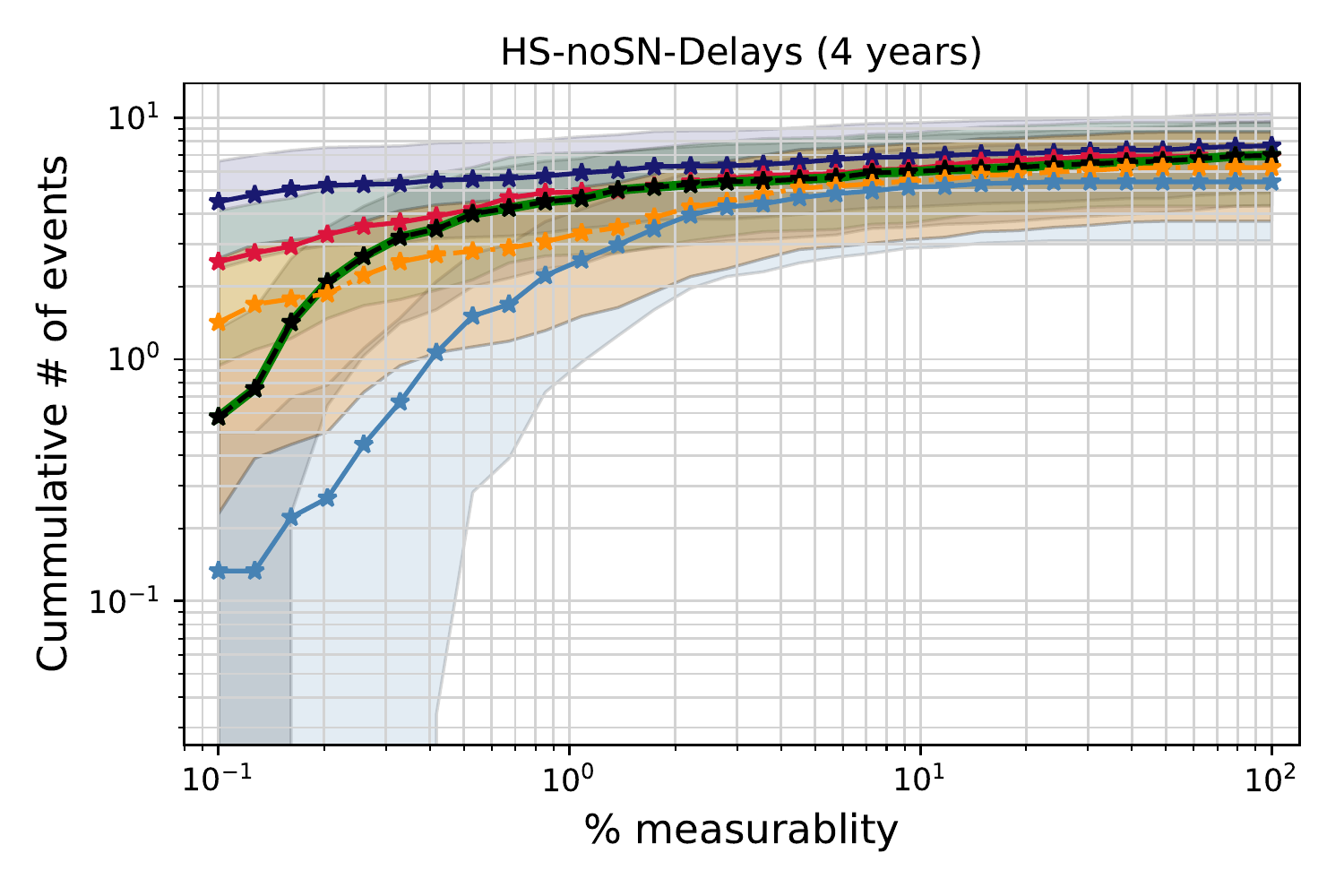} 
\caption{Same as in Fig.~\ref{fig:3-plus-param-lisa} but for the measurability with LISA of 1 or 2 QNM parameters in various combinations. }
\label{fig:3-minus-param-lisa}
\end{figure*}

\subsection{BH spectroscopy with LISA by measuring less than 3 parameters}
\label{sec:3-minus-param-lisa}
From the previous section, we see that for the HS models, a fraction of total events allow for precision measurement of mores than 3 QNM parameters and would allow for an unprecedented no-hair theorem test using the ringdown alone. In Fig.~\ref{fig:3-minus-param-lisa}, we focus on the events in the population that do not necessarily allow for the precision measurement of 3 QNM parameters -- either because the values of $q$ and of the spins do not
allow for sufficient excitation of the subdominant mode or because of low $\rho_{\rm rd}$. These events can still be used for stringent tests of GR, including the no-hair theorem, when combined with the information obtained from the inspiral.
In the context of inspiral-merger-ringdown tests, a ringdown detection at the optimal frequency would typically correspond to an inspiral-merger with even higher SNR. The typical situation is that the remnant mass and spin inferred from the inspiral-merger part of the signal are more accurate than those measured through the ringdown. In such a case, the accuracy of the test is limited by the measurability of the ringdown modes.

From Fig.~\ref{fig:3-minus-param-lisa}, we note that more than 100 events in the \emph{shortDelays} models would allow for $0.1 \%$ measurability of $f_{22}$ along, and about 10 events would allow for $0.1 \%$ measurability of the combination of $ \{ f_{22}, f_{33} \}$ . Even though these events do not allow for a no-hair theorem test using the ringdown alone, they can be used to perform a high-precision consistency check of GR. For the \emph{Delays} model, 10 events will allow for the estimation of  $f_{22}$ and $\sim 1$ event for the combined estimation of $ \{f_{22}, f_{33} \}$ with $0.1 \%$ measurability.

Statistically, we see that the measurability of the different QNM parameters for the population corresponding to \emph{shortDelays} models are as follows -- $\Delta f_{22} > \Delta f_{33}>\Delta \tau_{22}> \Delta f_{44}> \Delta f_{21}$. On the other hand, for the \emph{Delays} populations, we see that measurability of the QNM parameters goes as follows -- $ \Delta f_{22} > \Delta f_{33} \sim \Delta \tau_{22} > \Delta f_{44} \sim \Delta f_{21}$.

This difference in the trend can be understood by looking at Fig.~\ref{fig:assym} along with Fig.~\ref{fig:sigma}. Unlike the \emph{shortDelays} models, the \emph{Delays} models have binary BHs that cluster more around $q\approx1$. Therefore, a substantial fraction of signals produced by the \emph{Delays} models will have relatively low subdominant mode excitation and will follow the low $q$ measurability trend.

\section{Ringdown test with ET}
\label{sec:results-et}
The LS models seem unsuitable for spectroscopy with LISA, as they produce remnant BHs with lower masses, ringing at frequencies higher than the LISA bandwidth. Given  the low event rates for BH spectroscopy with LISA in the LS scenario, in this section we explore if ringdowns expected in some LS models can be accurately measured by the ET detector~\cite{Maggiore:2019uih} that will operate at higher frequencies~\cite{Hild:2010id}. However, throughout this discussion, the reader should bear in mind that ET is not optimized to detect MBH ringdowns, as it can be seen from the cumulative SNR plot in Fig.~\ref{fig:snr-cumu}. The ringdown frequency distribution in the LS scenario peaks in between the LISA and the ground-based GW detector (ET/CE/LIGO/Virgo) bandwidths, but the tail of the distribution extends above $\approx10\,{\rm Hz}$ (see Fig.~\ref{fig:Freq-tau-22}). Although we focus on MBH ringdown here, it should be stressed that the performance of third-generation detectors such as ET will far exceed the results presented in this work when spectroscopy is performed on stellar origin BH binaries.

\begin{figure*}[ht]
\includegraphics[width=0.49\textwidth]{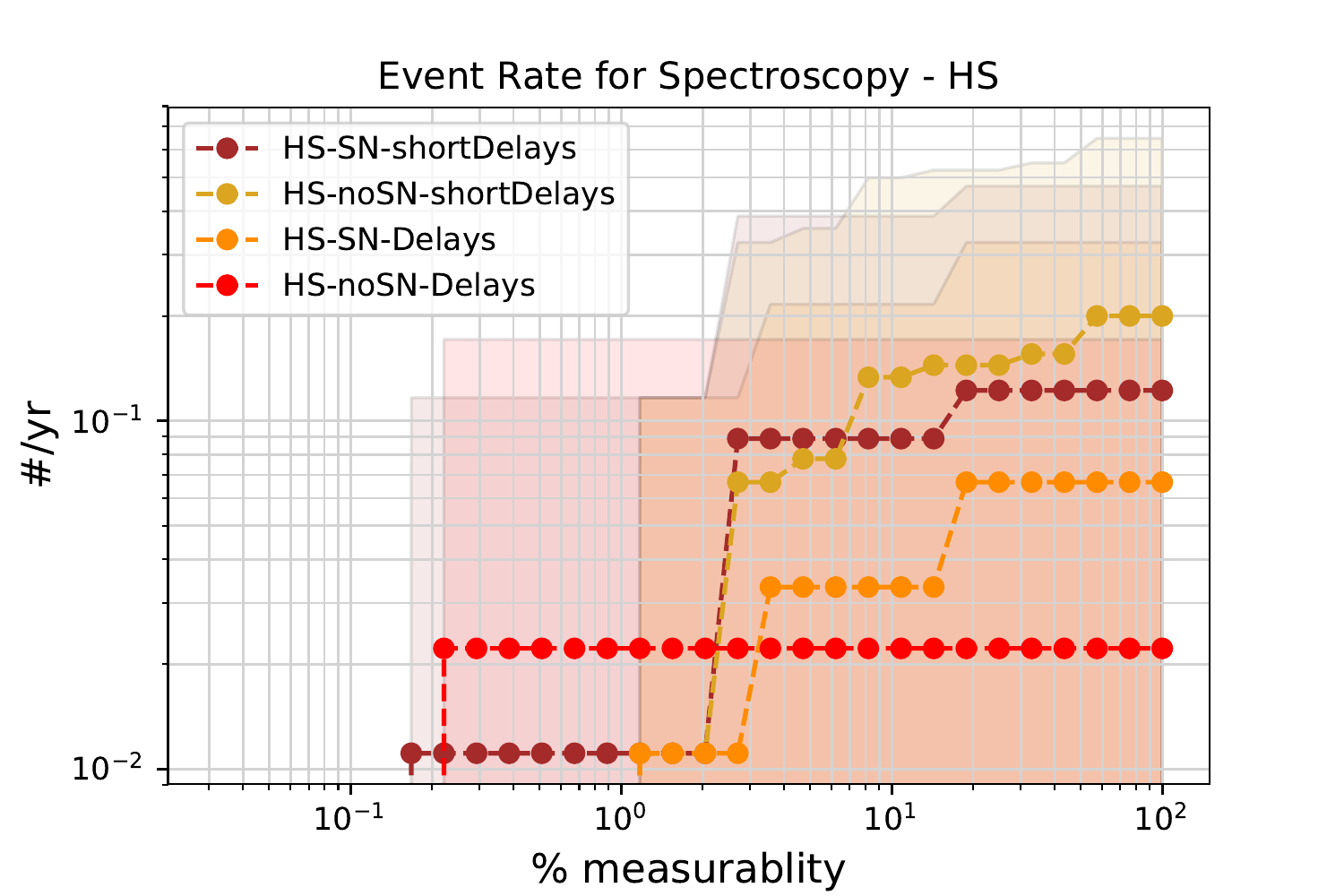}
\includegraphics[width=0.49\textwidth]{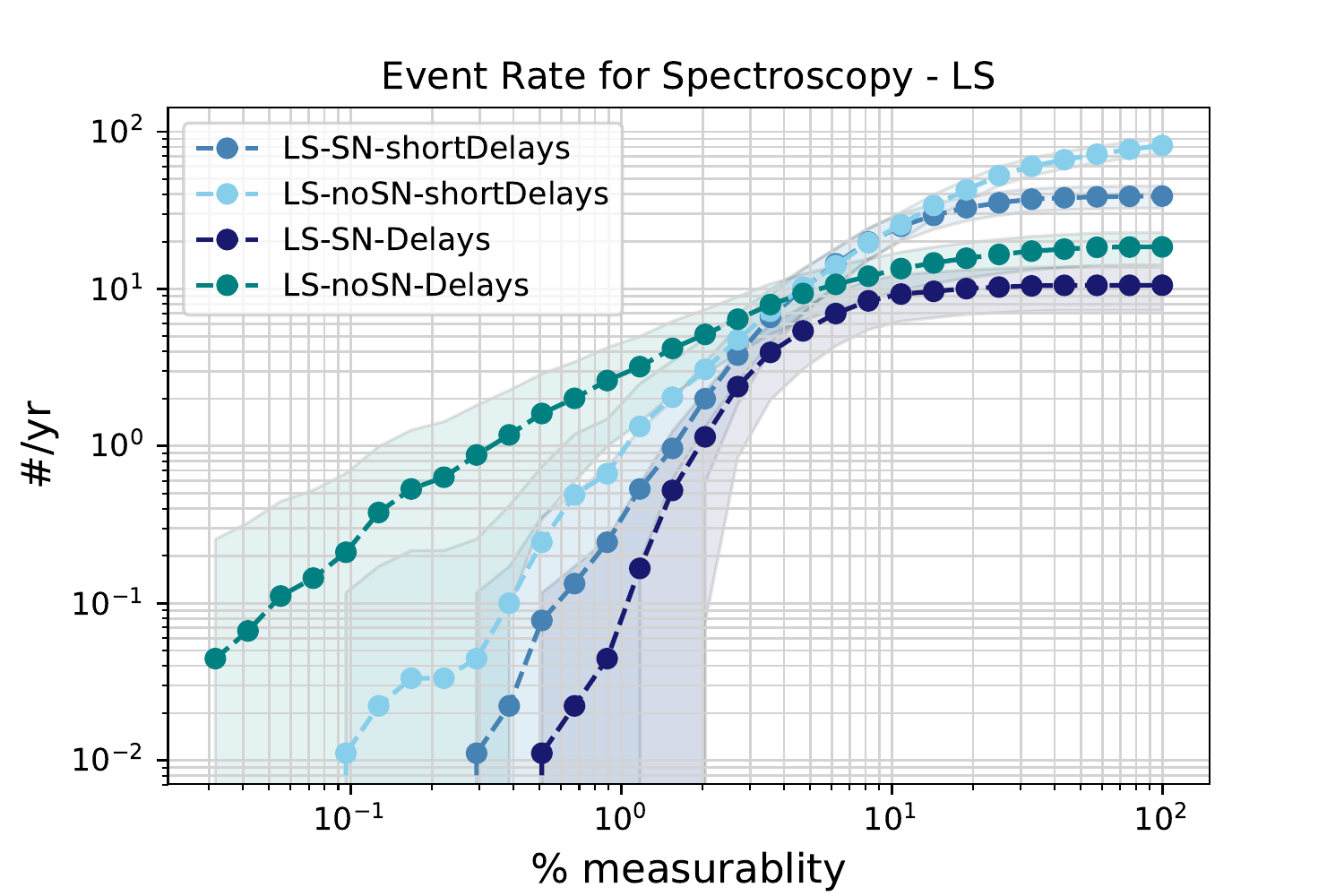}
\caption{
Same as Fig.~\ref{fig:No-hair-rate} but for ET. Rates of events for which \emph{any} 3 independent ringdown parameters can be resolved and measured by ET with a given precision for the HS (left panel) and LS (right panel) scenarios considered in this work. The shaded areas indicate the Poisson error in the estimation of the number of events for each model, while the solid curve indicates the average number of events. The most favorable MBH population models belong to the LS scenario with \emph{shortDelays} (green and pink curves in the right panel). }
\label{fig:No-hair-rate-et}
\end{figure*}

\begin{table*}%
\begin{tabular}{|c|c|c|c|c|}
\hline
\rowcolor[HTML]{C0C0C0} 
\textbf{Seed} & \textbf{Population Models } & \textbf{Rate ($0.1\%$ measurable)} & \textbf{Rate ($1\%$ measurable)} & \textbf{Rate ($10\%$ measurable)} \\ \hline \hline \hline
            & SN-shortDelays    & 0/yr    & 0.2/yr    & 18.2/yr  \\ \cline{2-5} 
\textbf{LS} & noSN-shortDelays   & 0/yr    & 0/yr    & 5.2/yr     \\ \cline{2-5} 
            & SN-Delays & 0/yr  & 0/yr & 8.3/yr  \\ \cline{2-5} 
            & noSN-Delays       & 0/yr & 0.3/yr & 5.6/yr   \\ \hline
\end{tabular}%

\caption{Same of Table~\ref{tab:no-hair-rate} but for ET and focusing on the LS scenario only. Rates of events for which \emph{any} 3 independent ringdown parameters can be resolved and measured by ET with $0.1,1,10\%$ precision for the 8 different MBH population scenarios considered in this work.
}
\label{tab:No-hair-rate-et}
\end{table*}

\subsection{Summary on prospects of no-hair theorem tests with ET}
\label{sec:rate-et}
HS models ring at frequencies that are much below the lower frequency sensitivity of ET. From Fig.~\ref{fig:No-hair-rate-et}, we note that the HS models are unsuitable for performing spectroscopy with ET, with less than 1 event in 10 years that might allow for  no-hair theorem tests with $10 \%$ measurability. On the other hand, with LS models, we expect to perform  no-hair theorem tests with at least $2 \%$ measurability in 1 event/yr, for all but the \emph{noSN-shortDelays} model. Moreover, LS models with \emph{SN-shortDelays} will have 20 events/yr with $\sim 10 \%$ measurability while the other LS models will have few to ten events at the same measurability level.  Table \ref{tab:No-hair-rate-et} quantitatively summarizes the prospects of  no-hair theorem tests with ET.

\subsection{BH spectroscopy with ET by measuring 3 or more QNM parameters}
\label{sec:3-plus-param-et}
\begin{figure*}
\hspace*{-1cm}
\includegraphics[width=0.45\textwidth]{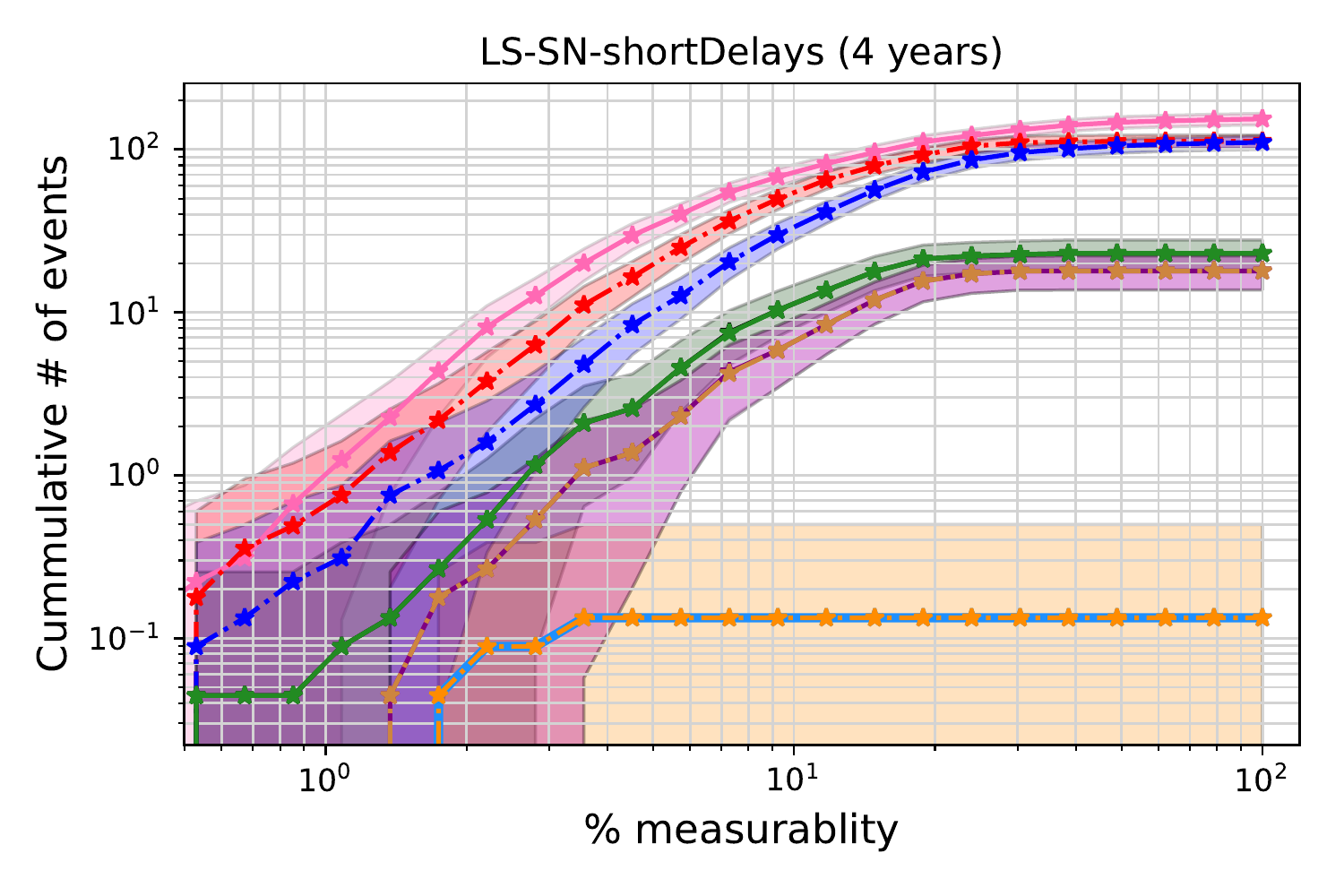}
\includegraphics[width=0.45\textwidth]{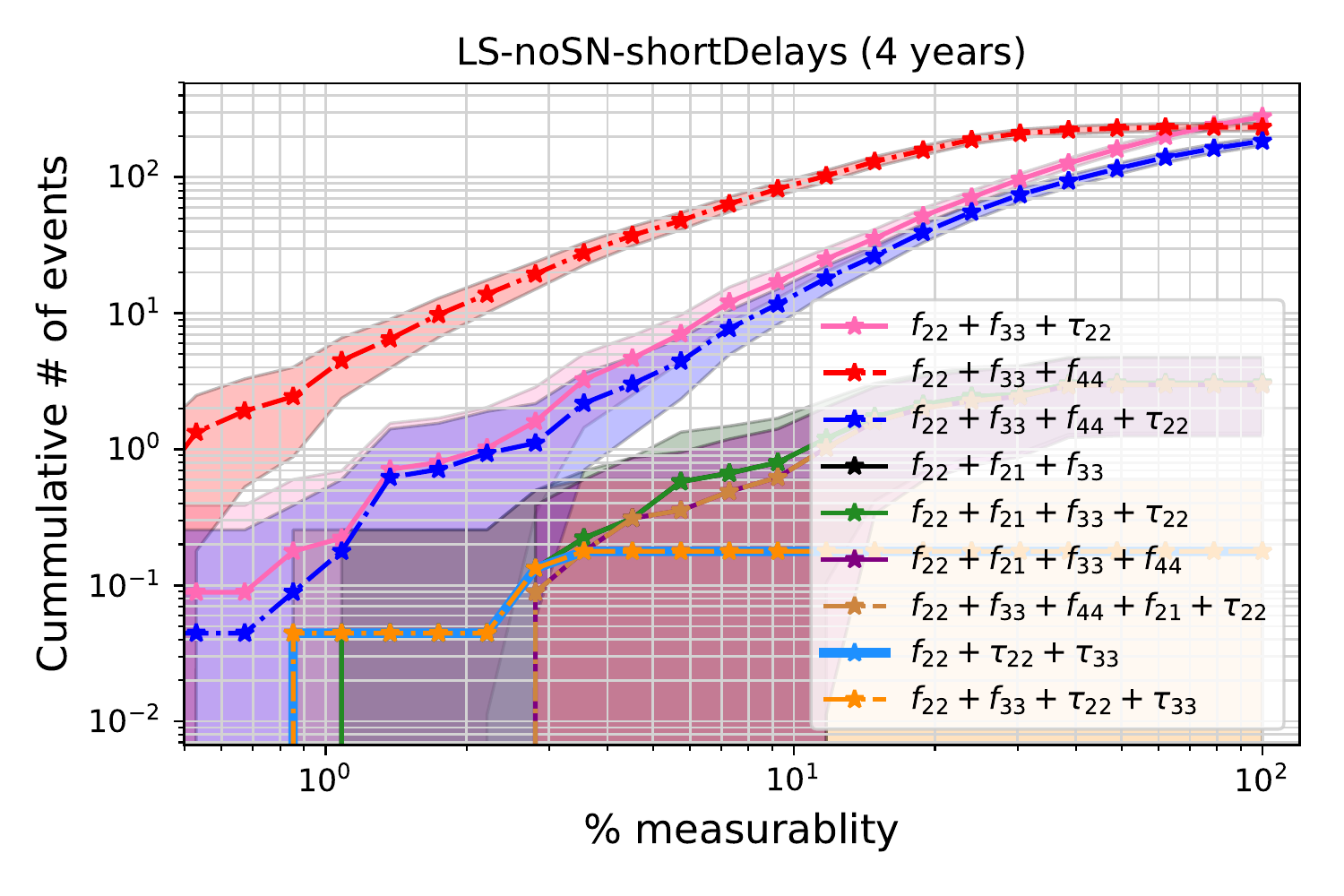} \\ 
\hspace*{-1cm}
\includegraphics[width=0.45\textwidth]{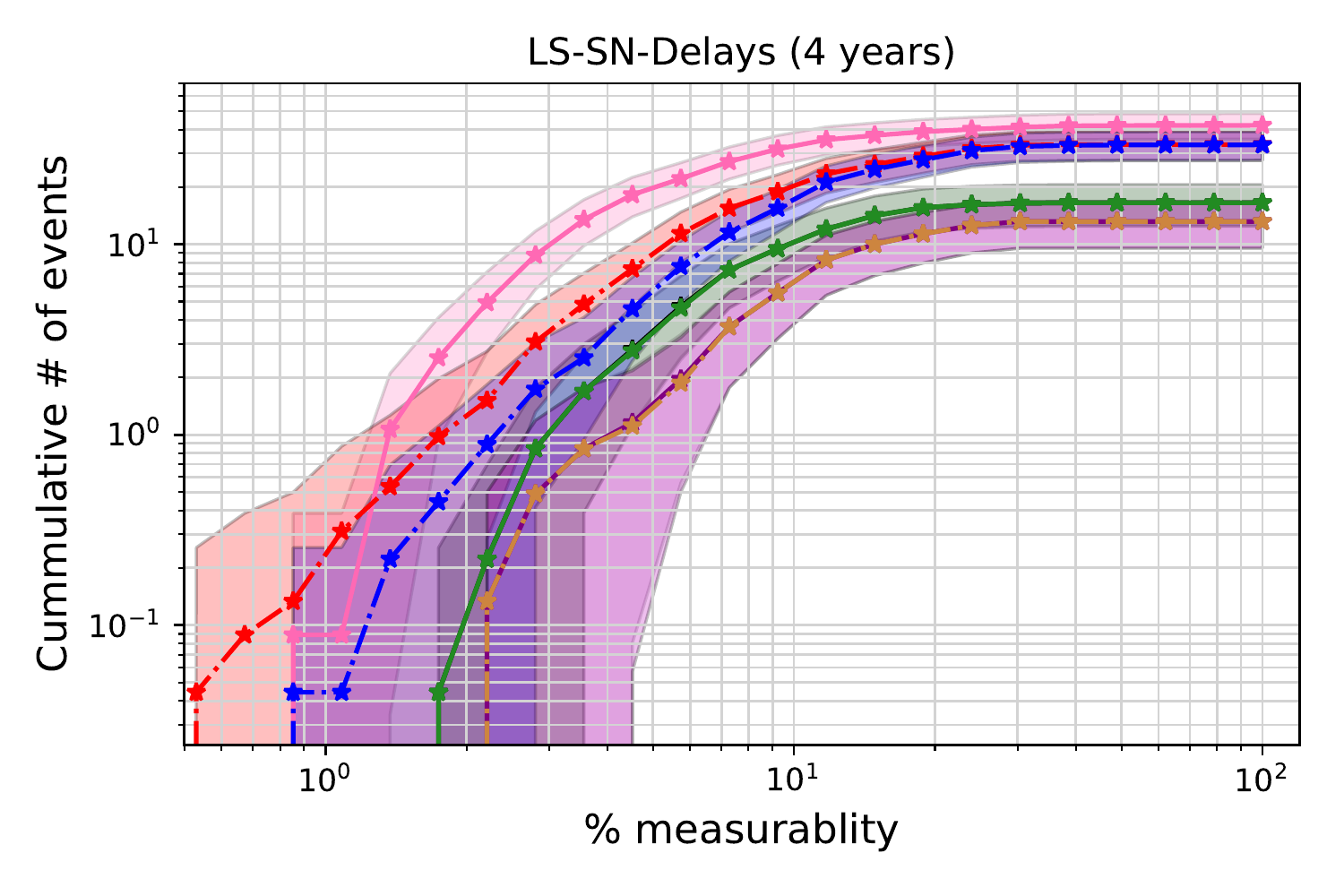}
\includegraphics[width=0.45\textwidth]{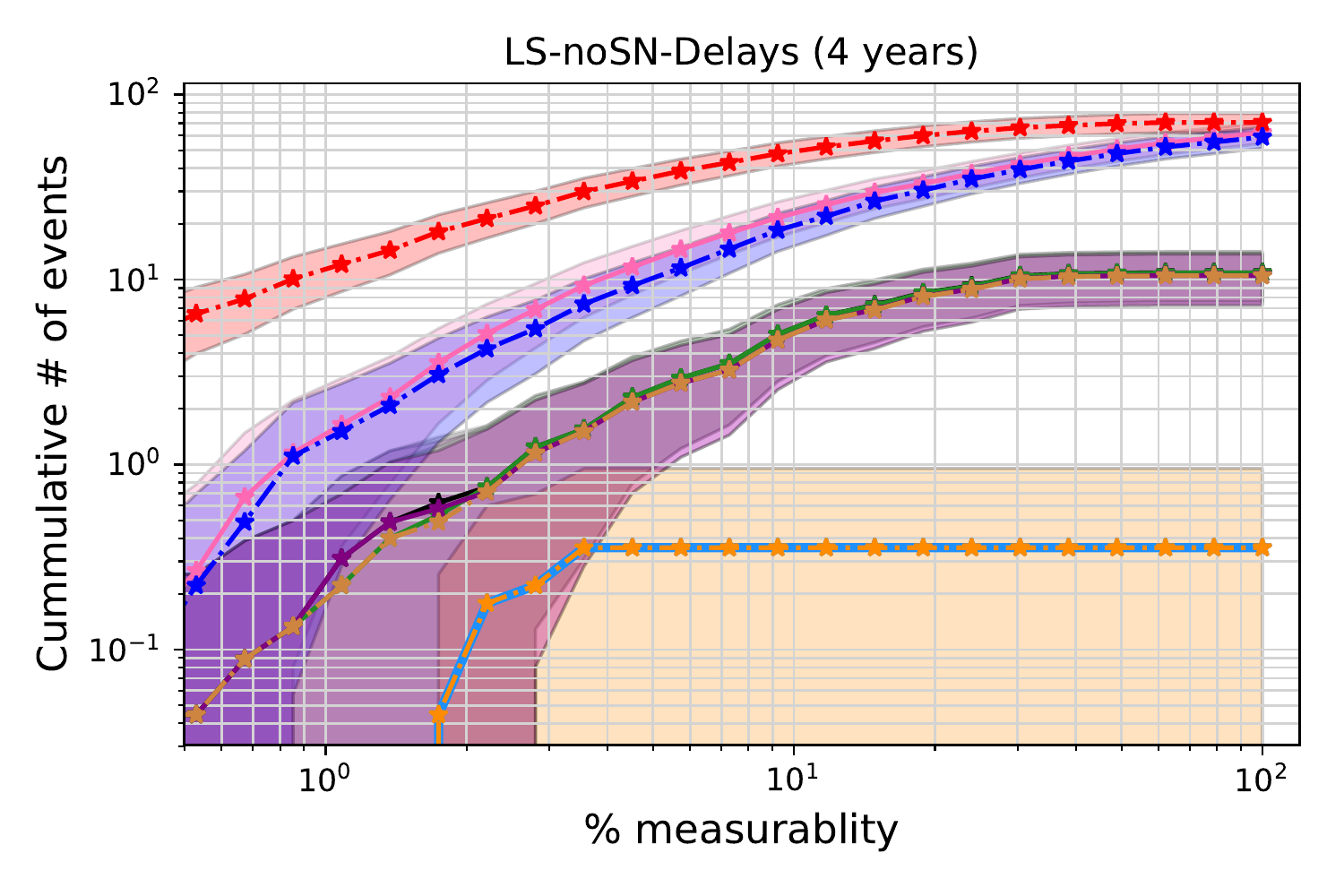}
\caption{Same as Fig.~\ref{fig:3-plus-param-lisa} but for ET and focusing on the LS scenario only. 
}
\label{fig:3-plus-param-et}
\end{figure*}

We explore the measurability for different combinations of the QNM parameters for the 4 LS models in Fig.~\ref{fig:3-plus-param-et} and discuss the results in this section. The results are presented for 4 years of ET data for the sake of comparison with the LISA results in the previous section, although the realistic lifespan of ET is expected to be longer. 

First, we focus on the \emph{SN models}. In this case, by comparing the pink, red and black curve in Fig.~\ref{fig:3-plus-param-et}, we see that the hierarchy of number of events allowing for measurability at a given level is as follows 
\begin{equation*}
\begin{aligned}
&\#(\{f_{22}, f_{33}, \tau_{22} \} )>\#(\{f_{22}, f_{33}, f_{44} \} )>\\
&\#(\{f_{22}, f_{21}, f_{33} \} )
\end{aligned}    
\end{equation*}
Note that measuring $f_{21}$ is the bottleneck in the latter combination. Also, by looking at the blue curve indicating the number of events that will allow for measurability of 4 QNM parameters $\{f_{22}, f_{33}, \tau_{22}, f_{44} \}$, we can infer that not all systems that allow for measuring the QNM parameter combination $ \{f_{22}, f_{33}, \tau_{22} \}$ will allow for $\{f_{22}, f_{33}, f_{44} \}$ combination simultaneously.  

We also note that the \emph{SN-shortDelays} model is best suited for ringdown analysis with ET. With both SN models, we expect $\sim 10$ events or more to allow for $5 \%$ measurability of $ \{f_{22}, f_{33}, \tau_{22} \}$ as well as of $\{f_{22}, f_{33}, f_{44} \}$. The intrinsic event rate of the \emph{SN-Delays} model is much smaller than that of the \emph{SN-shortDelays} model and this is reflected starkly into the number of expected events allowing for a higher level of measurability. For example, the \emph{SN-shortDelays} model predicts $\sim 80$ events
with $20 \%$ measurability for $\{ f_{22}, f_{33}, \tau_{22} \}$, while the \emph{SN-Delays} model predicts $\sim 40$ events. Furthermore, while 30 events of the \emph{SN-shortDelays} population allow for the estimation of QNM parameter combination $\{ f_{22}, f_{33}, \tau_{22}, f_{44} \}$  at $10 \%$ measurability, less than 20 events will allow for the same for \emph{SN-Delays}.

Let us now present the results for the \emph{noSN models}. In this case the hierarchy of number of events allowing for measurability at a given level is
\begin{equation*}
    \begin{aligned}
    &\#(\{f_{22}, f_{33}, f_{44} \})\gg\#(\{ f_{22}, f_{33}, \tau_{22}\})\sim\\
    &\#(\{f_{22}, f_{33}, f_{44}, \tau_{22}\})
    \end{aligned}
\end{equation*} For these models, $f_{44}$ is  measured significantly better than $\tau_{22} $ within our statistical sample.

Note that ringdowns from these populations are not the ideal candidates for the ET detector. About $\sim$20 events and 10 events allow for few-percent measurability of $\{f_{22}, f_{33}, f_{44}\}$ in the \emph{noSN-shortDelays} and \emph{noSN-Delays} models, respectively. At the less stringent level of $10 \%$ measurability of this QNM combination, we expect the \emph{noSN-shortDelays} model to have $\sim 100$ events in 4 years, whereas there are only $\sim 40-50$ events in the \emph{noSN-Delays} population. Moreover, in each of these models, $\sim 20$ events in 4-year data allow for measuring the QNM parameter combination $\{f_{22}, f_{33}, \tau_{22}\}$ at $10 \%$ level.

\subsection{BH spectroscopy with ET by measuring less than 3 parameters}
\label{sec:3-minus-param-et}
\begin{figure*}

\includegraphics[width=0.45\textwidth]{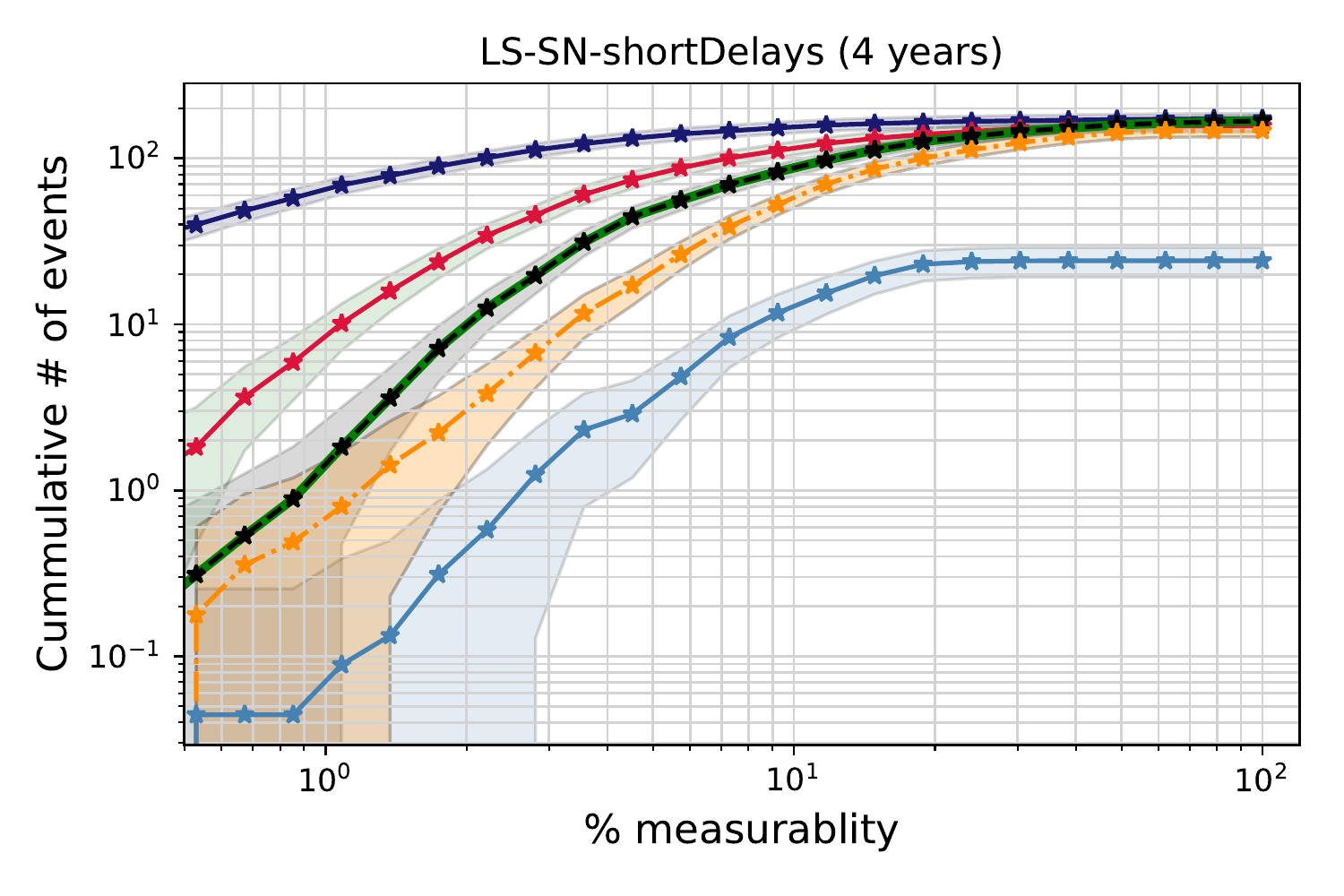}
\includegraphics[width=0.45\textwidth]{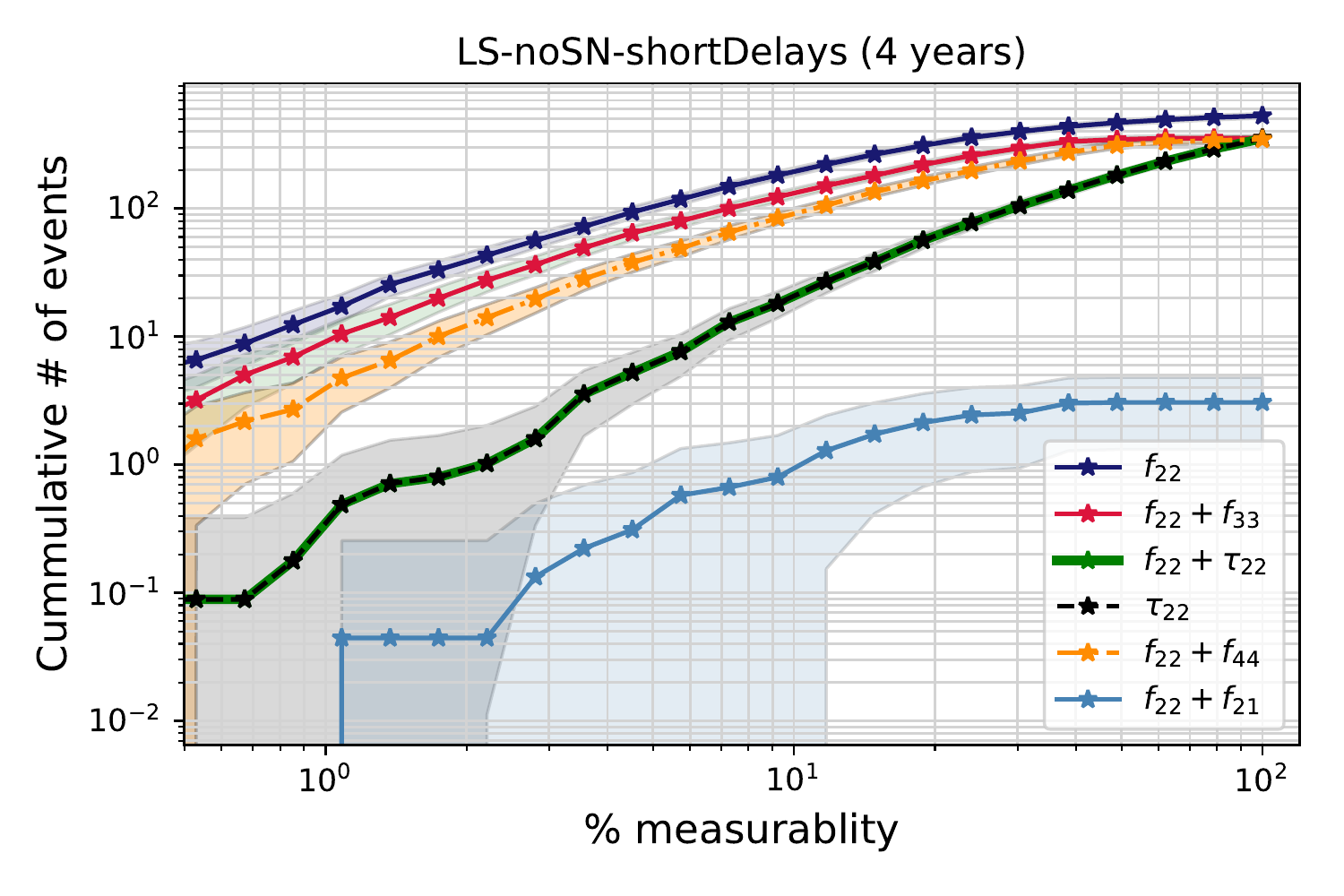} \\ 
\includegraphics[width=0.45\textwidth]{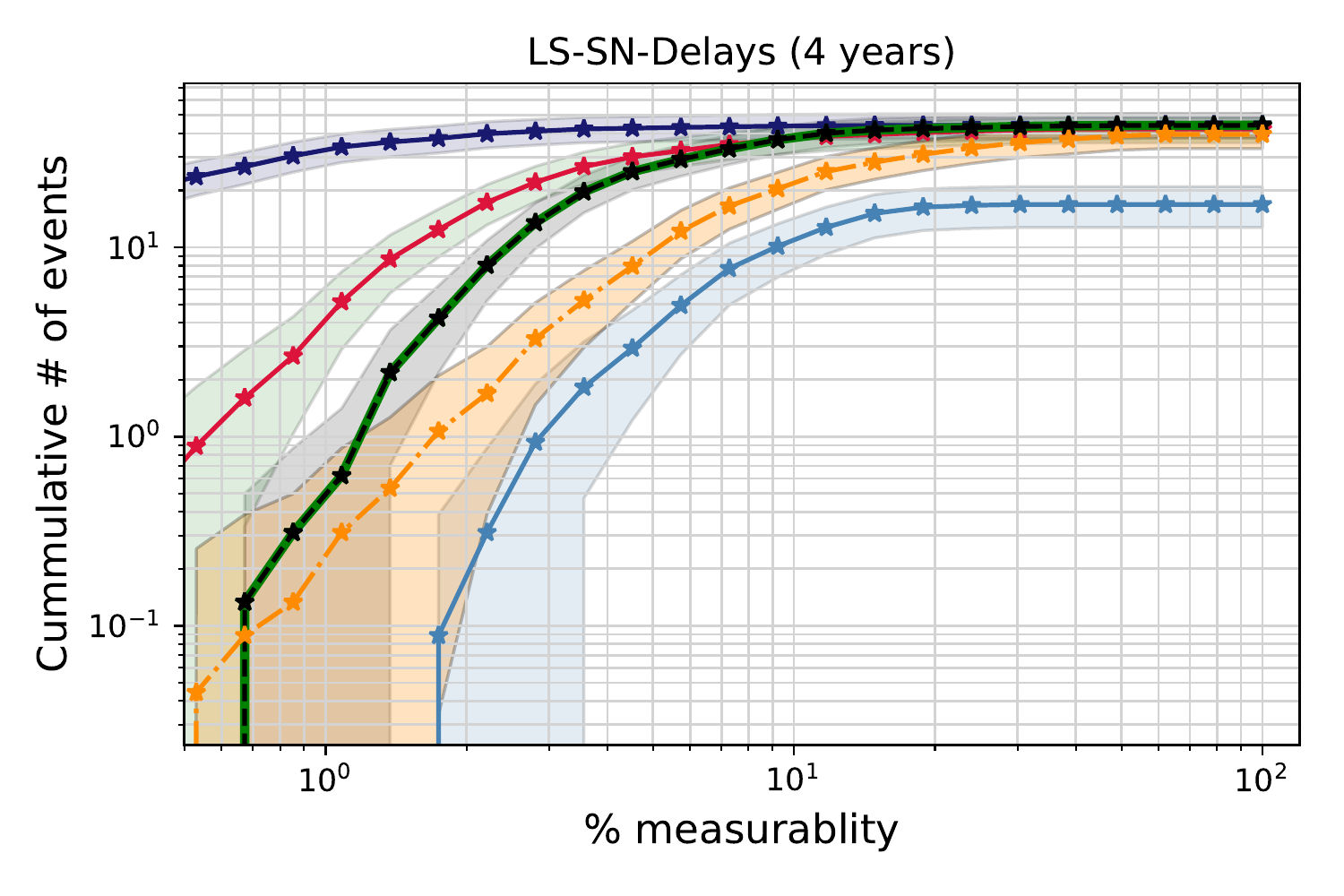}
\includegraphics[width=0.45\textwidth]{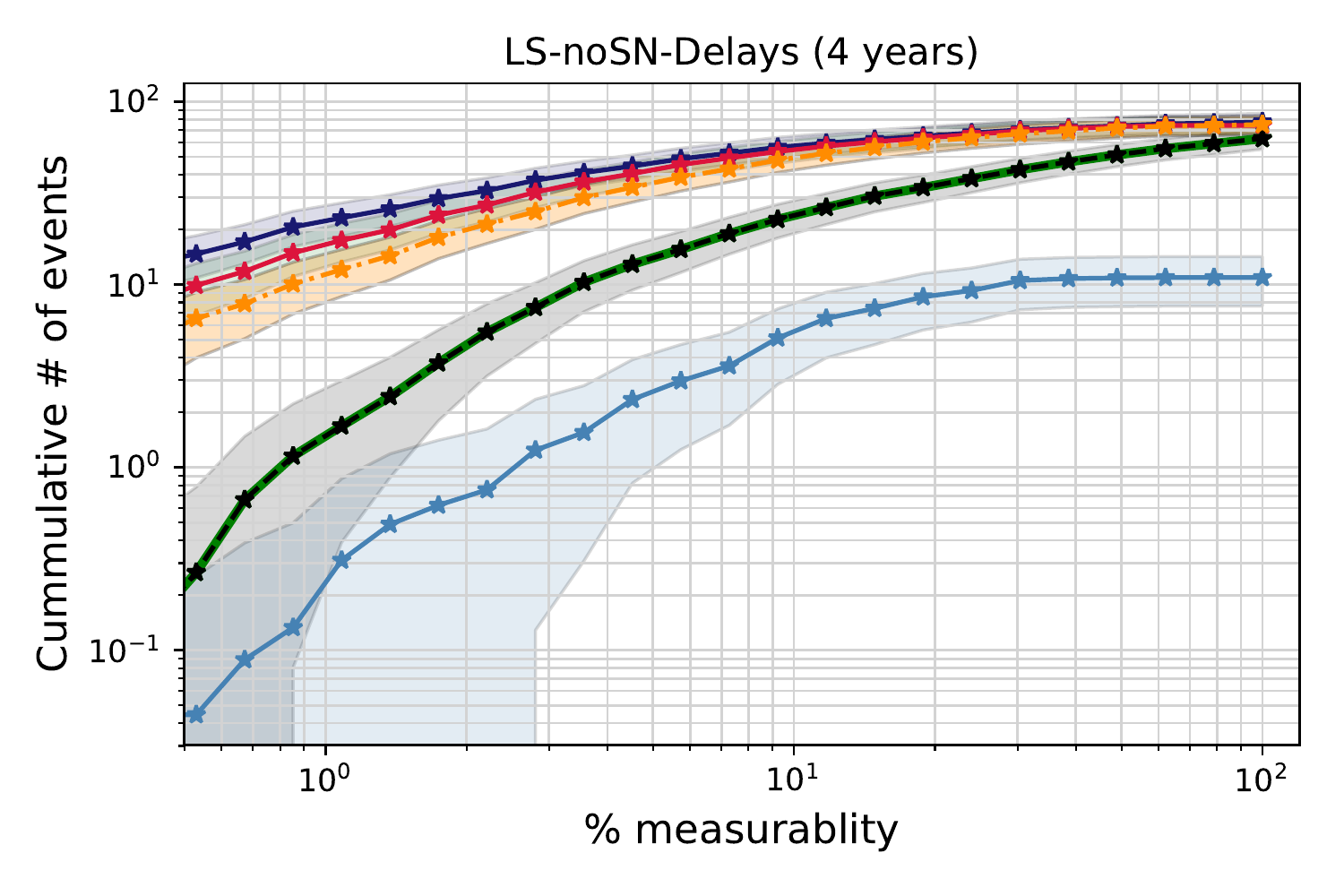} 

\caption{Same as Fig.~\ref{fig:3-minus-param-lisa} but for ET and focusing on the LS scenario only.}
\label{fig:3-minus-param-et}
\end{figure*}

Similarly to the study for LISA in Sec.~\ref{sec:3-minus-param-lisa}, in Fig.~\ref{fig:3-minus-param-et}, we focus on the events in the population that do not necessarily allow for the precision measurement of 3 QNM parameters. Again, the ringdown measurements of these events can be combined with the information obtained from the inspiral to perform tests of GR.\footnote{Note however that here the inspiral can be scarcely measured, because the ringdown frequency is at the lower end of the ET bandwidth. In this case, a multiwavelength approach~\cite{Sesana:2016ljz} would be needed, wherein the binary parameters can be measured by LISA detecting the inspiral and the subsequent ringdown could be measured by ET~\cite{Carson:2019kkh}.}

From Fig.~\ref{fig:3-minus-param-et}, we see that the trend of the number of events allowing for a certain level of measurability qualitatively changes between the \emph{SN} and \emph{noSN} models. For \emph{SN} models, the hierarchy goes as
\begin{equation*}
    \begin{aligned}
    \#(f_{22}) > \# (f_{33}) > \#(\tau_{22}) > \#(f_{44}) > \#(f_{21})  \,,
    \end{aligned}
\end{equation*}
while for the \emph{noSN} models it goes as
\begin{equation*}
    \begin{aligned}
    \#(f_{22}) > \#(f_{33}) > \#(f_{44}) > \#(\tau_{22}) > \#(f_{21})\,.
    \end{aligned}
\end{equation*}
Again this trend is consistent with the expectation we have by looking at the asymmetry of the progenitor binary system in Fig.~\ref{fig:assym} and relative measurement uncertainty in Fig.~\ref{fig:sigma}. 

From Fig.~\ref{fig:3-minus-param-et}, $\sim 1$ event within 4 years would allow for $0.1 \%$ measurability of $f_{22}$ for both the \emph{SN} models. Further, while the \emph{SN-shortDelays} model predicts $\sim 50$, 10 and 1 event in 4 years with $1 \%$ measurability of $f_{22}, \{ f_{22}, f_{33} \}$, and $\{f_{22}, \tau_{22} \}$, respectively, the \emph{SN-Delays} model predicts $\sim$ 30 and less than 10 with $1 \%$ measurability of $f_{22}$ and $\{ f_{22}, f_{33} \}$, respectively. 

For the \emph{noSN-shortDelays} model, we expect that $\sim 1$ event in 4 years would allow  for $0.2 \%$ measurability of $f_{22}$ and up to 10 events with subpercent measurability of $ \{f_{22}, f_{33} \}$. Moreover, approximately $100$ events will allow for $10 \%$ measurability of $\{ f_{22}, f_{33} \}$ and $\{ f_{22}, f_{44} \}$. On the other hand, for the \emph{noSN-Delays} model, $\sim 2$ events and up to few tens of events will respectively allow for $ 0.1 \%$ and $1 \%$ measurability for both $f_{22}$ and $f_{33}$. About 10 events will also allow for $1 \%$ measurability of $\{f_{22}, f_{44}\}$. Also, approximately $50$ events will allow for $10 \%$ measurability of $\{f_{22}, f_{33} \}$ and of $\{f_{22}, f_{44} \}$.

\section{Discussion}
\label{sec:conc}
One of LISA's main science goals is to probe the nature of massive compact objects and the nature of gravity through the ringdown of MBH merger remnants~\cite{LISA:2017pwj,Barausse:2020rsu}. While ringdown signals from MBHs with mass ${\cal O}(10^5)M_\odot$ are expected to have very large SNR in LISA, the actual rates for such events crucially depend on the underlying MBH populations. We have assessed LISA's ability to perform BH spectroscopy by measuring multiple ringdown modes in several scenarios that bracket current uncertainties in MBH population models. We found that the prospects for BH spectroscopy with LISA depend significantly on the underlying MBH population. 
In the HS scenario, we found that approximately ${\cal O}(100)$ (less than 1) events in LISA's effective operational time, i.e. 4 years of data, would allow for ringdown test at $1 \%$ (${\cal O}(0.1)\%$) precision for at least 3 independent QNM parameters.  Furthermore, the dominant mode frequency can be measured with ${\cal O}(0.1) \%$ precision in more than $100$ events in 4-year data. On the other hand, in the LS scenarios, only one model (\emph{noSN-shortDelays}) might allow for $1 \%$ precision measurement of 3 QNM parameters in $1-3$ events in 4-year data.

We argued that the unfavorable prospect is due to the smaller remnant mass predicted in the LS scenario compared to the HS one, which makes the ringdown SNR low in LISA bandwidth. Given the smaller remnant masses (i.e., higher QNM frequency) we explored the possibility that BH spectroscopy in the LS scenario might be performed with the next-generation ground-based interferometer ET. Interestingly, we found that a tail of the LS population will merge in the ET frequency window, allowing for 3-QNM spectroscopy at a few percent measurability in a few events/yr.
Furthermore, in the LS scenario ET could measure the fundamental mode alone at ${\cal O}(0.1)\%$ level in at least 1 event in 4 years, and perform a $2$-QNM measurement (useful for inspiral-merger-ringdown tests) at $1\%$ level in a handful of events. The exact numbers depend on the LS scenarios (see Figs.~\ref{fig:3-plus-param-et} and \ref{fig:3-minus-param-et}).  We note that the low-frequency end of ET's sensitivity curve is particularly relevant for these rates, since even the tail of the MBH population in the LS scenarios has support at higher masses compared to stellar-origin BHs. For this reason, we expect lower rates for CE, due to its expected lower sensitivity at low frequencies.

Clearly, the accuracy of these tests is much lower than what third-generation detectors will be able to do using stellar-origin BH binaries, but it is nevertheless interesting that a fraction of the population of light MBHs can be detected by the ground-based detectors. In particular, the light BHs predicted in the LS scenario call for a multi-wavelength GW approach~\cite{Sesana:2016ljz}: the low-frequency early inspiral can be detected by LISA and the subsequent ringdown could be measured by ET~\cite{Carson:2019kkh}. 
Nonetheless, if some LS scenarios turn out to be realized in nature, the best prospects for MBH spectroscopy would be reached in the deci-Hz band, for example by the proposed DECIGO~\cite{Kawamura:2011zz} or by similar concepts. We leave a detailed analysis of this case for a future work.

Another interesting result of our analysis is that BH spectroscopy with LISA can be performed beyond 3 QNM parameters (4 or even 5 parameters) with a high level of precision for favorable HS models. In fact, for a measurability threshold $\geq 1 \%$, the prospects of extracting 3 or 4 QNM parameters are similar; for instance, LISA will be able to measure at least 4 independent QNM parameters with less than $1\%$ uncertainty (in each parameter) for $\sim 100$ events. Additional independent measurements could be used to devise multiple null-hypothesis tests of the no-hair theorem. We have also considered the case of measuring $5$ QNM parameters of the first four dominant angular modes, those for which amplitude and phase fits are available. An interesting question is whether even more than 4-5 QNM parameters can be measured in the HS \emph{shortDelays} models. In future work, it would be interesting to explore the possibility of extracting the best 5 or more QNM parameters in this optimistic scenario and also to understand how many modes should the ringdown waveform include for optimal ringdown parameter estimation with LISA. This would require well-calibrated numerical-relativity fits of the amplitude ratios and the phase differences for several QNMs across the whole parameter space identified by the binary's mass ratio and individual spins. These results could eventually be used to select the optimal set of QNM parameters depending on the progenitor binary parameters as to inform tests of gravity with parametrized approaches~\cite{Maselli:2019mjd}.

Finally, we focused on BH spectroscopy with multiple QNMs from a \emph{single} source.
For particularly pessimistic scenarios (especially in the LS case) and in the absence of single golden event allowing for accurate measurability, stacking a statistical number of ringdowns~\cite{Brito:2018rfr,Yang:2017zxs,Carullo:2018sfu} with individually low measurability could be an alternative to performing no-hair theorem tests. For example, an event rate of $\approx 25$ events/yr with $10\%$ measurability can lead to an overall measurability of $1\%$ or better\footnote{Note that our criterion for $x\%$ measurability is that \emph{all} (say) 3 QNM parameters are measured with at least $x\%$ precision. This means that typically most of the QNM parameters under consideration are measured with greater precision, so stacking $N$ different events might lead to higher precision than the standard $N^{-1/2} x\%$ (assuming one is able to efficiently stack all $N$ events)} by stacking the $\approx 100$ events expected in 4 years. Assuming one is able to efficiently stack all these events, stacking might also be used in the HS scenario to reach sub$0.1\%$ precision.

\begin{acknowledgments}
The authors are grateful to Vishal Baibhav for discussion on EMOP energy and amplitude computations used in this work, and to Emanuele Berti and Vitor Cardoso for useful discussions. 
Numerical computations were performed at the Vera cluster of the Amaldi Research Center funded by the MIUR program ``Dipartimento di Eccellenza'' (CUP:~B81I18001170001).
S.B., C.P., and P.P. acknowledge the financial support provided under the European Union's H2020 ERC, Starting Grant agreement no.~DarkGRA--757480. We also acknowledge support under the MIUR PRIN and FARE programmes (GW- NEXT, CUP: B84I20000100001).
E.B. acknowledges financial support provided under the European Union's H2020 ERC Consolidator Grant ``GRavity from Astrophysical to Microscopic Scales'' grant agreement no. GRAMS-815673. This work was supported by the EU Horizon 2020 Research and Innovation Programme under the Marie Sklodowska-Curie Grant Agreement No. 101007855.

\end{acknowledgments}

\bibliography{0000.bib}

\begin{thebibliography}{120}%
\makeatletter
\providecommand \@ifxundefined [1]{%
 \@ifx{#1\undefined}
}%
\providecommand \@ifnum [1]{%
 \ifnum #1\expandafter \@firstoftwo
 \else \expandafter \@secondoftwo
 \fi
}%
\providecommand \@ifx [1]{%
 \ifx #1\expandafter \@firstoftwo
 \else \expandafter \@secondoftwo
 \fi
}%
\providecommand \natexlab [1]{#1}%
\providecommand \enquote  [1]{``#1''}%
\providecommand \bibnamefont  [1]{#1}%
\providecommand \bibfnamefont [1]{#1}%
\providecommand \citenamefont [1]{#1}%
\providecommand \href@noop [0]{\@secondoftwo}%
\providecommand \href [0]{\begingroup \@sanitize@url \@href}%
\providecommand \@href[1]{\@@startlink{#1}\@@href}%
\providecommand \@@href[1]{\endgroup#1\@@endlink}%
\providecommand \@sanitize@url [0]{\catcode `\\12\catcode `\$12\catcode
  `\&12\catcode `\#12\catcode `\^12\catcode `\_12\catcode `\%12\relax}%
\providecommand \@@startlink[1]{}%
\providecommand \@@endlink[0]{}%
\providecommand \url  [0]{\begingroup\@sanitize@url \@url }%
\providecommand \@url [1]{\endgroup\@href {#1}{\urlprefix }}%
\providecommand \urlprefix  [0]{URL }%
\providecommand \Eprint [0]{\href }%
\providecommand \doibase [0]{http://dx.doi.org/}%
\providecommand \selectlanguage [0]{\@gobble}%
\providecommand \bibinfo  [0]{\@secondoftwo}%
\providecommand \bibfield  [0]{\@secondoftwo}%
\providecommand \translation [1]{[#1]}%
\providecommand \BibitemOpen [0]{}%
\providecommand \bibitemStop [0]{}%
\providecommand \bibitemNoStop [0]{.\EOS\space}%
\providecommand \EOS [0]{\spacefactor3000\relax}%
\providecommand \BibitemShut  [1]{\csname bibitem#1\endcsname}%
\let\auto@bib@innerbib\@empty
\bibitem [{\citenamefont {{Chandrasekhar}}\ and\ \citenamefont
  {{Detweiler}}(1975)}]{QNM-Chandrasekhar}%
  \BibitemOpen
  \bibfield  {author} {\bibinfo {author} {\bibfnamefont {S.}~\bibnamefont
  {{Chandrasekhar}}}\ and\ \bibinfo {author} {\bibfnamefont {S.}~\bibnamefont
  {{Detweiler}}},\ }\href {\doibase 10.1098/rspa.1975.0112} {\bibfield
  {journal} {\bibinfo  {journal} {Proceedings of the Royal Society of London
  Series A}\ }\textbf {\bibinfo {volume} {344}},\ \bibinfo {pages} {441}
  (\bibinfo {year} {1975})}\BibitemShut {NoStop}%
\bibitem [{\citenamefont {{Vishveshwara}}(1970)}]{Vishveshwara}%
  \BibitemOpen
  \bibfield  {author} {\bibinfo {author} {\bibfnamefont {C.~V.}\ \bibnamefont
  {{Vishveshwara}}},\ }\href {\doibase 10.1038/227936a0} {\bibfield  {journal}
  {\bibinfo  {journal} {\nat}\ }\textbf {\bibinfo {volume} {227}},\ \bibinfo
  {pages} {936} (\bibinfo {year} {1970})}\BibitemShut {NoStop}%
\bibitem [{\citenamefont {Kokkotas}\ and\ \citenamefont
  {Schmidt}(1999)}]{Kokkotas-review}%
  \BibitemOpen
  \bibfield  {author} {\bibinfo {author} {\bibfnamefont {K.~D.}\ \bibnamefont
  {Kokkotas}}\ and\ \bibinfo {author} {\bibfnamefont {B.~G.}\ \bibnamefont
  {Schmidt}},\ }\href {\doibase 10.12942/lrr-1999-2} {\bibfield  {journal}
  {\bibinfo  {journal} {Living Rev. Rel.}\ }\textbf {\bibinfo {volume} {2}},\
  \bibinfo {pages} {2} (\bibinfo {year} {1999})},\ \Eprint
  {http://arxiv.org/abs/gr-qc/9909058} {arXiv:gr-qc/9909058} \BibitemShut
  {NoStop}%
\bibitem [{\citenamefont {Berti}\ \emph {et~al.}(2009)\citenamefont {Berti},
  \citenamefont {Cardoso},\ and\ \citenamefont {Starinets}}]{Berti:2009kk}%
  \BibitemOpen
  \bibfield  {author} {\bibinfo {author} {\bibfnamefont {E.}~\bibnamefont
  {Berti}}, \bibinfo {author} {\bibfnamefont {V.}~\bibnamefont {Cardoso}}, \
  and\ \bibinfo {author} {\bibfnamefont {A.~O.}\ \bibnamefont {Starinets}},\
  }\href {\doibase 10.1088/0264-9381/26/16/163001} {\bibfield  {journal}
  {\bibinfo  {journal} {Class. Quant. Grav.}\ }\textbf {\bibinfo {volume}
  {26}},\ \bibinfo {pages} {163001} (\bibinfo {year} {2009})},\ \Eprint
  {http://arxiv.org/abs/0905.2975} {arXiv:0905.2975 [gr-qc]} \BibitemShut
  {NoStop}%
\bibitem [{\citenamefont {Carter}(1971)}]{No-hair-original}%
  \BibitemOpen
  \bibfield  {author} {\bibinfo {author} {\bibfnamefont {B.}~\bibnamefont
  {Carter}},\ }\href {\doibase 10.1103/PhysRevLett.26.331} {\bibfield
  {journal} {\bibinfo  {journal} {Phys. Rev. Lett.}\ }\textbf {\bibinfo
  {volume} {26}},\ \bibinfo {pages} {331} (\bibinfo {year} {1971})}\BibitemShut
  {NoStop}%
\bibitem [{\citenamefont {Nollert}(1999)}]{QNM-review-first-tgr-proposal}%
  \BibitemOpen
  \bibfield  {author} {\bibinfo {author} {\bibfnamefont {H.-P.}\ \bibnamefont
  {Nollert}},\ }\href {\doibase 10.1088/0264-9381/16/12/201} {\ \textbf
  {\bibinfo {volume} {16}},\ \bibinfo {pages} {R159} (\bibinfo {year}
  {1999})}\BibitemShut {NoStop}%
\bibitem [{\citenamefont {Berti}\ \emph {et~al.}(2018)\citenamefont {Berti},
  \citenamefont {Yagi}, \citenamefont {Yang},\ and\ \citenamefont
  {Yunes}}]{Berti:2018vdi}%
  \BibitemOpen
  \bibfield  {author} {\bibinfo {author} {\bibfnamefont {E.}~\bibnamefont
  {Berti}}, \bibinfo {author} {\bibfnamefont {K.}~\bibnamefont {Yagi}},
  \bibinfo {author} {\bibfnamefont {H.}~\bibnamefont {Yang}}, \ and\ \bibinfo
  {author} {\bibfnamefont {N.}~\bibnamefont {Yunes}},\ }\href {\doibase
  10.1007/s10714-018-2372-6} {\bibfield  {journal} {\bibinfo  {journal} {Gen.
  Rel. Grav.}\ }\textbf {\bibinfo {volume} {50}},\ \bibinfo {pages} {49}
  (\bibinfo {year} {2018})},\ \Eprint {http://arxiv.org/abs/1801.03587}
  {arXiv:1801.03587 [gr-qc]} \BibitemShut {NoStop}%
\bibitem [{\citenamefont {Li}\ \emph {et~al.}(2012)\citenamefont {Li},
  \citenamefont {Del~Pozzo}, \citenamefont {Vitale}, \citenamefont {Van
  Den~Broeck}, \citenamefont {Agathos}, \citenamefont {Veitch}, \citenamefont
  {Grover}, \citenamefont {Sidery}, \citenamefont {Sturani},\ and\
  \citenamefont {Vecchio}}]{PhysRevD.85.082003}%
  \BibitemOpen
  \bibfield  {author} {\bibinfo {author} {\bibfnamefont {T.~G.~F.}\
  \bibnamefont {Li}}, \bibinfo {author} {\bibfnamefont {W.}~\bibnamefont
  {Del~Pozzo}}, \bibinfo {author} {\bibfnamefont {S.}~\bibnamefont {Vitale}},
  \bibinfo {author} {\bibfnamefont {C.}~\bibnamefont {Van Den~Broeck}},
  \bibinfo {author} {\bibfnamefont {M.}~\bibnamefont {Agathos}}, \bibinfo
  {author} {\bibfnamefont {J.}~\bibnamefont {Veitch}}, \bibinfo {author}
  {\bibfnamefont {K.}~\bibnamefont {Grover}}, \bibinfo {author} {\bibfnamefont
  {T.}~\bibnamefont {Sidery}}, \bibinfo {author} {\bibfnamefont
  {R.}~\bibnamefont {Sturani}}, \ and\ \bibinfo {author} {\bibfnamefont
  {A.}~\bibnamefont {Vecchio}},\ }\href {\doibase 10.1103/PhysRevD.85.082003}
  {\bibfield  {journal} {\bibinfo  {journal} {Phys. Rev. D}\ }\textbf {\bibinfo
  {volume} {85}},\ \bibinfo {pages} {082003} (\bibinfo {year}
  {2012})}\BibitemShut {NoStop}%
\bibitem [{\citenamefont {Maselli}\ \emph {et~al.}(2020)\citenamefont
  {Maselli}, \citenamefont {Pani}, \citenamefont {Gualtieri},\ and\
  \citenamefont {Berti}}]{Maselli:2019mjd}%
  \BibitemOpen
  \bibfield  {author} {\bibinfo {author} {\bibfnamefont {A.}~\bibnamefont
  {Maselli}}, \bibinfo {author} {\bibfnamefont {P.}~\bibnamefont {Pani}},
  \bibinfo {author} {\bibfnamefont {L.}~\bibnamefont {Gualtieri}}, \ and\
  \bibinfo {author} {\bibfnamefont {E.}~\bibnamefont {Berti}},\ }\href
  {\doibase 10.1103/PhysRevD.101.024043} {\bibfield  {journal} {\bibinfo
  {journal} {Phys. Rev. D}\ }\textbf {\bibinfo {volume} {101}},\ \bibinfo
  {pages} {024043} (\bibinfo {year} {2020})},\ \Eprint
  {http://arxiv.org/abs/1910.12893} {arXiv:1910.12893 [gr-qc]} \BibitemShut
  {NoStop}%
\bibitem [{\citenamefont {V\"olkel}\ and\ \citenamefont
  {Barausse}(2020)}]{Volkel:2020daa}%
  \BibitemOpen
  \bibfield  {author} {\bibinfo {author} {\bibfnamefont {S.~H.}\ \bibnamefont
  {V\"olkel}}\ and\ \bibinfo {author} {\bibfnamefont {E.}~\bibnamefont
  {Barausse}},\ }\href {\doibase 10.1103/PhysRevD.102.084025} {\bibfield
  {journal} {\bibinfo  {journal} {Phys. Rev. D}\ }\textbf {\bibinfo {volume}
  {102}},\ \bibinfo {pages} {084025} (\bibinfo {year} {2020})},\ \Eprint
  {http://arxiv.org/abs/2007.02986} {arXiv:2007.02986 [gr-qc]} \BibitemShut
  {NoStop}%
\bibitem [{\citenamefont {Berti}\ \emph {et~al.}(2015)\citenamefont {Berti}
  \emph {et~al.}}]{Berti:2015itd}%
  \BibitemOpen
  \bibfield  {author} {\bibinfo {author} {\bibfnamefont {E.}~\bibnamefont
  {Berti}} \emph {et~al.},\ }\href {\doibase 10.1088/0264-9381/32/24/243001}
  {\bibfield  {journal} {\bibinfo  {journal} {Class. Quant. Grav.}\ }\textbf
  {\bibinfo {volume} {32}},\ \bibinfo {pages} {243001} (\bibinfo {year}
  {2015})},\ \Eprint {http://arxiv.org/abs/1501.07274} {arXiv:1501.07274
  [gr-qc]} \BibitemShut {NoStop}%
\bibitem [{\citenamefont {Yagi}\ and\ \citenamefont
  {Stein}(2016)}]{Yagi:2016jml}%
  \BibitemOpen
  \bibfield  {author} {\bibinfo {author} {\bibfnamefont {K.}~\bibnamefont
  {Yagi}}\ and\ \bibinfo {author} {\bibfnamefont {L.~C.}\ \bibnamefont
  {Stein}},\ }\href {\doibase 10.1088/0264-9381/33/5/054001} {\bibfield
  {journal} {\bibinfo  {journal} {Class. Quant. Grav.}\ }\textbf {\bibinfo
  {volume} {33}},\ \bibinfo {pages} {054001} (\bibinfo {year} {2016})},\
  \Eprint {http://arxiv.org/abs/1602.02413} {arXiv:1602.02413 [gr-qc]}
  \BibitemShut {NoStop}%
\bibitem [{\citenamefont {Barausse}\ and\ \citenamefont
  {Sotiriou}(2008)}]{Barausse:2008xv}%
  \BibitemOpen
  \bibfield  {author} {\bibinfo {author} {\bibfnamefont {E.}~\bibnamefont
  {Barausse}}\ and\ \bibinfo {author} {\bibfnamefont {T.~P.}\ \bibnamefont
  {Sotiriou}},\ }\href {\doibase 10.1103/PhysRevLett.101.099001} {\bibfield
  {journal} {\bibinfo  {journal} {Phys. Rev. Lett.}\ }\textbf {\bibinfo
  {volume} {101}},\ \bibinfo {pages} {099001} (\bibinfo {year} {2008})},\
  \Eprint {http://arxiv.org/abs/0803.3433} {arXiv:0803.3433 [gr-qc]}
  \BibitemShut {NoStop}%
\bibitem [{\citenamefont {Pani}\ and\ \citenamefont
  {Cardoso}(2009)}]{Pani:2009wy}%
  \BibitemOpen
  \bibfield  {author} {\bibinfo {author} {\bibfnamefont {P.}~\bibnamefont
  {Pani}}\ and\ \bibinfo {author} {\bibfnamefont {V.}~\bibnamefont {Cardoso}},\
  }\href {\doibase 10.1103/PhysRevD.79.084031} {\bibfield  {journal} {\bibinfo
  {journal} {Phys. Rev. D}\ }\textbf {\bibinfo {volume} {79}},\ \bibinfo
  {pages} {084031} (\bibinfo {year} {2009})},\ \Eprint
  {http://arxiv.org/abs/0902.1569} {arXiv:0902.1569 [gr-qc]} \BibitemShut
  {NoStop}%
\bibitem [{\citenamefont {Barausse}\ \emph {et~al.}(2014)\citenamefont
  {Barausse}, \citenamefont {Cardoso},\ and\ \citenamefont
  {Pani}}]{Barausse:2014tra}%
  \BibitemOpen
  \bibfield  {author} {\bibinfo {author} {\bibfnamefont {E.}~\bibnamefont
  {Barausse}}, \bibinfo {author} {\bibfnamefont {V.}~\bibnamefont {Cardoso}}, \
  and\ \bibinfo {author} {\bibfnamefont {P.}~\bibnamefont {Pani}},\ }\href
  {\doibase 10.1103/PhysRevD.89.104059} {\bibfield  {journal} {\bibinfo
  {journal} {Phys. Rev. D}\ }\textbf {\bibinfo {volume} {89}},\ \bibinfo
  {pages} {104059} (\bibinfo {year} {2014})},\ \Eprint
  {http://arxiv.org/abs/1404.7149} {arXiv:1404.7149 [gr-qc]} \BibitemShut
  {NoStop}%
\bibitem [{\citenamefont {Cardoso}\ and\ \citenamefont
  {Pani}(2019)}]{Cardoso:2019rvt}%
  \BibitemOpen
  \bibfield  {author} {\bibinfo {author} {\bibfnamefont {V.}~\bibnamefont
  {Cardoso}}\ and\ \bibinfo {author} {\bibfnamefont {P.}~\bibnamefont {Pani}},\
  }\href {\doibase 10.1007/s41114-019-0020-4} {\bibfield  {journal} {\bibinfo
  {journal} {Living Rev. Rel.}\ }\textbf {\bibinfo {volume} {22}},\ \bibinfo
  {pages} {4} (\bibinfo {year} {2019})},\ \Eprint
  {http://arxiv.org/abs/1904.05363} {arXiv:1904.05363 [gr-qc]} \BibitemShut
  {NoStop}%
\bibitem [{\citenamefont {Cardoso}\ \emph {et~al.}(2016)\citenamefont
  {Cardoso}, \citenamefont {Hopper}, \citenamefont {Macedo}, \citenamefont
  {Palenzuela},\ and\ \citenamefont {Pani}}]{Cardoso:2016oxy}%
  \BibitemOpen
  \bibfield  {author} {\bibinfo {author} {\bibfnamefont {V.}~\bibnamefont
  {Cardoso}}, \bibinfo {author} {\bibfnamefont {S.}~\bibnamefont {Hopper}},
  \bibinfo {author} {\bibfnamefont {C.~F.~B.}\ \bibnamefont {Macedo}}, \bibinfo
  {author} {\bibfnamefont {C.}~\bibnamefont {Palenzuela}}, \ and\ \bibinfo
  {author} {\bibfnamefont {P.}~\bibnamefont {Pani}},\ }\href {\doibase
  10.1103/PhysRevD.94.084031} {\bibfield  {journal} {\bibinfo  {journal} {Phys.
  Rev. D}\ }\textbf {\bibinfo {volume} {94}},\ \bibinfo {pages} {084031}
  (\bibinfo {year} {2016})},\ \Eprint {http://arxiv.org/abs/1608.08637}
  {arXiv:1608.08637 [gr-qc]} \BibitemShut {NoStop}%
\bibitem [{\citenamefont {{Maggio}}\ \emph {et~al.}(2021)\citenamefont
  {{Maggio}}, \citenamefont {{Pani}},\ and\ \citenamefont
  {{Raposo}}}]{2021arXiv210506410M}%
  \BibitemOpen
  \bibfield  {author} {\bibinfo {author} {\bibfnamefont {E.}~\bibnamefont
  {{Maggio}}}, \bibinfo {author} {\bibfnamefont {P.}~\bibnamefont {{Pani}}}, \
  and\ \bibinfo {author} {\bibfnamefont {G.}~\bibnamefont {{Raposo}}},\
  }\href@noop {} {\bibfield  {journal} {\bibinfo  {journal} {arXiv e-prints}\
  ,\ \bibinfo {eid} {arXiv:2105.06410}} (\bibinfo {year} {2021})},\ \Eprint
  {http://arxiv.org/abs/2105.06410} {arXiv:2105.06410 [gr-qc]} \BibitemShut
  {NoStop}%
\bibitem [{\citenamefont {Maggio}\ \emph {et~al.}(2020)\citenamefont {Maggio},
  \citenamefont {Buoninfante}, \citenamefont {Mazumdar},\ and\ \citenamefont
  {Pani}}]{Maggio:2020jml}%
  \BibitemOpen
  \bibfield  {author} {\bibinfo {author} {\bibfnamefont {E.}~\bibnamefont
  {Maggio}}, \bibinfo {author} {\bibfnamefont {L.}~\bibnamefont {Buoninfante}},
  \bibinfo {author} {\bibfnamefont {A.}~\bibnamefont {Mazumdar}}, \ and\
  \bibinfo {author} {\bibfnamefont {P.}~\bibnamefont {Pani}},\ }\href {\doibase
  10.1103/PhysRevD.102.064053} {\bibfield  {journal} {\bibinfo  {journal}
  {Phys. Rev. D}\ }\textbf {\bibinfo {volume} {102}},\ \bibinfo {pages}
  {064053} (\bibinfo {year} {2020})},\ \Eprint
  {http://arxiv.org/abs/2006.14628} {arXiv:2006.14628 [gr-qc]} \BibitemShut
  {NoStop}%
\bibitem [{\citenamefont {{Bustillo}}\ \emph {et~al.}(2021)\citenamefont
  {{Bustillo}}, \citenamefont {{Lasky}},\ and\ \citenamefont
  {{Thrane}}}]{2021PhRvD.103b4041B}%
  \BibitemOpen
  \bibfield  {author} {\bibinfo {author} {\bibfnamefont {J.~C.}\ \bibnamefont
  {{Bustillo}}}, \bibinfo {author} {\bibfnamefont {P.~D.}\ \bibnamefont
  {{Lasky}}}, \ and\ \bibinfo {author} {\bibfnamefont {E.}~\bibnamefont
  {{Thrane}}},\ }\href {\doibase 10.1103/PhysRevD.103.024041} {\bibfield
  {journal} {\bibinfo  {journal} {\prd}\ }\textbf {\bibinfo {volume} {103}},\
  \bibinfo {eid} {024041} (\bibinfo {year} {2021})}\BibitemShut {NoStop}%
\bibitem [{\citenamefont {{Cardoso}}\ and\ \citenamefont
  {{Pani}}(2019)}]{2019LRR....22....4C}%
  \BibitemOpen
  \bibfield  {author} {\bibinfo {author} {\bibfnamefont {V.}~\bibnamefont
  {{Cardoso}}}\ and\ \bibinfo {author} {\bibfnamefont {P.}~\bibnamefont
  {{Pani}}},\ }\href {\doibase 10.1007/s41114-019-0020-4} {\bibfield  {journal}
  {\bibinfo  {journal} {Living Reviews in Relativity}\ }\textbf {\bibinfo
  {volume} {22}},\ \bibinfo {eid} {4} (\bibinfo {year} {2019})},\ \Eprint
  {http://arxiv.org/abs/1904.05363} {arXiv:1904.05363 [gr-qc]} \BibitemShut
  {NoStop}%
\bibitem [{\citenamefont {Abbott}\ \emph {et~al.}(2016)\citenamefont {Abbott}
  \emph {et~al.}}]{TGR-gw150914}%
  \BibitemOpen
  \bibfield  {author} {\bibinfo {author} {\bibfnamefont {B.~P.}\ \bibnamefont
  {Abbott}} \emph {et~al.} (\bibinfo {collaboration} {LIGO Scientific,
  Virgo}),\ }\href {\doibase 10.1103/PhysRevLett.116.221101} {\bibfield
  {journal} {\bibinfo  {journal} {Phys. Rev. Lett.}\ }\textbf {\bibinfo
  {volume} {116}},\ \bibinfo {pages} {221101} (\bibinfo {year} {2016})},\
  \bibinfo {note} {[Erratum: Phys.Rev.Lett. 121, 129902 (2018)]},\ \Eprint
  {http://arxiv.org/abs/1602.03841} {arXiv:1602.03841 [gr-qc]} \BibitemShut
  {NoStop}%
\bibitem [{\citenamefont {Ghosh}\ \emph {et~al.}(2021)\citenamefont {Ghosh},
  \citenamefont {Brito},\ and\ \citenamefont {Buonanno}}]{Ghosh:2021mrv}%
  \BibitemOpen
  \bibfield  {author} {\bibinfo {author} {\bibfnamefont {A.}~\bibnamefont
  {Ghosh}}, \bibinfo {author} {\bibfnamefont {R.}~\bibnamefont {Brito}}, \ and\
  \bibinfo {author} {\bibfnamefont {A.}~\bibnamefont {Buonanno}},\ }\href
  {\doibase 10.1103/PhysRevD.103.124041} {\bibfield  {journal} {\bibinfo
  {journal} {Phys. Rev. D}\ }\textbf {\bibinfo {volume} {103}},\ \bibinfo
  {pages} {124041} (\bibinfo {year} {2021})},\ \Eprint
  {http://arxiv.org/abs/2104.01906} {arXiv:2104.01906 [gr-qc]} \BibitemShut
  {NoStop}%
\bibitem [{\citenamefont {Ghosh}\ \emph {et~al.}(2016)\citenamefont {Ghosh}
  \emph {et~al.}}]{Ghosh:2016qgn}%
  \BibitemOpen
  \bibfield  {author} {\bibinfo {author} {\bibfnamefont {A.}~\bibnamefont
  {Ghosh}} \emph {et~al.},\ }\href {\doibase 10.1103/PhysRevD.94.021101}
  {\bibfield  {journal} {\bibinfo  {journal} {Phys. Rev. D}\ }\textbf {\bibinfo
  {volume} {94}},\ \bibinfo {pages} {021101} (\bibinfo {year} {2016})},\
  \Eprint {http://arxiv.org/abs/1602.02453} {arXiv:1602.02453 [gr-qc]}
  \BibitemShut {NoStop}%
\bibitem [{\citenamefont {Abbott}\ \emph {et~al.}(2019)\citenamefont {Abbott}
  \emph {et~al.}}]{LIGOScientific:2019fpa}%
  \BibitemOpen
  \bibfield  {author} {\bibinfo {author} {\bibfnamefont {B.~P.}\ \bibnamefont
  {Abbott}} \emph {et~al.} (\bibinfo {collaboration} {LIGO Scientific,
  Virgo}),\ }\href {\doibase 10.1103/PhysRevD.100.104036} {\bibfield  {journal}
  {\bibinfo  {journal} {Phys. Rev. D}\ }\textbf {\bibinfo {volume} {100}},\
  \bibinfo {pages} {104036} (\bibinfo {year} {2019})},\ \Eprint
  {http://arxiv.org/abs/1903.04467} {arXiv:1903.04467 [gr-qc]} \BibitemShut
  {NoStop}%
\bibitem [{\citenamefont {Abbott}\ \emph
  {et~al.}(2021{\natexlab{a}})\citenamefont {Abbott} \emph
  {et~al.}}]{LIGOScientific:2020tif}%
  \BibitemOpen
  \bibfield  {author} {\bibinfo {author} {\bibfnamefont {R.}~\bibnamefont
  {Abbott}} \emph {et~al.} (\bibinfo {collaboration} {LIGO Scientific,
  Virgo}),\ }\href {\doibase 10.1103/PhysRevD.103.122002} {\bibfield  {journal}
  {\bibinfo  {journal} {Phys. Rev. D}\ }\textbf {\bibinfo {volume} {103}},\
  \bibinfo {pages} {122002} (\bibinfo {year} {2021}{\natexlab{a}})},\ \Eprint
  {http://arxiv.org/abs/2010.14529} {arXiv:2010.14529 [gr-qc]} \BibitemShut
  {NoStop}%
\bibitem [{\citenamefont {Abbott}\ \emph
  {et~al.}(2021{\natexlab{b}})\citenamefont {Abbott} \emph
  {et~al.}}]{LIGOScientific:2021sio}%
  \BibitemOpen
  \bibfield  {author} {\bibinfo {author} {\bibfnamefont {R.}~\bibnamefont
  {Abbott}} \emph {et~al.} (\bibinfo {collaboration} {LIGO Scientific, VIRGO,
  KAGRA}),\ }\href@noop {} {\  (\bibinfo {year} {2021}{\natexlab{b}})},\
  \Eprint {http://arxiv.org/abs/2112.06861} {arXiv:2112.06861 [gr-qc]}
  \BibitemShut {NoStop}%
\bibitem [{\citenamefont {Berti}\ \emph {et~al.}(2006)\citenamefont {Berti},
  \citenamefont {Cardoso},\ and\ \citenamefont {Will}}]{Berti:2005ys}%
  \BibitemOpen
  \bibfield  {author} {\bibinfo {author} {\bibfnamefont {E.}~\bibnamefont
  {Berti}}, \bibinfo {author} {\bibfnamefont {V.}~\bibnamefont {Cardoso}}, \
  and\ \bibinfo {author} {\bibfnamefont {C.~M.}\ \bibnamefont {Will}},\ }\href
  {\doibase 10.1103/PhysRevD.73.064030} {\bibfield  {journal} {\bibinfo
  {journal} {Phys. Rev. D}\ }\textbf {\bibinfo {volume} {73}},\ \bibinfo
  {pages} {064030} (\bibinfo {year} {2006})},\ \Eprint
  {http://arxiv.org/abs/gr-qc/0512160} {arXiv:gr-qc/0512160} \BibitemShut
  {NoStop}%
\bibitem [{\citenamefont {{Gossan}}\ \emph {et~al.}(2012)\citenamefont
  {{Gossan}}, \citenamefont {{Veitch}},\ and\ \citenamefont
  {{Sathyaprakash}}}]{gossan-et-al}%
  \BibitemOpen
  \bibfield  {author} {\bibinfo {author} {\bibfnamefont {S.}~\bibnamefont
  {{Gossan}}}, \bibinfo {author} {\bibfnamefont {J.}~\bibnamefont {{Veitch}}},
  \ and\ \bibinfo {author} {\bibfnamefont {B.~S.}\ \bibnamefont
  {{Sathyaprakash}}},\ }\href {\doibase 10.1103/PhysRevD.85.124056} {\bibfield
  {journal} {\bibinfo  {journal} {\prd}\ }\textbf {\bibinfo {volume} {85}},\
  \bibinfo {eid} {124056} (\bibinfo {year} {2012})},\ \Eprint
  {http://arxiv.org/abs/1111.5819} {arXiv:1111.5819 [gr-qc]} \BibitemShut
  {NoStop}%
\bibitem [{\citenamefont {Dreyer}\ \emph {et~al.}(2004)\citenamefont {Dreyer},
  \citenamefont {Kelly}, \citenamefont {Krishnan}, \citenamefont {Finn},
  \citenamefont {Garrison},\ and\ \citenamefont
  {Lopez-Aleman}}]{RD-TGR-first-proposal}%
  \BibitemOpen
  \bibfield  {author} {\bibinfo {author} {\bibfnamefont {O.}~\bibnamefont
  {Dreyer}}, \bibinfo {author} {\bibfnamefont {B.~J.}\ \bibnamefont {Kelly}},
  \bibinfo {author} {\bibfnamefont {B.}~\bibnamefont {Krishnan}}, \bibinfo
  {author} {\bibfnamefont {L.~S.}\ \bibnamefont {Finn}}, \bibinfo {author}
  {\bibfnamefont {D.}~\bibnamefont {Garrison}}, \ and\ \bibinfo {author}
  {\bibfnamefont {R.}~\bibnamefont {Lopez-Aleman}},\ }\href {\doibase
  10.1088/0264-9381/21/4/003} {\bibfield  {journal} {\bibinfo  {journal}
  {Class. Quant. Grav.}\ }\textbf {\bibinfo {volume} {21}},\ \bibinfo {pages}
  {787} (\bibinfo {year} {2004})},\ \Eprint
  {http://arxiv.org/abs/gr-qc/0309007} {arXiv:gr-qc/0309007} \BibitemShut
  {NoStop}%
\bibitem [{\citenamefont {Isi}\ \emph {et~al.}(2019)\citenamefont {Isi},
  \citenamefont {Giesler}, \citenamefont {Farr}, \citenamefont {Scheel},\ and\
  \citenamefont {Teukolsky}}]{Isi:2019aib}%
  \BibitemOpen
  \bibfield  {author} {\bibinfo {author} {\bibfnamefont {M.}~\bibnamefont
  {Isi}}, \bibinfo {author} {\bibfnamefont {M.}~\bibnamefont {Giesler}},
  \bibinfo {author} {\bibfnamefont {W.~M.}\ \bibnamefont {Farr}}, \bibinfo
  {author} {\bibfnamefont {M.~A.}\ \bibnamefont {Scheel}}, \ and\ \bibinfo
  {author} {\bibfnamefont {S.~A.}\ \bibnamefont {Teukolsky}},\ }\href {\doibase
  10.1103/PhysRevLett.123.111102} {\bibfield  {journal} {\bibinfo  {journal}
  {Phys. Rev. Lett.}\ }\textbf {\bibinfo {volume} {123}},\ \bibinfo {pages}
  {111102} (\bibinfo {year} {2019})},\ \Eprint
  {http://arxiv.org/abs/1905.00869} {arXiv:1905.00869 [gr-qc]} \BibitemShut
  {NoStop}%
\bibitem [{\citenamefont {{Capano}}\ and\ \citenamefont
  {{Nitz}}(2020)}]{2020PhRvD.102l4070C}%
  \BibitemOpen
  \bibfield  {author} {\bibinfo {author} {\bibfnamefont {C.~D.}\ \bibnamefont
  {{Capano}}}\ and\ \bibinfo {author} {\bibfnamefont {A.~H.}\ \bibnamefont
  {{Nitz}}},\ }\href {\doibase 10.1103/PhysRevD.102.124070} {\bibfield
  {journal} {\bibinfo  {journal} {\prd}\ }\textbf {\bibinfo {volume} {102}},\
  \bibinfo {eid} {124070} (\bibinfo {year} {2020})},\ \Eprint
  {http://arxiv.org/abs/2008.02248} {arXiv:2008.02248 [gr-qc]} \BibitemShut
  {NoStop}%
\bibitem [{\citenamefont {Bhagwat}\ \emph {et~al.}(2020)\citenamefont
  {Bhagwat}, \citenamefont {Forteza}, \citenamefont {Pani},\ and\ \citenamefont
  {Ferrari}}]{Bhagwat:2019dtm}%
  \BibitemOpen
  \bibfield  {author} {\bibinfo {author} {\bibfnamefont {S.}~\bibnamefont
  {Bhagwat}}, \bibinfo {author} {\bibfnamefont {X.~J.}\ \bibnamefont
  {Forteza}}, \bibinfo {author} {\bibfnamefont {P.}~\bibnamefont {Pani}}, \
  and\ \bibinfo {author} {\bibfnamefont {V.}~\bibnamefont {Ferrari}},\ }\href
  {\doibase 10.1103/PhysRevD.101.044033} {\bibfield  {journal} {\bibinfo
  {journal} {Phys. Rev. D}\ }\textbf {\bibinfo {volume} {101}},\ \bibinfo
  {pages} {044033} (\bibinfo {year} {2020})},\ \Eprint
  {http://arxiv.org/abs/1910.08708} {arXiv:1910.08708 [gr-qc]} \BibitemShut
  {NoStop}%
\bibitem [{\citenamefont {Ota}\ and\ \citenamefont
  {Chirenti}(2020)}]{Ota:2019bzl}%
  \BibitemOpen
  \bibfield  {author} {\bibinfo {author} {\bibfnamefont {I.}~\bibnamefont
  {Ota}}\ and\ \bibinfo {author} {\bibfnamefont {C.}~\bibnamefont {Chirenti}},\
  }\href {\doibase 10.1103/PhysRevD.101.104005} {\bibfield  {journal} {\bibinfo
   {journal} {Phys. Rev. D}\ }\textbf {\bibinfo {volume} {101}},\ \bibinfo
  {pages} {104005} (\bibinfo {year} {2020})},\ \Eprint
  {http://arxiv.org/abs/1911.00440} {arXiv:1911.00440 [gr-qc]} \BibitemShut
  {NoStop}%
\bibitem [{\citenamefont {{Bhagwat}}\ \emph
  {et~al.}(2020{\natexlab{a}})\citenamefont {{Bhagwat}}, \citenamefont
  {{Forteza}}, \citenamefont {{Pani}},\ and\ \citenamefont
  {{Ferrari}}}]{2020PhRvD.101d4033B}%
  \BibitemOpen
  \bibfield  {author} {\bibinfo {author} {\bibfnamefont {S.}~\bibnamefont
  {{Bhagwat}}}, \bibinfo {author} {\bibfnamefont {X.~J.}\ \bibnamefont
  {{Forteza}}}, \bibinfo {author} {\bibfnamefont {P.}~\bibnamefont {{Pani}}}, \
  and\ \bibinfo {author} {\bibfnamefont {V.}~\bibnamefont {{Ferrari}}},\ }\href
  {\doibase 10.1103/PhysRevD.101.044033} {\bibfield  {journal} {\bibinfo
  {journal} {\prd}\ }\textbf {\bibinfo {volume} {101}},\ \bibinfo {eid}
  {044033} (\bibinfo {year} {2020}{\natexlab{a}})},\ \Eprint
  {http://arxiv.org/abs/1910.08708} {arXiv:1910.08708 [gr-qc]} \BibitemShut
  {NoStop}%
\bibitem [{\citenamefont {Jim\'enez~Forteza}\ \emph {et~al.}(2020)\citenamefont
  {Jim\'enez~Forteza}, \citenamefont {Bhagwat}, \citenamefont {Pani},\ and\
  \citenamefont {Ferrari}}]{JimenezForteza:2020cve}%
  \BibitemOpen
  \bibfield  {author} {\bibinfo {author} {\bibfnamefont {X.}~\bibnamefont
  {Jim\'enez~Forteza}}, \bibinfo {author} {\bibfnamefont {S.}~\bibnamefont
  {Bhagwat}}, \bibinfo {author} {\bibfnamefont {P.}~\bibnamefont {Pani}}, \
  and\ \bibinfo {author} {\bibfnamefont {V.}~\bibnamefont {Ferrari}},\ }\href
  {\doibase 10.1103/PhysRevD.102.044053} {\bibfield  {journal} {\bibinfo
  {journal} {Phys. Rev. D}\ }\textbf {\bibinfo {volume} {102}},\ \bibinfo
  {pages} {044053} (\bibinfo {year} {2020})},\ \Eprint
  {http://arxiv.org/abs/2005.03260} {arXiv:2005.03260 [gr-qc]} \BibitemShut
  {NoStop}%
\bibitem [{\citenamefont {Berti}\ \emph {et~al.}(2016)\citenamefont {Berti},
  \citenamefont {Sesana}, \citenamefont {Barausse}, \citenamefont {Cardoso},\
  and\ \citenamefont {Belczynski}}]{Berti:2016lat}%
  \BibitemOpen
  \bibfield  {author} {\bibinfo {author} {\bibfnamefont {E.}~\bibnamefont
  {Berti}}, \bibinfo {author} {\bibfnamefont {A.}~\bibnamefont {Sesana}},
  \bibinfo {author} {\bibfnamefont {E.}~\bibnamefont {Barausse}}, \bibinfo
  {author} {\bibfnamefont {V.}~\bibnamefont {Cardoso}}, \ and\ \bibinfo
  {author} {\bibfnamefont {K.}~\bibnamefont {Belczynski}},\ }\href {\doibase
  10.1103/PhysRevLett.117.101102} {\bibfield  {journal} {\bibinfo  {journal}
  {Phys. Rev. Lett.}\ }\textbf {\bibinfo {volume} {117}},\ \bibinfo {pages}
  {101102} (\bibinfo {year} {2016})},\ \Eprint
  {http://arxiv.org/abs/1605.09286} {arXiv:1605.09286 [gr-qc]} \BibitemShut
  {NoStop}%
\bibitem [{\citenamefont {Barack}\ \emph {et~al.}(2019)\citenamefont {Barack}
  \emph {et~al.}}]{Barack:2018yly}%
  \BibitemOpen
  \bibfield  {author} {\bibinfo {author} {\bibfnamefont {L.}~\bibnamefont
  {Barack}} \emph {et~al.},\ }\href {\doibase 10.1088/1361-6382/ab0587}
  {\bibfield  {journal} {\bibinfo  {journal} {Class. Quant. Grav.}\ }\textbf
  {\bibinfo {volume} {36}},\ \bibinfo {pages} {143001} (\bibinfo {year}
  {2019})},\ \Eprint {http://arxiv.org/abs/1806.05195} {arXiv:1806.05195
  [gr-qc]} \BibitemShut {NoStop}%
\bibitem [{\citenamefont {Amaro-Seoane}\ \emph {et~al.}(2017)\citenamefont
  {Amaro-Seoane} \emph {et~al.}}]{LISA:2017pwj}%
  \BibitemOpen
  \bibfield  {author} {\bibinfo {author} {\bibfnamefont {P.}~\bibnamefont
  {Amaro-Seoane}} \emph {et~al.} (\bibinfo {collaboration} {LISA}),\
  }\href@noop {} {\  (\bibinfo {year} {2017})},\ \Eprint
  {http://arxiv.org/abs/1702.00786} {arXiv:1702.00786 [astro-ph.IM]}
  \BibitemShut {NoStop}%
\bibitem [{\citenamefont {Kalogera}\ \emph {et~al.}(2021)\citenamefont
  {Kalogera} \emph {et~al.}}]{Kalogera:2021bya}%
  \BibitemOpen
  \bibfield  {author} {\bibinfo {author} {\bibfnamefont {V.}~\bibnamefont
  {Kalogera}} \emph {et~al.},\ }\href@noop {} {\  (\bibinfo {year} {2021})},\
  \Eprint {http://arxiv.org/abs/2111.06990} {arXiv:2111.06990 [gr-qc]}
  \BibitemShut {NoStop}%
\bibitem [{\citenamefont {Abbott}\ \emph {et~al.}(2017)\citenamefont {Abbott}
  \emph {et~al.}}]{Evans:2016mbw}%
  \BibitemOpen
  \bibfield  {author} {\bibinfo {author} {\bibfnamefont {B.~P.}\ \bibnamefont
  {Abbott}} \emph {et~al.} (\bibinfo {collaboration} {LIGO Scientific}),\
  }\href {\doibase 10.1088/1361-6382/aa51f4} {\bibfield  {journal} {\bibinfo
  {journal} {Class. Quant. Grav.}\ }\textbf {\bibinfo {volume} {34}},\ \bibinfo
  {pages} {044001} (\bibinfo {year} {2017})},\ \Eprint
  {http://arxiv.org/abs/1607.08697} {arXiv:1607.08697 [astro-ph.IM]}
  \BibitemShut {NoStop}%
\bibitem [{\citenamefont {Essick}\ \emph {et~al.}(2017)\citenamefont {Essick},
  \citenamefont {Vitale},\ and\ \citenamefont {Evans}}]{Essick:2017wyl}%
  \BibitemOpen
  \bibfield  {author} {\bibinfo {author} {\bibfnamefont {R.}~\bibnamefont
  {Essick}}, \bibinfo {author} {\bibfnamefont {S.}~\bibnamefont {Vitale}}, \
  and\ \bibinfo {author} {\bibfnamefont {M.}~\bibnamefont {Evans}},\ }\href
  {\doibase 10.1103/PhysRevD.96.084004} {\bibfield  {journal} {\bibinfo
  {journal} {Phys. Rev.}\ }\textbf {\bibinfo {volume} {D96}},\ \bibinfo {pages}
  {084004} (\bibinfo {year} {2017})},\ \Eprint
  {http://arxiv.org/abs/1708.06843} {arXiv:1708.06843 [gr-qc]} \BibitemShut
  {NoStop}%
\bibitem [{\citenamefont {Hild}\ \emph {et~al.}(2011)\citenamefont {Hild} \emph
  {et~al.}}]{Hild:2010id}%
  \BibitemOpen
  \bibfield  {author} {\bibinfo {author} {\bibfnamefont {S.}~\bibnamefont
  {Hild}} \emph {et~al.},\ }\href {\doibase 10.1088/0264-9381/28/9/094013}
  {\bibfield  {journal} {\bibinfo  {journal} {Class. Quant. Grav.}\ }\textbf
  {\bibinfo {volume} {28}},\ \bibinfo {pages} {094013} (\bibinfo {year}
  {2011})},\ \Eprint {http://arxiv.org/abs/1012.0908} {arXiv:1012.0908 [gr-qc]}
  \BibitemShut {NoStop}%
\bibitem [{\citenamefont {Maggiore}\ \emph {et~al.}(2020)\citenamefont
  {Maggiore} \emph {et~al.}}]{Maggiore:2019uih}%
  \BibitemOpen
  \bibfield  {author} {\bibinfo {author} {\bibfnamefont {M.}~\bibnamefont
  {Maggiore}} \emph {et~al.},\ }\href {\doibase 10.1088/1475-7516/2020/03/050}
  {\bibfield  {journal} {\bibinfo  {journal} {JCAP}\ }\textbf {\bibinfo
  {volume} {03}},\ \bibinfo {pages} {050} (\bibinfo {year} {2020})},\ \Eprint
  {http://arxiv.org/abs/1912.02622} {arXiv:1912.02622 [astro-ph.CO]}
  \BibitemShut {NoStop}%
\bibitem [{\citenamefont {{Klein}}\ \emph {et~al.}(2016)\citenamefont
  {{Klein}}, \citenamefont {{Barausse}}, \citenamefont {{Sesana}},
  \citenamefont {{Petiteau}}, \citenamefont {{Berti}}, \citenamefont {{Babak}},
  \citenamefont {{Gair}}, \citenamefont {{Aoudia}}, \citenamefont {{Hinder}},
  \citenamefont {{Ohme}},\ and\ \citenamefont
  {{Wardell}}}]{2016PhRvD..93b4003K}%
  \BibitemOpen
  \bibfield  {author} {\bibinfo {author} {\bibfnamefont {A.}~\bibnamefont
  {{Klein}}}, \bibinfo {author} {\bibfnamefont {E.}~\bibnamefont {{Barausse}}},
  \bibinfo {author} {\bibfnamefont {A.}~\bibnamefont {{Sesana}}}, \bibinfo
  {author} {\bibfnamefont {A.}~\bibnamefont {{Petiteau}}}, \bibinfo {author}
  {\bibfnamefont {E.}~\bibnamefont {{Berti}}}, \bibinfo {author} {\bibfnamefont
  {S.}~\bibnamefont {{Babak}}}, \bibinfo {author} {\bibfnamefont
  {J.}~\bibnamefont {{Gair}}}, \bibinfo {author} {\bibfnamefont
  {S.}~\bibnamefont {{Aoudia}}}, \bibinfo {author} {\bibfnamefont
  {I.}~\bibnamefont {{Hinder}}}, \bibinfo {author} {\bibfnamefont
  {F.}~\bibnamefont {{Ohme}}}, \ and\ \bibinfo {author} {\bibfnamefont
  {B.}~\bibnamefont {{Wardell}}},\ }\href {\doibase 10.1103/PhysRevD.93.024003}
  {\bibfield  {journal} {\bibinfo  {journal} {\prd}\ }\textbf {\bibinfo
  {volume} {93}},\ \bibinfo {eid} {024003} (\bibinfo {year} {2016})},\ \Eprint
  {http://arxiv.org/abs/1511.05581} {arXiv:1511.05581 [gr-qc]} \BibitemShut
  {NoStop}%
\bibitem [{\citenamefont {Barausse}\ \emph
  {et~al.}(2020{\natexlab{a}})\citenamefont {Barausse} \emph
  {et~al.}}]{Barausse:2020rsu}%
  \BibitemOpen
  \bibfield  {author} {\bibinfo {author} {\bibfnamefont {E.}~\bibnamefont
  {Barausse}} \emph {et~al.},\ }\href {\doibase 10.1007/s10714-020-02691-1}
  {\bibfield  {journal} {\bibinfo  {journal} {Gen. Rel. Grav.}\ }\textbf
  {\bibinfo {volume} {52}},\ \bibinfo {pages} {81} (\bibinfo {year}
  {2020}{\natexlab{a}})},\ \Eprint {http://arxiv.org/abs/2001.09793}
  {arXiv:2001.09793 [gr-qc]} \BibitemShut {NoStop}%
\bibitem [{\citenamefont {Sesana}\ \emph {et~al.}(2021)\citenamefont {Sesana}
  \emph {et~al.}}]{Sesana:2019vho}%
  \BibitemOpen
  \bibfield  {author} {\bibinfo {author} {\bibfnamefont {A.}~\bibnamefont
  {Sesana}} \emph {et~al.},\ }\href {\doibase 10.1007/s10686-021-09709-9}
  {\bibfield  {journal} {\bibinfo  {journal} {Exper. Astron.}\ }\textbf
  {\bibinfo {volume} {51}},\ \bibinfo {pages} {1333} (\bibinfo {year}
  {2021})},\ \Eprint {http://arxiv.org/abs/1908.11391} {arXiv:1908.11391
  [astro-ph.IM]} \BibitemShut {NoStop}%
\bibitem [{\citenamefont {Baibhav}\ \emph {et~al.}(2021)\citenamefont {Baibhav}
  \emph {et~al.}}]{Baibhav:2019rsa}%
  \BibitemOpen
  \bibfield  {author} {\bibinfo {author} {\bibfnamefont {V.}~\bibnamefont
  {Baibhav}} \emph {et~al.},\ }\href {\doibase 10.1007/s10686-021-09741-9}
  {\bibfield  {journal} {\bibinfo  {journal} {Exper. Astron.}\ }\textbf
  {\bibinfo {volume} {51}},\ \bibinfo {pages} {1385} (\bibinfo {year}
  {2021})},\ \Eprint {http://arxiv.org/abs/1908.11390} {arXiv:1908.11390
  [astro-ph.HE]} \BibitemShut {NoStop}%
\bibitem [{\citenamefont {Barausse}\ \emph
  {et~al.}(2020{\natexlab{b}})\citenamefont {Barausse}, \citenamefont
  {Dvorkin}, \citenamefont {Tremmel}, \citenamefont {Volonteri},\ and\
  \citenamefont {Bonetti}}]{Barausse:2020mdt}%
  \BibitemOpen
  \bibfield  {author} {\bibinfo {author} {\bibfnamefont {E.}~\bibnamefont
  {Barausse}}, \bibinfo {author} {\bibfnamefont {I.}~\bibnamefont {Dvorkin}},
  \bibinfo {author} {\bibfnamefont {M.}~\bibnamefont {Tremmel}}, \bibinfo
  {author} {\bibfnamefont {M.}~\bibnamefont {Volonteri}}, \ and\ \bibinfo
  {author} {\bibfnamefont {M.}~\bibnamefont {Bonetti}},\ }\href {\doibase
  10.3847/1538-4357/abba7f} {\bibfield  {journal} {\bibinfo  {journal}
  {Astrophys. J.}\ }\textbf {\bibinfo {volume} {904}},\ \bibinfo {pages} {16}
  (\bibinfo {year} {2020}{\natexlab{b}})},\ \Eprint
  {http://arxiv.org/abs/2006.03065} {arXiv:2006.03065 [astro-ph.GA]}
  \BibitemShut {NoStop}%
\bibitem [{\citenamefont {Barausse}\ and\ \citenamefont
  {Lapi}(2020)}]{Barausse:2020gbp}%
  \BibitemOpen
  \bibfield  {author} {\bibinfo {author} {\bibfnamefont {E.}~\bibnamefont
  {Barausse}}\ and\ \bibinfo {author} {\bibfnamefont {A.}~\bibnamefont
  {Lapi}},\ }\href@noop {} {\  (\bibinfo {year} {2020})},\ \Eprint
  {http://arxiv.org/abs/2011.01994} {arXiv:2011.01994 [astro-ph.GA]}
  \BibitemShut {NoStop}%
\bibitem [{\citenamefont {{Baibhav}}\ and\ \citenamefont
  {{Berti}}(2019)}]{2019PhRvD..99b4005B}%
  \BibitemOpen
  \bibfield  {author} {\bibinfo {author} {\bibfnamefont {V.}~\bibnamefont
  {{Baibhav}}}\ and\ \bibinfo {author} {\bibfnamefont {E.}~\bibnamefont
  {{Berti}}},\ }\href {\doibase 10.1103/PhysRevD.99.024005} {\bibfield
  {journal} {\bibinfo  {journal} {\prd}\ }\textbf {\bibinfo {volume} {99}},\
  \bibinfo {eid} {024005} (\bibinfo {year} {2019})},\ \Eprint
  {http://arxiv.org/abs/1809.03500} {arXiv:1809.03500 [gr-qc]} \BibitemShut
  {NoStop}%
\bibitem [{\citenamefont {{Baibhav}}\ \emph {et~al.}(2020)\citenamefont
  {{Baibhav}}, \citenamefont {{Berti}},\ and\ \citenamefont
  {{Cardoso}}}]{2020PhRvD.101h4053B}%
  \BibitemOpen
  \bibfield  {author} {\bibinfo {author} {\bibfnamefont {V.}~\bibnamefont
  {{Baibhav}}}, \bibinfo {author} {\bibfnamefont {E.}~\bibnamefont {{Berti}}},
  \ and\ \bibinfo {author} {\bibfnamefont {V.}~\bibnamefont {{Cardoso}}},\
  }\href {\doibase 10.1103/PhysRevD.101.084053} {\bibfield  {journal} {\bibinfo
   {journal} {\prd}\ }\textbf {\bibinfo {volume} {101}},\ \bibinfo {eid}
  {084053} (\bibinfo {year} {2020})},\ \Eprint
  {http://arxiv.org/abs/2001.10011} {arXiv:2001.10011 [gr-qc]} \BibitemShut
  {NoStop}%
\bibitem [{\citenamefont {{Bhagwat}}\ \emph
  {et~al.}(2020{\natexlab{b}})\citenamefont {{Bhagwat}}, \citenamefont
  {{Cabero}}, \citenamefont {{Capano}}, \citenamefont {{Krishnan}},\ and\
  \citenamefont {{Brown}}}]{2020PhRvD.102b4023B}%
  \BibitemOpen
  \bibfield  {author} {\bibinfo {author} {\bibfnamefont {S.}~\bibnamefont
  {{Bhagwat}}}, \bibinfo {author} {\bibfnamefont {M.}~\bibnamefont {{Cabero}}},
  \bibinfo {author} {\bibfnamefont {C.~D.}\ \bibnamefont {{Capano}}}, \bibinfo
  {author} {\bibfnamefont {B.}~\bibnamefont {{Krishnan}}}, \ and\ \bibinfo
  {author} {\bibfnamefont {D.~A.}\ \bibnamefont {{Brown}}},\ }\href {\doibase
  10.1103/PhysRevD.102.024023} {\bibfield  {journal} {\bibinfo  {journal}
  {\prd}\ }\textbf {\bibinfo {volume} {102}},\ \bibinfo {eid} {024023}
  (\bibinfo {year} {2020}{\natexlab{b}})},\ \Eprint
  {http://arxiv.org/abs/1910.13203} {arXiv:1910.13203 [gr-qc]} \BibitemShut
  {NoStop}%
\bibitem [{\citenamefont {Berti}\ \emph
  {et~al.}(2007{\natexlab{a}})\citenamefont {Berti}, \citenamefont {Cardoso},
  \citenamefont {Cardoso},\ and\ \citenamefont {Cavaglia}}]{Berti:2007zu}%
  \BibitemOpen
  \bibfield  {author} {\bibinfo {author} {\bibfnamefont {E.}~\bibnamefont
  {Berti}}, \bibinfo {author} {\bibfnamefont {J.}~\bibnamefont {Cardoso}},
  \bibinfo {author} {\bibfnamefont {V.}~\bibnamefont {Cardoso}}, \ and\
  \bibinfo {author} {\bibfnamefont {M.}~\bibnamefont {Cavaglia}},\ }\href
  {\doibase 10.1103/PhysRevD.76.104044} {\bibfield  {journal} {\bibinfo
  {journal} {Phys. Rev. D}\ }\textbf {\bibinfo {volume} {76}},\ \bibinfo
  {pages} {104044} (\bibinfo {year} {2007}{\natexlab{a}})},\ \Eprint
  {http://arxiv.org/abs/0707.1202} {arXiv:0707.1202 [gr-qc]} \BibitemShut
  {NoStop}%
\bibitem [{\citenamefont {{Bhagwat}}\ \emph {et~al.}(2016)\citenamefont
  {{Bhagwat}}, \citenamefont {{Brown}},\ and\ \citenamefont
  {{Ballmer}}}]{2016PhRvD..94h4024B}%
  \BibitemOpen
  \bibfield  {author} {\bibinfo {author} {\bibfnamefont {S.}~\bibnamefont
  {{Bhagwat}}}, \bibinfo {author} {\bibfnamefont {D.~A.}\ \bibnamefont
  {{Brown}}}, \ and\ \bibinfo {author} {\bibfnamefont {S.~W.}\ \bibnamefont
  {{Ballmer}}},\ }\href {\doibase 10.1103/PhysRevD.94.084024} {\bibfield
  {journal} {\bibinfo  {journal} {\prd}\ }\textbf {\bibinfo {volume} {94}},\
  \bibinfo {eid} {084024} (\bibinfo {year} {2016})},\ \Eprint
  {http://arxiv.org/abs/1607.07845} {arXiv:1607.07845 [gr-qc]} \BibitemShut
  {NoStop}%
\bibitem [{\citenamefont {Ota}\ and\ \citenamefont
  {Chirenti}(2021)}]{Ota:2021ypb}%
  \BibitemOpen
  \bibfield  {author} {\bibinfo {author} {\bibfnamefont {I.}~\bibnamefont
  {Ota}}\ and\ \bibinfo {author} {\bibfnamefont {C.}~\bibnamefont {Chirenti}},\
  }\href@noop {} {\  (\bibinfo {year} {2021})},\ \Eprint
  {http://arxiv.org/abs/2108.01774} {arXiv:2108.01774 [gr-qc]} \BibitemShut
  {NoStop}%
\bibitem [{\citenamefont {{Barausse}}(2012)}]{EB2012}%
  \BibitemOpen
  \bibfield  {author} {\bibinfo {author} {\bibfnamefont {E.}~\bibnamefont
  {{Barausse}}},\ }\href {\doibase 10.1111/j.1365-2966.2012.21057.x} {\bibfield
   {journal} {\bibinfo  {journal} {\textit{Monthly Notices of the Royal
  Astronomical Society}}\ }\textbf {\bibinfo {volume} {423}},\ \bibinfo {pages}
  {2533} (\bibinfo {year} {2012})},\ \Eprint {http://arxiv.org/abs/1201.5888}
  {arXiv:1201.5888} \BibitemShut {NoStop}%
\bibitem [{\citenamefont {{Sesana}}\ \emph {et~al.}(2014)\citenamefont
  {{Sesana}}, \citenamefont {{Barausse}}, \citenamefont {{Dotti}},\ and\
  \citenamefont {{Rossi}}}]{sesana2014}%
  \BibitemOpen
  \bibfield  {author} {\bibinfo {author} {\bibfnamefont {A.}~\bibnamefont
  {{Sesana}}}, \bibinfo {author} {\bibfnamefont {E.}~\bibnamefont
  {{Barausse}}}, \bibinfo {author} {\bibfnamefont {M.}~\bibnamefont {{Dotti}}},
  \ and\ \bibinfo {author} {\bibfnamefont {E.~M.}\ \bibnamefont {{Rossi}}},\
  }\href {\doibase 10.1088/0004-637X/794/2/104} {\bibfield  {journal} {\bibinfo
   {journal} {\textit{The Astrophysical Journal}}\ }\textbf {\bibinfo {volume}
  {794}},\ \bibinfo {eid} {104} (\bibinfo {year} {2014})},\ \Eprint
  {http://arxiv.org/abs/1402.7088} {arXiv:1402.7088} \BibitemShut {NoStop}%
\bibitem [{\citenamefont {{Antonini}}\ \emph
  {et~al.}(2015{\natexlab{a}})\citenamefont {{Antonini}}, \citenamefont
  {{Barausse}},\ and\ \citenamefont {{Silk}}}]{antonini1}%
  \BibitemOpen
  \bibfield  {author} {\bibinfo {author} {\bibfnamefont {F.}~\bibnamefont
  {{Antonini}}}, \bibinfo {author} {\bibfnamefont {E.}~\bibnamefont
  {{Barausse}}}, \ and\ \bibinfo {author} {\bibfnamefont {J.}~\bibnamefont
  {{Silk}}},\ }\href {\doibase 10.1088/2041-8205/806/1/L8} {\bibfield
  {journal} {\bibinfo  {journal} {\textit{The Astrophysical Journal Letters}}\
  }\textbf {\bibinfo {volume} {806}},\ \bibinfo {eid} {L8} (\bibinfo {year}
  {2015}{\natexlab{a}})},\ \Eprint {http://arxiv.org/abs/1504.04033}
  {arXiv:1504.04033} \BibitemShut {NoStop}%
\bibitem [{\citenamefont {{Antonini}}\ \emph
  {et~al.}(2015{\natexlab{b}})\citenamefont {{Antonini}}, \citenamefont
  {{Barausse}},\ and\ \citenamefont {{Silk}}}]{antonini2}%
  \BibitemOpen
  \bibfield  {author} {\bibinfo {author} {\bibfnamefont {F.}~\bibnamefont
  {{Antonini}}}, \bibinfo {author} {\bibfnamefont {E.}~\bibnamefont
  {{Barausse}}}, \ and\ \bibinfo {author} {\bibfnamefont {J.}~\bibnamefont
  {{Silk}}},\ }\href {\doibase 10.1088/0004-637X/812/1/72} {\bibfield
  {journal} {\bibinfo  {journal} {\textit{The Astrophysical Journal}}\ }\textbf
  {\bibinfo {volume} {812}},\ \bibinfo {eid} {72} (\bibinfo {year}
  {2015}{\natexlab{b}})},\ \Eprint {http://arxiv.org/abs/1506.02050}
  {arXiv:1506.02050} \BibitemShut {NoStop}%
\bibitem [{\citenamefont {{Bonetti}}\ \emph
  {et~al.}(2018{\natexlab{a}})\citenamefont {{Bonetti}}, \citenamefont
  {{Haardt}}, \citenamefont {{Sesana}},\ and\ \citenamefont
  {{Barausse}}}]{BonettiII}%
  \BibitemOpen
  \bibfield  {author} {\bibinfo {author} {\bibfnamefont {M.}~\bibnamefont
  {{Bonetti}}}, \bibinfo {author} {\bibfnamefont {F.}~\bibnamefont {{Haardt}}},
  \bibinfo {author} {\bibfnamefont {A.}~\bibnamefont {{Sesana}}}, \ and\
  \bibinfo {author} {\bibfnamefont {E.}~\bibnamefont {{Barausse}}},\ }\href
  {\doibase 10.1093/mnras/sty896} {\bibfield  {journal} {\bibinfo  {journal}
  {\textit{Monthly Notices of the Royal Astronomical Society}}\ }\textbf
  {\bibinfo {volume} {477}},\ \bibinfo {pages} {3910} (\bibinfo {year}
  {2018}{\natexlab{a}})},\ \Eprint {http://arxiv.org/abs/1709.06088}
  {arXiv:1709.06088} \BibitemShut {NoStop}%
\bibitem [{\citenamefont {{Bonetti}}\ \emph {et~al.}(2019)\citenamefont
  {{Bonetti}}, \citenamefont {{Sesana}}, \citenamefont {{Haardt}},
  \citenamefont {{Barausse}},\ and\ \citenamefont {{Colpi}}}]{bonetti2019}%
  \BibitemOpen
  \bibfield  {author} {\bibinfo {author} {\bibfnamefont {M.}~\bibnamefont
  {{Bonetti}}}, \bibinfo {author} {\bibfnamefont {A.}~\bibnamefont {{Sesana}}},
  \bibinfo {author} {\bibfnamefont {F.}~\bibnamefont {{Haardt}}}, \bibinfo
  {author} {\bibfnamefont {E.}~\bibnamefont {{Barausse}}}, \ and\ \bibinfo
  {author} {\bibfnamefont {M.}~\bibnamefont {{Colpi}}},\ }\href {\doibase
  10.1093/mnras/stz903} {\bibfield  {journal} {\bibinfo  {journal}
  {\textit{Monthly Notices of the Royal Astronomical Society}}\ }\textbf
  {\bibinfo {volume} {486}},\ \bibinfo {pages} {4044} (\bibinfo {year}
  {2019})},\ \Eprint {http://arxiv.org/abs/1812.01011} {arXiv:1812.01011
  [astro-ph.GA]} \BibitemShut {NoStop}%
\bibitem [{\citenamefont {{Dekel}}\ and\ \citenamefont
  {{Birnboim}}(2006)}]{Dekel2006}%
  \BibitemOpen
  \bibfield  {author} {\bibinfo {author} {\bibfnamefont {A.}~\bibnamefont
  {{Dekel}}}\ and\ \bibinfo {author} {\bibfnamefont {Y.}~\bibnamefont
  {{Birnboim}}},\ }\href {\doibase 10.1111/j.1365-2966.2006.10145.x} {\bibfield
   {journal} {\bibinfo  {journal} {\textit{Monthly Notices of the Royal
  Astronomical Society}}\ }\textbf {\bibinfo {volume} {368}},\ \bibinfo {pages}
  {2} (\bibinfo {year} {2006})},\ \Eprint
  {http://arxiv.org/abs/astro-ph/0412300} {astro-ph/0412300} \BibitemShut
  {NoStop}%
\bibitem [{\citenamefont {{Cattaneo}}\ \emph {et~al.}(2006)\citenamefont
  {{Cattaneo}}, \citenamefont {{Dekel}}, \citenamefont {{Devriendt}},
  \citenamefont {{Guiderdoni}},\ and\ \citenamefont
  {{Blaizot}}}]{Cattaneo2006}%
  \BibitemOpen
  \bibfield  {author} {\bibinfo {author} {\bibfnamefont {A.}~\bibnamefont
  {{Cattaneo}}}, \bibinfo {author} {\bibfnamefont {A.}~\bibnamefont {{Dekel}}},
  \bibinfo {author} {\bibfnamefont {J.}~\bibnamefont {{Devriendt}}}, \bibinfo
  {author} {\bibfnamefont {B.}~\bibnamefont {{Guiderdoni}}}, \ and\ \bibinfo
  {author} {\bibfnamefont {J.}~\bibnamefont {{Blaizot}}},\ }\href {\doibase
  10.1111/j.1365-2966.2006.10608.x} {\bibfield  {journal} {\bibinfo  {journal}
  {\textit{Monthly Notices of the Royal Astronomical Society}}\ }\textbf
  {\bibinfo {volume} {370}},\ \bibinfo {pages} {1651} (\bibinfo {year}
  {2006})},\ \Eprint {http://arxiv.org/abs/astro-ph/0601295} {astro-ph/0601295}
  \BibitemShut {NoStop}%
\bibitem [{\citenamefont {{Dekel}}\ \emph {et~al.}(2009)\citenamefont
  {{Dekel}}, \citenamefont {{Birnboim}}, \citenamefont {{Engel}}, \citenamefont
  {{Freundlich}}, \citenamefont {{Goerdt}}, \citenamefont {{Mumcuoglu}},
  \citenamefont {{Neistein}}, \citenamefont {{Pichon}}, \citenamefont
  {{Teyssier}},\ and\ \citenamefont {{Zinger}}}]{Dekel2009}%
  \BibitemOpen
  \bibfield  {author} {\bibinfo {author} {\bibfnamefont {A.}~\bibnamefont
  {{Dekel}}}, \bibinfo {author} {\bibfnamefont {Y.}~\bibnamefont {{Birnboim}}},
  \bibinfo {author} {\bibfnamefont {G.}~\bibnamefont {{Engel}}}, \bibinfo
  {author} {\bibfnamefont {J.}~\bibnamefont {{Freundlich}}}, \bibinfo {author}
  {\bibfnamefont {T.}~\bibnamefont {{Goerdt}}}, \bibinfo {author}
  {\bibfnamefont {M.}~\bibnamefont {{Mumcuoglu}}}, \bibinfo {author}
  {\bibfnamefont {E.}~\bibnamefont {{Neistein}}}, \bibinfo {author}
  {\bibfnamefont {C.}~\bibnamefont {{Pichon}}}, \bibinfo {author}
  {\bibfnamefont {R.}~\bibnamefont {{Teyssier}}}, \ and\ \bibinfo {author}
  {\bibfnamefont {E.}~\bibnamefont {{Zinger}}},\ }\href {\doibase
  10.1038/nature07648} {\bibfield  {journal} {\bibinfo  {journal}
  {\textit{Nature}}\ }\textbf {\bibinfo {volume} {457}},\ \bibinfo {pages}
  {451} (\bibinfo {year} {2009})},\ \Eprint {http://arxiv.org/abs/0808.0553}
  {arXiv:0808.0553} \BibitemShut {NoStop}%
\bibitem [{\citenamefont {{Mo}}\ \emph {et~al.}(1998)\citenamefont {{Mo}},
  \citenamefont {{Mao}},\ and\ \citenamefont {{White}}}]{1998MNRAS.295..319M}%
  \BibitemOpen
  \bibfield  {author} {\bibinfo {author} {\bibfnamefont {H.~J.}\ \bibnamefont
  {{Mo}}}, \bibinfo {author} {\bibfnamefont {S.}~\bibnamefont {{Mao}}}, \ and\
  \bibinfo {author} {\bibfnamefont {S.~D.~M.}\ \bibnamefont {{White}}},\ }\href
  {\doibase 10.1046/j.1365-8711.1998.01227.x} {\bibfield  {journal} {\bibinfo
  {journal} {\textit{Monthly Notices of the Royal Astronomical Society}}\
  }\textbf {\bibinfo {volume} {295}},\ \bibinfo {pages} {319} (\bibinfo {year}
  {1998})},\ \Eprint {http://arxiv.org/abs/astro-ph/9707093}
  {arXiv:astro-ph/9707093 [astro-ph]} \BibitemShut {NoStop}%
\bibitem [{\citenamefont {Granato}\ \emph {et~al.}(2004)\citenamefont
  {Granato}, \citenamefont {De~Zotti}, \citenamefont {Silva}, \citenamefont
  {Bressan},\ and\ \citenamefont {Danese}}]{Granato:2003ch}%
  \BibitemOpen
  \bibfield  {author} {\bibinfo {author} {\bibfnamefont {G.~L.}\ \bibnamefont
  {Granato}}, \bibinfo {author} {\bibfnamefont {G.}~\bibnamefont {De~Zotti}},
  \bibinfo {author} {\bibfnamefont {L.}~\bibnamefont {Silva}}, \bibinfo
  {author} {\bibfnamefont {A.}~\bibnamefont {Bressan}}, \ and\ \bibinfo
  {author} {\bibfnamefont {L.}~\bibnamefont {Danese}},\ }\href {\doibase
  10.1086/379875} {\bibfield  {journal} {\bibinfo  {journal} {Astrophys. J.}\
  }\textbf {\bibinfo {volume} {600}},\ \bibinfo {pages} {580} (\bibinfo {year}
  {2004})},\ \Eprint {http://arxiv.org/abs/astro-ph/0307202}
  {arXiv:astro-ph/0307202} \BibitemShut {NoStop}%
\bibitem [{\citenamefont {{Springel}}\ and\ \citenamefont
  {{Hernquist}}(2002)}]{fb1}%
  \BibitemOpen
  \bibfield  {author} {\bibinfo {author} {\bibfnamefont {V.}~\bibnamefont
  {{Springel}}}\ and\ \bibinfo {author} {\bibfnamefont {L.}~\bibnamefont
  {{Hernquist}}},\ }\href {\doibase 10.1046/j.1365-8711.2002.05445.x}
  {\bibfield  {journal} {\bibinfo  {journal} {\textit{Monthly Notices of the
  Royal Astronomical Society}}\ }\textbf {\bibinfo {volume} {333}},\ \bibinfo
  {pages} {649} (\bibinfo {year} {2002})},\ \Eprint
  {http://arxiv.org/abs/astro-ph/0111016} {arXiv:astro-ph/0111016 [astro-ph]}
  \BibitemShut {NoStop}%
\bibitem [{\citenamefont {{Fujita}}\ \emph {et~al.}(2004)\citenamefont
  {{Fujita}}, \citenamefont {{Mac Low}}, \citenamefont {{Ferrara}},\ and\
  \citenamefont {{Meiksin}}}]{fb2}%
  \BibitemOpen
  \bibfield  {author} {\bibinfo {author} {\bibfnamefont {A.}~\bibnamefont
  {{Fujita}}}, \bibinfo {author} {\bibfnamefont {M.-M.}\ \bibnamefont {{Mac
  Low}}}, \bibinfo {author} {\bibfnamefont {A.}~\bibnamefont {{Ferrara}}}, \
  and\ \bibinfo {author} {\bibfnamefont {A.}~\bibnamefont {{Meiksin}}},\ }\href
  {\doibase 10.1086/422861} {\bibfield  {journal} {\bibinfo  {journal}
  {\textit{The Astrophysical Journal}}\ }\textbf {\bibinfo {volume} {613}},\
  \bibinfo {pages} {159} (\bibinfo {year} {2004})},\ \Eprint
  {http://arxiv.org/abs/astro-ph/0405611} {arXiv:astro-ph/0405611 [astro-ph]}
  \BibitemShut {NoStop}%
\bibitem [{\citenamefont {{Rasera}}\ and\ \citenamefont
  {{Teyssier}}(2006)}]{fb3}%
  \BibitemOpen
  \bibfield  {author} {\bibinfo {author} {\bibfnamefont {Y.}~\bibnamefont
  {{Rasera}}}\ and\ \bibinfo {author} {\bibfnamefont {R.}~\bibnamefont
  {{Teyssier}}},\ }\href {\doibase 10.1051/0004-6361:20053116} {\bibfield
  {journal} {\bibinfo  {journal} {\textit{Astronomy and Astrophysics}}\
  }\textbf {\bibinfo {volume} {445}},\ \bibinfo {pages} {1} (\bibinfo {year}
  {2006})},\ \Eprint {http://arxiv.org/abs/astro-ph/0505473}
  {arXiv:astro-ph/0505473 [astro-ph]} \BibitemShut {NoStop}%
\bibitem [{\citenamefont {{Croton}}\ \emph {et~al.}(2006)\citenamefont
  {{Croton}}, \citenamefont {{Springel}}, \citenamefont {{White}},
  \citenamefont {{De Lucia}}, \citenamefont {{Frenk}}, \citenamefont {{Gao}},
  \citenamefont {{Jenkins}}, \citenamefont {{Kauffmann}}, \citenamefont
  {{Navarro}},\ and\ \citenamefont {{Yoshida}}}]{Croton2006}%
  \BibitemOpen
  \bibfield  {author} {\bibinfo {author} {\bibfnamefont {D.~J.}\ \bibnamefont
  {{Croton}}}, \bibinfo {author} {\bibfnamefont {V.}~\bibnamefont
  {{Springel}}}, \bibinfo {author} {\bibfnamefont {S.~D.~M.}\ \bibnamefont
  {{White}}}, \bibinfo {author} {\bibfnamefont {G.}~\bibnamefont {{De Lucia}}},
  \bibinfo {author} {\bibfnamefont {C.~S.}\ \bibnamefont {{Frenk}}}, \bibinfo
  {author} {\bibfnamefont {L.}~\bibnamefont {{Gao}}}, \bibinfo {author}
  {\bibfnamefont {A.}~\bibnamefont {{Jenkins}}}, \bibinfo {author}
  {\bibfnamefont {G.}~\bibnamefont {{Kauffmann}}}, \bibinfo {author}
  {\bibfnamefont {J.~F.}\ \bibnamefont {{Navarro}}}, \ and\ \bibinfo {author}
  {\bibfnamefont {N.}~\bibnamefont {{Yoshida}}},\ }\href {\doibase
  10.1111/j.1365-2966.2005.09675.x} {\bibfield  {journal} {\bibinfo  {journal}
  {\textit{Monthly Notices of the Royal Astronomical Society}}\ }\textbf
  {\bibinfo {volume} {365}},\ \bibinfo {pages} {11} (\bibinfo {year} {2006})},\
  \Eprint {http://arxiv.org/abs/astro-ph/0508046} {astro-ph/0508046}
  \BibitemShut {NoStop}%
\bibitem [{\citenamefont {{Hopkins}}\ \emph {et~al.}(2008)\citenamefont
  {{Hopkins}}, \citenamefont {{Cox}}, \citenamefont {{Kere{\v{s}}}},\ and\
  \citenamefont {{Hernquist}}}]{2008ApJS..175..390H}%
  \BibitemOpen
  \bibfield  {author} {\bibinfo {author} {\bibfnamefont {P.~F.}\ \bibnamefont
  {{Hopkins}}}, \bibinfo {author} {\bibfnamefont {T.~J.}\ \bibnamefont
  {{Cox}}}, \bibinfo {author} {\bibfnamefont {D.}~\bibnamefont
  {{Kere{\v{s}}}}}, \ and\ \bibinfo {author} {\bibfnamefont {L.}~\bibnamefont
  {{Hernquist}}},\ }\href {\doibase 10.1086/524363} {\bibfield  {journal}
  {\bibinfo  {journal} {\textit{The Astrophysical Journal}s}\ }\textbf
  {\bibinfo {volume} {175}},\ \bibinfo {pages} {390} (\bibinfo {year}
  {2008})},\ \Eprint {http://arxiv.org/abs/0706.1246} {arXiv:0706.1246
  [astro-ph]} \BibitemShut {NoStop}%
\bibitem [{\citenamefont {{Bower}}\ \emph {et~al.}(2006)\citenamefont
  {{Bower}}, \citenamefont {{Benson}}, \citenamefont {{Malbon}}, \citenamefont
  {{Helly}}, \citenamefont {{Frenk}}, \citenamefont {{Baugh}}, \citenamefont
  {{Cole}},\ and\ \citenamefont {{Lacey}}}]{2006MNRAS.370..645B}%
  \BibitemOpen
  \bibfield  {author} {\bibinfo {author} {\bibfnamefont {R.~G.}\ \bibnamefont
  {{Bower}}}, \bibinfo {author} {\bibfnamefont {A.~J.}\ \bibnamefont
  {{Benson}}}, \bibinfo {author} {\bibfnamefont {R.}~\bibnamefont {{Malbon}}},
  \bibinfo {author} {\bibfnamefont {J.~C.}\ \bibnamefont {{Helly}}}, \bibinfo
  {author} {\bibfnamefont {C.~S.}\ \bibnamefont {{Frenk}}}, \bibinfo {author}
  {\bibfnamefont {C.~M.}\ \bibnamefont {{Baugh}}}, \bibinfo {author}
  {\bibfnamefont {S.}~\bibnamefont {{Cole}}}, \ and\ \bibinfo {author}
  {\bibfnamefont {C.~G.}\ \bibnamefont {{Lacey}}},\ }\href {\doibase
  10.1111/j.1365-2966.2006.10519.x} {\bibfield  {journal} {\bibinfo  {journal}
  {\textit{Monthly Notices of the Royal Astronomical Society}}\ }\textbf
  {\bibinfo {volume} {370}},\ \bibinfo {pages} {645} (\bibinfo {year}
  {2006})},\ \Eprint {http://arxiv.org/abs/astro-ph/0511338}
  {arXiv:astro-ph/0511338 [astro-ph]} \BibitemShut {NoStop}%
\bibitem [{\citenamefont {{Habouzit}}\ \emph {et~al.}(2017)\citenamefont
  {{Habouzit}}, \citenamefont {{Volonteri}},\ and\ \citenamefont
  {{Dubois}}}]{habouzit}%
  \BibitemOpen
  \bibfield  {author} {\bibinfo {author} {\bibfnamefont {M.}~\bibnamefont
  {{Habouzit}}}, \bibinfo {author} {\bibfnamefont {M.}~\bibnamefont
  {{Volonteri}}}, \ and\ \bibinfo {author} {\bibfnamefont {Y.}~\bibnamefont
  {{Dubois}}},\ }\href {\doibase 10.1093/mnras/stx666} {\bibfield  {journal}
  {\bibinfo  {journal} {\textit{Monthly Notices of the Royal Astronomical
  Society}}\ }\textbf {\bibinfo {volume} {468}},\ \bibinfo {pages} {3935}
  (\bibinfo {year} {2017})},\ \Eprint {http://arxiv.org/abs/1605.09394}
  {arXiv:1605.09394 [astro-ph.GA]} \BibitemShut {NoStop}%
\bibitem [{\citenamefont {{Press}}\ and\ \citenamefont
  {{Schechter}}(1974)}]{PS}%
  \BibitemOpen
  \bibfield  {author} {\bibinfo {author} {\bibfnamefont {W.~H.}\ \bibnamefont
  {{Press}}}\ and\ \bibinfo {author} {\bibfnamefont {P.}~\bibnamefont
  {{Schechter}}},\ }\href {\doibase 10.1086/152650} {\bibfield  {journal}
  {\bibinfo  {journal} {\textit{The Astrophysical Journal}}\ }\textbf {\bibinfo
  {volume} {187}},\ \bibinfo {pages} {425} (\bibinfo {year}
  {1974})}\BibitemShut {NoStop}%
\bibitem [{\citenamefont {{Parkinson}}\ \emph {et~al.}(2008)\citenamefont
  {{Parkinson}}, \citenamefont {{Cole}},\ and\ \citenamefont
  {{Helly}}}]{Parkinson2008}%
  \BibitemOpen
  \bibfield  {author} {\bibinfo {author} {\bibfnamefont {H.}~\bibnamefont
  {{Parkinson}}}, \bibinfo {author} {\bibfnamefont {S.}~\bibnamefont {{Cole}}},
  \ and\ \bibinfo {author} {\bibfnamefont {J.}~\bibnamefont {{Helly}}},\ }\href
  {\doibase 10.1111/j.1365-2966.2007.12517.x} {\bibfield  {journal} {\bibinfo
  {journal} {\textit{Monthly Notices of the Royal Astronomical Society}}\
  }\textbf {\bibinfo {volume} {383}},\ \bibinfo {pages} {557} (\bibinfo {year}
  {2008})},\ \Eprint {http://arxiv.org/abs/0708.1382} {arXiv:0708.1382}
  \BibitemShut {NoStop}%
\bibitem [{\citenamefont {{Boylan-Kolchin}}\ \emph {et~al.}(2008)\citenamefont
  {{Boylan-Kolchin}}, \citenamefont {{Ma}},\ and\ \citenamefont
  {{Quataert}}}]{boylankolchin}%
  \BibitemOpen
  \bibfield  {author} {\bibinfo {author} {\bibfnamefont {M.}~\bibnamefont
  {{Boylan-Kolchin}}}, \bibinfo {author} {\bibfnamefont {C.-P.}\ \bibnamefont
  {{Ma}}}, \ and\ \bibinfo {author} {\bibfnamefont {E.}~\bibnamefont
  {{Quataert}}},\ }\href {\doibase 10.1111/j.1365-2966.2007.12530.x} {\bibfield
   {journal} {\bibinfo  {journal} {\textit{Monthly Notices of the Royal
  Astronomical Society}}\ }\textbf {\bibinfo {volume} {383}},\ \bibinfo {pages}
  {93} (\bibinfo {year} {2008})},\ \Eprint {http://arxiv.org/abs/0707.2960}
  {arXiv:0707.2960} \BibitemShut {NoStop}%
\bibitem [{\citenamefont {{Taffoni}}\ \emph {et~al.}(2003)\citenamefont
  {{Taffoni}}, \citenamefont {{Mayer}}, \citenamefont {{Colpi}},\ and\
  \citenamefont {{Governato}}}]{taffoni}%
  \BibitemOpen
  \bibfield  {author} {\bibinfo {author} {\bibfnamefont {G.}~\bibnamefont
  {{Taffoni}}}, \bibinfo {author} {\bibfnamefont {L.}~\bibnamefont {{Mayer}}},
  \bibinfo {author} {\bibfnamefont {M.}~\bibnamefont {{Colpi}}}, \ and\
  \bibinfo {author} {\bibfnamefont {F.}~\bibnamefont {{Governato}}},\ }\href
  {\doibase 10.1046/j.1365-8711.2003.06395.x} {\bibfield  {journal} {\bibinfo
  {journal} {\textit{Monthly Notices of the Royal Astronomical Society}}\
  }\textbf {\bibinfo {volume} {341}},\ \bibinfo {pages} {434} (\bibinfo {year}
  {2003})},\ \Eprint {http://arxiv.org/abs/astro-ph/0301271} {astro-ph/0301271}
  \BibitemShut {NoStop}%
\bibitem [{\citenamefont {{Binney}}\ and\ \citenamefont
  {{Tremaine}}(2008)}]{binneytremaine}%
  \BibitemOpen
  \bibfield  {author} {\bibinfo {author} {\bibfnamefont {J.}~\bibnamefont
  {{Binney}}}\ and\ \bibinfo {author} {\bibfnamefont {S.}~\bibnamefont
  {{Tremaine}}},\ }\href@noop {} {\emph {\bibinfo {title} {Galactic Dynamics:
  Second Edition, by James Binney and Scott Tremaine.~ISBN 978-0-691-13026-2
  (HB).~Published by Princeton University Press, Princeton, NJ USA, 2008.}}}\
  (\bibinfo  {publisher} {Princeton University Press},\ \bibinfo {year}
  {2008})\BibitemShut {NoStop}%
\bibitem [{\citenamefont {{Dosopoulou}}\ and\ \citenamefont
  {{Antonini}}(2017)}]{2017ApJ...840...31D}%
  \BibitemOpen
  \bibfield  {author} {\bibinfo {author} {\bibfnamefont {F.}~\bibnamefont
  {{Dosopoulou}}}\ and\ \bibinfo {author} {\bibfnamefont {F.}~\bibnamefont
  {{Antonini}}},\ }\href {\doibase 10.3847/1538-4357/aa6b58} {\bibfield
  {journal} {\bibinfo  {journal} {\textit{The Astrophysical Journal}}\ }\textbf
  {\bibinfo {volume} {840}},\ \bibinfo {eid} {31} (\bibinfo {year} {2017})},\
  \Eprint {http://arxiv.org/abs/1611.06573} {arXiv:1611.06573 [astro-ph.GA]}
  \BibitemShut {NoStop}%
\bibitem [{\citenamefont {{Tremmel}}\ \emph {et~al.}(2018)\citenamefont
  {{Tremmel}}, \citenamefont {{Governato}}, \citenamefont {{Volonteri}},
  \citenamefont {{Quinn}},\ and\ \citenamefont {{Pontzen}}}]{changa}%
  \BibitemOpen
  \bibfield  {author} {\bibinfo {author} {\bibfnamefont {M.}~\bibnamefont
  {{Tremmel}}}, \bibinfo {author} {\bibfnamefont {F.}~\bibnamefont
  {{Governato}}}, \bibinfo {author} {\bibfnamefont {M.}~\bibnamefont
  {{Volonteri}}}, \bibinfo {author} {\bibfnamefont {T.~R.}\ \bibnamefont
  {{Quinn}}}, \ and\ \bibinfo {author} {\bibfnamefont {A.}~\bibnamefont
  {{Pontzen}}},\ }\href {\doibase 10.1093/mnras/sty139} {\bibfield  {journal}
  {\bibinfo  {journal} {\textit{Monthly Notices of the Royal Astronomical
  Society}}\ }\textbf {\bibinfo {volume} {475}},\ \bibinfo {pages} {4967}
  (\bibinfo {year} {2018})},\ \Eprint {http://arxiv.org/abs/1708.07126}
  {arXiv:1708.07126} \BibitemShut {NoStop}%
\bibitem [{\citenamefont {{MacFadyen}}\ and\ \citenamefont
  {{Milosavljevi{\'c}}}(2008)}]{MacFadyen2008}%
  \BibitemOpen
  \bibfield  {author} {\bibinfo {author} {\bibfnamefont {A.~I.}\ \bibnamefont
  {{MacFadyen}}}\ and\ \bibinfo {author} {\bibfnamefont {M.}~\bibnamefont
  {{Milosavljevi{\'c}}}},\ }\href {\doibase 10.1086/523869} {\bibfield
  {journal} {\bibinfo  {journal} {\textit{The Astrophysical Journal}}\ }\textbf
  {\bibinfo {volume} {672}},\ \bibinfo {eid} {83-93} (\bibinfo {year}
  {2008})},\ \Eprint {http://arxiv.org/abs/astro-ph/0607467} {astro-ph/0607467}
  \BibitemShut {NoStop}%
\bibitem [{\citenamefont {{Cuadra}}\ \emph {et~al.}(2009)\citenamefont
  {{Cuadra}}, \citenamefont {{Armitage}}, \citenamefont {{Alexander}},\ and\
  \citenamefont {{Begelman}}}]{Cuadra2009}%
  \BibitemOpen
  \bibfield  {author} {\bibinfo {author} {\bibfnamefont {J.}~\bibnamefont
  {{Cuadra}}}, \bibinfo {author} {\bibfnamefont {P.~J.}\ \bibnamefont
  {{Armitage}}}, \bibinfo {author} {\bibfnamefont {R.~D.}\ \bibnamefont
  {{Alexander}}}, \ and\ \bibinfo {author} {\bibfnamefont {M.~C.}\ \bibnamefont
  {{Begelman}}},\ }\href {\doibase 10.1111/j.1365-2966.2008.14147.x} {\bibfield
   {journal} {\bibinfo  {journal} {\textit{Monthly Notices of the Royal
  Astronomical Society}}\ }\textbf {\bibinfo {volume} {393}},\ \bibinfo {pages}
  {1423} (\bibinfo {year} {2009})},\ \Eprint {http://arxiv.org/abs/0809.0311}
  {arXiv:0809.0311} \BibitemShut {NoStop}%
\bibitem [{\citenamefont {{Lodato}}\ \emph {et~al.}(2009)\citenamefont
  {{Lodato}}, \citenamefont {{Nayakshin}}, \citenamefont {{King}},\ and\
  \citenamefont {{Pringle}}}]{Lodato2009}%
  \BibitemOpen
  \bibfield  {author} {\bibinfo {author} {\bibfnamefont {G.}~\bibnamefont
  {{Lodato}}}, \bibinfo {author} {\bibfnamefont {S.}~\bibnamefont
  {{Nayakshin}}}, \bibinfo {author} {\bibfnamefont {A.~R.}\ \bibnamefont
  {{King}}}, \ and\ \bibinfo {author} {\bibfnamefont {J.~E.}\ \bibnamefont
  {{Pringle}}},\ }\href {\doibase 10.1111/j.1365-2966.2009.15179.x} {\bibfield
  {journal} {\bibinfo  {journal} {\textit{Monthly Notices of the Royal
  Astronomical Society}}\ }\textbf {\bibinfo {volume} {398}},\ \bibinfo {pages}
  {1392} (\bibinfo {year} {2009})},\ \Eprint {http://arxiv.org/abs/0906.0737}
  {arXiv:0906.0737} \BibitemShut {NoStop}%
\bibitem [{\citenamefont {{Roedig}}\ \emph {et~al.}(2011)\citenamefont
  {{Roedig}}, \citenamefont {{Dotti}}, \citenamefont {{Sesana}}, \citenamefont
  {{Cuadra}},\ and\ \citenamefont {{Colpi}}}]{Roedig2011}%
  \BibitemOpen
  \bibfield  {author} {\bibinfo {author} {\bibfnamefont {C.}~\bibnamefont
  {{Roedig}}}, \bibinfo {author} {\bibfnamefont {M.}~\bibnamefont {{Dotti}}},
  \bibinfo {author} {\bibfnamefont {A.}~\bibnamefont {{Sesana}}}, \bibinfo
  {author} {\bibfnamefont {J.}~\bibnamefont {{Cuadra}}}, \ and\ \bibinfo
  {author} {\bibfnamefont {M.}~\bibnamefont {{Colpi}}},\ }\href {\doibase
  10.1111/j.1365-2966.2011.18927.x} {\bibfield  {journal} {\bibinfo  {journal}
  {\textit{Monthly Notices of the Royal Astronomical Society}}\ }\textbf
  {\bibinfo {volume} {415}},\ \bibinfo {pages} {3033} (\bibinfo {year}
  {2011})},\ \Eprint {http://arxiv.org/abs/1104.3868} {arXiv:1104.3868}
  \BibitemShut {NoStop}%
\bibitem [{\citenamefont {{Nixon}}\ \emph {et~al.}(2011)\citenamefont
  {{Nixon}}, \citenamefont {{Cossins}}, \citenamefont {{King}},\ and\
  \citenamefont {{Pringle}}}]{Nixon2011}%
  \BibitemOpen
  \bibfield  {author} {\bibinfo {author} {\bibfnamefont {C.~J.}\ \bibnamefont
  {{Nixon}}}, \bibinfo {author} {\bibfnamefont {P.~J.}\ \bibnamefont
  {{Cossins}}}, \bibinfo {author} {\bibfnamefont {A.~R.}\ \bibnamefont
  {{King}}}, \ and\ \bibinfo {author} {\bibfnamefont {J.~E.}\ \bibnamefont
  {{Pringle}}},\ }\href {\doibase 10.1111/j.1365-2966.2010.17952.x} {\bibfield
  {journal} {\bibinfo  {journal} {\textit{Monthly Notices of the Royal
  Astronomical Society}}\ }\textbf {\bibinfo {volume} {412}},\ \bibinfo {pages}
  {1591} (\bibinfo {year} {2011})},\ \Eprint {http://arxiv.org/abs/1011.1914}
  {arXiv:1011.1914 [astro-ph.HE]} \BibitemShut {NoStop}%
\bibitem [{\citenamefont {{Duffell}}\ \emph {et~al.}(2019)\citenamefont
  {{Duffell}}, \citenamefont {{D'Orazio}}, \citenamefont {{Derdzinski}},
  \citenamefont {{Haiman}}, \citenamefont {{MacFadyen}}, \citenamefont
  {{Rosen}},\ and\ \citenamefont {{Zrake}}}]{Duffel2019}%
  \BibitemOpen
  \bibfield  {author} {\bibinfo {author} {\bibfnamefont {P.~C.}\ \bibnamefont
  {{Duffell}}}, \bibinfo {author} {\bibfnamefont {D.}~\bibnamefont
  {{D'Orazio}}}, \bibinfo {author} {\bibfnamefont {A.}~\bibnamefont
  {{Derdzinski}}}, \bibinfo {author} {\bibfnamefont {Z.}~\bibnamefont
  {{Haiman}}}, \bibinfo {author} {\bibfnamefont {A.}~\bibnamefont
  {{MacFadyen}}}, \bibinfo {author} {\bibfnamefont {A.~L.}\ \bibnamefont
  {{Rosen}}}, \ and\ \bibinfo {author} {\bibfnamefont {J.}~\bibnamefont
  {{Zrake}}},\ }\href@noop {} {\bibfield  {journal} {\bibinfo  {journal} {arXiv
  e-prints}\ ,\ \bibinfo {eid} {arXiv:1911.05506}} (\bibinfo {year} {2019})},\
  \Eprint {http://arxiv.org/abs/1911.05506} {arXiv:1911.05506 [astro-ph.SR]}
  \BibitemShut {NoStop}%
\bibitem [{\citenamefont {{Mu{\~n}oz}}\ \emph {et~al.}(2019)\citenamefont
  {{Mu{\~n}oz}}, \citenamefont {{Miranda}},\ and\ \citenamefont
  {{Lai}}}]{Munoz2019}%
  \BibitemOpen
  \bibfield  {author} {\bibinfo {author} {\bibfnamefont {D.~J.}\ \bibnamefont
  {{Mu{\~n}oz}}}, \bibinfo {author} {\bibfnamefont {R.}~\bibnamefont
  {{Miranda}}}, \ and\ \bibinfo {author} {\bibfnamefont {D.}~\bibnamefont
  {{Lai}}},\ }\href {\doibase 10.3847/1538-4357/aaf867} {\bibfield  {journal}
  {\bibinfo  {journal} {\textit{The Astrophysical Journal}}\ }\textbf {\bibinfo
  {volume} {871}},\ \bibinfo {eid} {84} (\bibinfo {year} {2019})},\ \Eprint
  {http://arxiv.org/abs/1810.04676} {arXiv:1810.04676 [astro-ph.HE]}
  \BibitemShut {NoStop}%
\bibitem [{\citenamefont {{Quinlan}}(1996)}]{Quinlan1996}%
  \BibitemOpen
  \bibfield  {author} {\bibinfo {author} {\bibfnamefont {G.~D.}\ \bibnamefont
  {{Quinlan}}},\ }\href {\doibase 10.1016/S1384-1076(96)00003-6} {\bibfield
  {journal} {\bibinfo  {journal} {\textit{Nature Astronomy}}\ }\textbf
  {\bibinfo {volume} {1}},\ \bibinfo {pages} {35} (\bibinfo {year} {1996})},\
  \Eprint {http://arxiv.org/abs/astro-ph/9601092} {astro-ph/9601092}
  \BibitemShut {NoStop}%
\bibitem [{\citenamefont {{Sesana}}\ and\ \citenamefont
  {{Khan}}(2015)}]{Sesana2015}%
  \BibitemOpen
  \bibfield  {author} {\bibinfo {author} {\bibfnamefont {A.}~\bibnamefont
  {{Sesana}}}\ and\ \bibinfo {author} {\bibfnamefont {F.~M.}\ \bibnamefont
  {{Khan}}},\ }\href {\doibase 10.1093/mnrasl/slv131} {\bibfield  {journal}
  {\bibinfo  {journal} {\textit{Monthly Notices of the Royal Astronomical
  Society}}\ }\textbf {\bibinfo {volume} {454}},\ \bibinfo {pages} {L66}
  (\bibinfo {year} {2015})},\ \Eprint {http://arxiv.org/abs/1505.02062}
  {arXiv:1505.02062} \BibitemShut {NoStop}%
\bibitem [{\citenamefont {{Hoffman}}\ and\ \citenamefont
  {{Loeb}}(2007)}]{Hoffman2007}%
  \BibitemOpen
  \bibfield  {author} {\bibinfo {author} {\bibfnamefont {L.}~\bibnamefont
  {{Hoffman}}}\ and\ \bibinfo {author} {\bibfnamefont {A.}~\bibnamefont
  {{Loeb}}},\ }\href {\doibase 10.1111/j.1365-2966.2007.11694.x} {\bibfield
  {journal} {\bibinfo  {journal} {\textit{Monthly Notices of the Royal
  Astronomical Society}}\ }\textbf {\bibinfo {volume} {377}},\ \bibinfo {pages}
  {957} (\bibinfo {year} {2007})},\ \Eprint
  {http://arxiv.org/abs/astro-ph/0612517} {astro-ph/0612517} \BibitemShut
  {NoStop}%
\bibitem [{\citenamefont {{Bonetti}}\ \emph {et~al.}(2016)\citenamefont
  {{Bonetti}}, \citenamefont {{Haardt}}, \citenamefont {{Sesana}},\ and\
  \citenamefont {{Barausse}}}]{BonettiI}%
  \BibitemOpen
  \bibfield  {author} {\bibinfo {author} {\bibfnamefont {M.}~\bibnamefont
  {{Bonetti}}}, \bibinfo {author} {\bibfnamefont {F.}~\bibnamefont {{Haardt}}},
  \bibinfo {author} {\bibfnamefont {A.}~\bibnamefont {{Sesana}}}, \ and\
  \bibinfo {author} {\bibfnamefont {E.}~\bibnamefont {{Barausse}}},\ }\href
  {\doibase 10.1093/mnras/stw1590} {\bibfield  {journal} {\bibinfo  {journal}
  {\textit{Monthly Notices of the Royal Astronomical Society}}\ }\textbf
  {\bibinfo {volume} {461}},\ \bibinfo {pages} {4419} (\bibinfo {year}
  {2016})},\ \Eprint {http://arxiv.org/abs/1604.08770} {arXiv:1604.08770}
  \BibitemShut {NoStop}%
\bibitem [{\citenamefont {{Bonetti}}\ \emph
  {et~al.}(2018{\natexlab{b}})\citenamefont {{Bonetti}}, \citenamefont
  {{Sesana}}, \citenamefont {{Barausse}},\ and\ \citenamefont
  {{Haardt}}}]{BonettiIII}%
  \BibitemOpen
  \bibfield  {author} {\bibinfo {author} {\bibfnamefont {M.}~\bibnamefont
  {{Bonetti}}}, \bibinfo {author} {\bibfnamefont {A.}~\bibnamefont {{Sesana}}},
  \bibinfo {author} {\bibfnamefont {E.}~\bibnamefont {{Barausse}}}, \ and\
  \bibinfo {author} {\bibfnamefont {F.}~\bibnamefont {{Haardt}}},\ }\href
  {\doibase 10.1093/mnras/sty874} {\bibfield  {journal} {\bibinfo  {journal}
  {\textit{Monthly Notices of the Royal Astronomical Society}}\ }\textbf
  {\bibinfo {volume} {477}},\ \bibinfo {pages} {2599} (\bibinfo {year}
  {2018}{\natexlab{b}})},\ \Eprint {http://arxiv.org/abs/1709.06095}
  {arXiv:1709.06095} \BibitemShut {NoStop}%
\bibitem [{\citenamefont {Barausse}\ \emph {et~al.}(2012)\citenamefont
  {Barausse}, \citenamefont {Morozova},\ and\ \citenamefont
  {Rezzolla}}]{Barausse:2012qz}%
  \BibitemOpen
  \bibfield  {author} {\bibinfo {author} {\bibfnamefont {E.}~\bibnamefont
  {Barausse}}, \bibinfo {author} {\bibfnamefont {V.}~\bibnamefont {Morozova}},
  \ and\ \bibinfo {author} {\bibfnamefont {L.}~\bibnamefont {Rezzolla}},\
  }\href {\doibase 10.1088/0004-637X/758/1/63} {\bibfield  {journal} {\bibinfo
  {journal} {Astrophys. J.}\ }\textbf {\bibinfo {volume} {758}},\ \bibinfo
  {pages} {63} (\bibinfo {year} {2012})},\ \bibinfo {note} {[Erratum:
  Astrophys.J. 786, 76 (2014)]},\ \Eprint {http://arxiv.org/abs/1206.3803}
  {arXiv:1206.3803 [gr-qc]} \BibitemShut {NoStop}%
\bibitem [{\citenamefont {Hofmann}\ \emph {et~al.}(2016)\citenamefont
  {Hofmann}, \citenamefont {Barausse},\ and\ \citenamefont
  {Rezzolla}}]{Hofmann:2016yih}%
  \BibitemOpen
  \bibfield  {author} {\bibinfo {author} {\bibfnamefont {F.}~\bibnamefont
  {Hofmann}}, \bibinfo {author} {\bibfnamefont {E.}~\bibnamefont {Barausse}}, \
  and\ \bibinfo {author} {\bibfnamefont {L.}~\bibnamefont {Rezzolla}},\ }\href
  {\doibase 10.3847/2041-8205/825/2/L19} {\bibfield  {journal} {\bibinfo
  {journal} {Astrophys. J. Lett.}\ }\textbf {\bibinfo {volume} {825}},\
  \bibinfo {pages} {L19} (\bibinfo {year} {2016})},\ \Eprint
  {http://arxiv.org/abs/1605.01938} {arXiv:1605.01938 [gr-qc]} \BibitemShut
  {NoStop}%
\bibitem [{\citenamefont {{van Meter}}\ \emph {et~al.}(2010)\citenamefont {{van
  Meter}}, \citenamefont {{Miller}}, \citenamefont {{Baker}}, \citenamefont
  {{Boggs}},\ and\ \citenamefont {{Kelly}}}]{kick}%
  \BibitemOpen
  \bibfield  {author} {\bibinfo {author} {\bibfnamefont {J.~R.}\ \bibnamefont
  {{van Meter}}}, \bibinfo {author} {\bibfnamefont {M.~C.}\ \bibnamefont
  {{Miller}}}, \bibinfo {author} {\bibfnamefont {J.~G.}\ \bibnamefont
  {{Baker}}}, \bibinfo {author} {\bibfnamefont {W.~D.}\ \bibnamefont
  {{Boggs}}}, \ and\ \bibinfo {author} {\bibfnamefont {B.~J.}\ \bibnamefont
  {{Kelly}}},\ }\href {\doibase 10.1088/0004-637X/719/2/1427} {\bibfield
  {journal} {\bibinfo  {journal} {\apj}\ }\textbf {\bibinfo {volume} {719}},\
  \bibinfo {pages} {1427} (\bibinfo {year} {2010})},\ \Eprint
  {http://arxiv.org/abs/1003.3865} {arXiv:1003.3865 [astro-ph.HE]} \BibitemShut
  {NoStop}%
\bibitem [{\citenamefont {{Madau}}\ and\ \citenamefont
  {{Rees}}(2001)}]{Madau2001}%
  \BibitemOpen
  \bibfield  {author} {\bibinfo {author} {\bibfnamefont {P.}~\bibnamefont
  {{Madau}}}\ and\ \bibinfo {author} {\bibfnamefont {M.~J.}\ \bibnamefont
  {{Rees}}},\ }\href {\doibase 10.1086/319848} {\bibfield  {journal} {\bibinfo
  {journal} {\textit{The Astrophysical Journal Letters}}\ }\textbf {\bibinfo
  {volume} {551}},\ \bibinfo {pages} {L27} (\bibinfo {year} {2001})},\ \Eprint
  {http://arxiv.org/abs/astro-ph/0101223} {astro-ph/0101223} \BibitemShut
  {NoStop}%
\bibitem [{\citenamefont {{Volonteri}}\ \emph {et~al.}(2008)\citenamefont
  {{Volonteri}}, \citenamefont {{Lodato}},\ and\ \citenamefont
  {{Natarajan}}}]{Volonteri2008}%
  \BibitemOpen
  \bibfield  {author} {\bibinfo {author} {\bibfnamefont {M.}~\bibnamefont
  {{Volonteri}}}, \bibinfo {author} {\bibfnamefont {G.}~\bibnamefont
  {{Lodato}}}, \ and\ \bibinfo {author} {\bibfnamefont {P.}~\bibnamefont
  {{Natarajan}}},\ }\href {\doibase 10.1111/j.1365-2966.2007.12589.x}
  {\bibfield  {journal} {\bibinfo  {journal} {\textit{Monthly Notices of the
  Royal Astronomical Society}}\ }\textbf {\bibinfo {volume} {383}},\ \bibinfo
  {pages} {1079} (\bibinfo {year} {2008})},\ \Eprint
  {http://arxiv.org/abs/0709.0529} {arXiv:0709.0529} \BibitemShut {NoStop}%
\bibitem [{\citenamefont {Toubiana}\ \emph {et~al.}(2021)\citenamefont
  {Toubiana}, \citenamefont {Wong}, \citenamefont {Babak}, \citenamefont
  {Barausse}, \citenamefont {Berti}, \citenamefont {Gair}, \citenamefont
  {Marsat},\ and\ \citenamefont {Taylor}}]{Toubiana:2021iuw}%
  \BibitemOpen
  \bibfield  {author} {\bibinfo {author} {\bibfnamefont {A.}~\bibnamefont
  {Toubiana}}, \bibinfo {author} {\bibfnamefont {K.~W.~K.}\ \bibnamefont
  {Wong}}, \bibinfo {author} {\bibfnamefont {S.}~\bibnamefont {Babak}},
  \bibinfo {author} {\bibfnamefont {E.}~\bibnamefont {Barausse}}, \bibinfo
  {author} {\bibfnamefont {E.}~\bibnamefont {Berti}}, \bibinfo {author}
  {\bibfnamefont {J.~R.}\ \bibnamefont {Gair}}, \bibinfo {author}
  {\bibfnamefont {S.}~\bibnamefont {Marsat}}, \ and\ \bibinfo {author}
  {\bibfnamefont {S.~R.}\ \bibnamefont {Taylor}},\ }\href {\doibase
  10.1103/PhysRevD.104.083027} {\  (\bibinfo {year} {2021}),\
  10.1103/PhysRevD.104.083027},\ \Eprint {http://arxiv.org/abs/2106.13819}
  {arXiv:2106.13819 [gr-qc]} \BibitemShut {NoStop}%
\bibitem [{\citenamefont {Babak}\ \emph {et~al.}(2021)\citenamefont {Babak},
  \citenamefont {Petiteau},\ and\ \citenamefont {Hewitson}}]{Babak:2021mhe}%
  \BibitemOpen
  \bibfield  {author} {\bibinfo {author} {\bibfnamefont {S.}~\bibnamefont
  {Babak}}, \bibinfo {author} {\bibfnamefont {A.}~\bibnamefont {Petiteau}}, \
  and\ \bibinfo {author} {\bibfnamefont {M.}~\bibnamefont {Hewitson}},\
  }\href@noop {} {\  (\bibinfo {year} {2021})},\ \Eprint
  {http://arxiv.org/abs/2108.01167} {arXiv:2108.01167 [astro-ph.IM]}
  \BibitemShut {NoStop}%
\bibitem [{\citenamefont {Robson}\ \emph {et~al.}(2019)\citenamefont {Robson},
  \citenamefont {Cornish},\ and\ \citenamefont {Liu}}]{Robson:2018ifk}%
  \BibitemOpen
  \bibfield  {author} {\bibinfo {author} {\bibfnamefont {T.}~\bibnamefont
  {Robson}}, \bibinfo {author} {\bibfnamefont {N.~J.}\ \bibnamefont {Cornish}},
  \ and\ \bibinfo {author} {\bibfnamefont {C.}~\bibnamefont {Liu}},\ }\href
  {\doibase 10.1088/1361-6382/ab1101} {\bibfield  {journal} {\bibinfo
  {journal} {Class. Quant. Grav.}\ }\textbf {\bibinfo {volume} {36}},\ \bibinfo
  {pages} {105011} (\bibinfo {year} {2019})},\ \Eprint
  {http://arxiv.org/abs/1803.01944} {arXiv:1803.01944 [astro-ph.HE]}
  \BibitemShut {NoStop}%
\bibitem [{\citenamefont {Berti}\ \emph
  {et~al.}(2007{\natexlab{b}})\citenamefont {Berti}, \citenamefont {Cardoso},
  \citenamefont {Gonzalez}, \citenamefont {Sperhake}, \citenamefont {Hannam},
  \citenamefont {Husa},\ and\ \citenamefont {Bruegmann}}]{Berti:2007fi}%
  \BibitemOpen
  \bibfield  {author} {\bibinfo {author} {\bibfnamefont {E.}~\bibnamefont
  {Berti}}, \bibinfo {author} {\bibfnamefont {V.}~\bibnamefont {Cardoso}},
  \bibinfo {author} {\bibfnamefont {J.~A.}\ \bibnamefont {Gonzalez}}, \bibinfo
  {author} {\bibfnamefont {U.}~\bibnamefont {Sperhake}}, \bibinfo {author}
  {\bibfnamefont {M.}~\bibnamefont {Hannam}}, \bibinfo {author} {\bibfnamefont
  {S.}~\bibnamefont {Husa}}, \ and\ \bibinfo {author} {\bibfnamefont
  {B.}~\bibnamefont {Bruegmann}},\ }\href {\doibase 10.1103/PhysRevD.76.064034}
  {\bibfield  {journal} {\bibinfo  {journal} {Phys. Rev. D}\ }\textbf {\bibinfo
  {volume} {76}},\ \bibinfo {pages} {064034} (\bibinfo {year}
  {2007}{\natexlab{b}})},\ \Eprint {http://arxiv.org/abs/gr-qc/0703053}
  {arXiv:gr-qc/0703053} \BibitemShut {NoStop}%
\bibitem [{\citenamefont {Baibhav}\ \emph {et~al.}(2018)\citenamefont
  {Baibhav}, \citenamefont {Berti}, \citenamefont {Cardoso},\ and\
  \citenamefont {Khanna}}]{Baibhav:2017jhs}%
  \BibitemOpen
  \bibfield  {author} {\bibinfo {author} {\bibfnamefont {V.}~\bibnamefont
  {Baibhav}}, \bibinfo {author} {\bibfnamefont {E.}~\bibnamefont {Berti}},
  \bibinfo {author} {\bibfnamefont {V.}~\bibnamefont {Cardoso}}, \ and\
  \bibinfo {author} {\bibfnamefont {G.}~\bibnamefont {Khanna}},\ }\href
  {\doibase 10.1103/PhysRevD.97.044048} {\bibfield  {journal} {\bibinfo
  {journal} {Phys. Rev. D}\ }\textbf {\bibinfo {volume} {97}},\ \bibinfo
  {pages} {044048} (\bibinfo {year} {2018})},\ \Eprint
  {http://arxiv.org/abs/1710.02156} {arXiv:1710.02156 [gr-qc]} \BibitemShut
  {NoStop}%
\bibitem [{\citenamefont {Ajith}\ \emph {et~al.}(2011)\citenamefont {Ajith}
  \emph {et~al.}}]{Ajith:2009bn}%
  \BibitemOpen
  \bibfield  {author} {\bibinfo {author} {\bibfnamefont {P.}~\bibnamefont
  {Ajith}} \emph {et~al.},\ }\href {\doibase 10.1103/PhysRevLett.106.241101}
  {\bibfield  {journal} {\bibinfo  {journal} {Phys. Rev. Lett.}\ }\textbf
  {\bibinfo {volume} {106}},\ \bibinfo {pages} {241101} (\bibinfo {year}
  {2011})},\ \Eprint {http://arxiv.org/abs/0909.2867} {arXiv:0909.2867 [gr-qc]}
  \BibitemShut {NoStop}%
\bibitem [{\citenamefont {Kamaretsos}\ \emph
  {et~al.}(2012{\natexlab{a}})\citenamefont {Kamaretsos}, \citenamefont
  {Hannam}, \citenamefont {Husa},\ and\ \citenamefont
  {Sathyaprakash}}]{q-from-rd}%
  \BibitemOpen
  \bibfield  {author} {\bibinfo {author} {\bibfnamefont {I.}~\bibnamefont
  {Kamaretsos}}, \bibinfo {author} {\bibfnamefont {M.}~\bibnamefont {Hannam}},
  \bibinfo {author} {\bibfnamefont {S.}~\bibnamefont {Husa}}, \ and\ \bibinfo
  {author} {\bibfnamefont {B.~S.}\ \bibnamefont {Sathyaprakash}},\ }\href
  {\doibase 10.1103/PhysRevD.85.024018} {\bibfield  {journal} {\bibinfo
  {journal} {Phys. Rev. D}\ }\textbf {\bibinfo {volume} {85}},\ \bibinfo
  {pages} {024018} (\bibinfo {year} {2012}{\natexlab{a}})}\BibitemShut
  {NoStop}%
\bibitem [{\citenamefont {Fisher}(1922)}]{TheFisher}%
  \BibitemOpen
  \bibfield  {author} {\bibinfo {author} {\bibfnamefont {R.~A.}\ \bibnamefont
  {Fisher}},\ }\href@noop {} {\bibfield  {journal} {\bibinfo  {journal}
  {Philosophical Transactions of the Royal Society of London,~A}\ }\textbf
  {\bibinfo {volume} {222}},\ \bibinfo {pages} {309} (\bibinfo {year}
  {1922})}\BibitemShut {NoStop}%
\bibitem [{\citenamefont {Vallisneri}(2008)}]{Use-and-abuse-of-FM}%
  \BibitemOpen
  \bibfield  {author} {\bibinfo {author} {\bibfnamefont {M.}~\bibnamefont
  {Vallisneri}},\ }\href {\doibase 10.1103/PhysRevD.77.042001} {\bibfield
  {journal} {\bibinfo  {journal} {Phys. Rev. D}\ }\textbf {\bibinfo {volume}
  {77}},\ \bibinfo {pages} {042001} (\bibinfo {year} {2008})},\ \Eprint
  {http://arxiv.org/abs/gr-qc/0703086} {arXiv:gr-qc/0703086} \BibitemShut
  {NoStop}%
\bibitem [{\citenamefont {Meurer}\ \emph {et~al.}(2017)\citenamefont {Meurer},
  \citenamefont {Smith}, \citenamefont {Paprocki}, \citenamefont
  {\v{C}ert\'{i}k}, \citenamefont {Kirpichev}, \citenamefont {Rocklin},
  \citenamefont {Kumar}, \citenamefont {Ivanov}, \citenamefont {Moore},
  \citenamefont {Singh}, \citenamefont {Rathnayake}, \citenamefont {Vig},
  \citenamefont {Granger}, \citenamefont {Muller}, \citenamefont {Bonazzi},
  \citenamefont {Gupta}, \citenamefont {Vats}, \citenamefont {Johansson},
  \citenamefont {Pedregosa}, \citenamefont {Curry}, \citenamefont {Terrel},
  \citenamefont {Rou\v{c}ka}, \citenamefont {Saboo}, \citenamefont {Fernando},
  \citenamefont {Kulal}, \citenamefont {Cimrman},\ and\ \citenamefont
  {Scopatz}}]{SYMPY}%
  \BibitemOpen
  \bibfield  {author} {\bibinfo {author} {\bibfnamefont {A.}~\bibnamefont
  {Meurer}}, \bibinfo {author} {\bibfnamefont {C.~P.}\ \bibnamefont {Smith}},
  \bibinfo {author} {\bibfnamefont {M.}~\bibnamefont {Paprocki}}, \bibinfo
  {author} {\bibfnamefont {O.}~\bibnamefont {\v{C}ert\'{i}k}}, \bibinfo
  {author} {\bibfnamefont {S.~B.}\ \bibnamefont {Kirpichev}}, \bibinfo {author}
  {\bibfnamefont {M.}~\bibnamefont {Rocklin}}, \bibinfo {author} {\bibfnamefont
  {A.}~\bibnamefont {Kumar}}, \bibinfo {author} {\bibfnamefont
  {S.}~\bibnamefont {Ivanov}}, \bibinfo {author} {\bibfnamefont {J.~K.}\
  \bibnamefont {Moore}}, \bibinfo {author} {\bibfnamefont {S.}~\bibnamefont
  {Singh}}, \bibinfo {author} {\bibfnamefont {T.}~\bibnamefont {Rathnayake}},
  \bibinfo {author} {\bibfnamefont {S.}~\bibnamefont {Vig}}, \bibinfo {author}
  {\bibfnamefont {B.~E.}\ \bibnamefont {Granger}}, \bibinfo {author}
  {\bibfnamefont {R.~P.}\ \bibnamefont {Muller}}, \bibinfo {author}
  {\bibfnamefont {F.}~\bibnamefont {Bonazzi}}, \bibinfo {author} {\bibfnamefont
  {H.}~\bibnamefont {Gupta}}, \bibinfo {author} {\bibfnamefont
  {S.}~\bibnamefont {Vats}}, \bibinfo {author} {\bibfnamefont {F.}~\bibnamefont
  {Johansson}}, \bibinfo {author} {\bibfnamefont {F.}~\bibnamefont
  {Pedregosa}}, \bibinfo {author} {\bibfnamefont {M.~J.}\ \bibnamefont
  {Curry}}, \bibinfo {author} {\bibfnamefont {A.~R.}\ \bibnamefont {Terrel}},
  \bibinfo {author} {\bibfnamefont {v.}~\bibnamefont {Rou\v{c}ka}}, \bibinfo
  {author} {\bibfnamefont {A.}~\bibnamefont {Saboo}}, \bibinfo {author}
  {\bibfnamefont {I.}~\bibnamefont {Fernando}}, \bibinfo {author}
  {\bibfnamefont {S.}~\bibnamefont {Kulal}}, \bibinfo {author} {\bibfnamefont
  {R.}~\bibnamefont {Cimrman}}, \ and\ \bibinfo {author} {\bibfnamefont
  {A.}~\bibnamefont {Scopatz}},\ }\href {\doibase 10.7717/peerj-cs.103}
  {\bibfield  {journal} {\bibinfo  {journal} {PeerJ Computer Science}\ }\textbf
  {\bibinfo {volume} {3}},\ \bibinfo {pages} {e103} (\bibinfo {year}
  {2017})}\BibitemShut {NoStop}%
\bibitem [{\citenamefont {Berti}\ and\ \citenamefont {Cardoso}()}]{Berti:data}%
  \BibitemOpen
  \bibfield  {author} {\bibinfo {author} {\bibfnamefont {E.}~\bibnamefont
  {Berti}}\ and\ \bibinfo {author} {\bibfnamefont {V.}~\bibnamefont
  {Cardoso}},\ }\href@noop {} {\enquote {\bibinfo {title} {Kerr qnm
  frequencies},}\ }\bibinfo {howpublished}
  {\url{https://pages.jh.edu/~eberti2/ringdown/},
  \url{https://centra.tecnico.ulisboa.pt/network/grit/files/ringdown/}}\BibitemShut
  {NoStop}%
\bibitem [{\citenamefont {London}\ \emph {et~al.}(2014)\citenamefont {London},
  \citenamefont {Shoemaker},\ and\ \citenamefont {Healy}}]{London:2014cma}%
  \BibitemOpen
  \bibfield  {author} {\bibinfo {author} {\bibfnamefont {L.}~\bibnamefont
  {London}}, \bibinfo {author} {\bibfnamefont {D.}~\bibnamefont {Shoemaker}}, \
  and\ \bibinfo {author} {\bibfnamefont {J.}~\bibnamefont {Healy}},\ }\href
  {\doibase 10.1103/PhysRevD.90.124032} {\bibfield  {journal} {\bibinfo
  {journal} {Phys. Rev. D}\ }\textbf {\bibinfo {volume} {90}},\ \bibinfo
  {pages} {124032} (\bibinfo {year} {2014})},\ \bibinfo {note} {[Erratum:
  Phys.Rev.D 94, 069902 (2016)]},\ \Eprint {http://arxiv.org/abs/1404.3197}
  {arXiv:1404.3197 [gr-qc]} \BibitemShut {NoStop}%
\bibitem [{\citenamefont {Kamaretsos}\ \emph
  {et~al.}(2012{\natexlab{b}})\citenamefont {Kamaretsos}, \citenamefont
  {Hannam},\ and\ \citenamefont {Sathyaprakash}}]{Kamaretsos:2012bs}%
  \BibitemOpen
  \bibfield  {author} {\bibinfo {author} {\bibfnamefont {I.}~\bibnamefont
  {Kamaretsos}}, \bibinfo {author} {\bibfnamefont {M.}~\bibnamefont {Hannam}},
  \ and\ \bibinfo {author} {\bibfnamefont {B.}~\bibnamefont {Sathyaprakash}},\
  }\href {\doibase 10.1103/PhysRevLett.109.141102} {\bibfield  {journal}
  {\bibinfo  {journal} {Phys. Rev. Lett.}\ }\textbf {\bibinfo {volume} {109}},\
  \bibinfo {pages} {141102} (\bibinfo {year} {2012}{\natexlab{b}})},\ \Eprint
  {http://arxiv.org/abs/1207.0399} {arXiv:1207.0399 [gr-qc]} \BibitemShut
  {NoStop}%
\bibitem [{\citenamefont {Pan}\ \emph {et~al.}(2011)\citenamefont {Pan},
  \citenamefont {Buonanno}, \citenamefont {Boyle}, \citenamefont {Buchman},
  \citenamefont {Kidder}, \citenamefont {Pfeiffer},\ and\ \citenamefont
  {Scheel}}]{Pan:2011gk}%
  \BibitemOpen
  \bibfield  {author} {\bibinfo {author} {\bibfnamefont {Y.}~\bibnamefont
  {Pan}}, \bibinfo {author} {\bibfnamefont {A.}~\bibnamefont {Buonanno}},
  \bibinfo {author} {\bibfnamefont {M.}~\bibnamefont {Boyle}}, \bibinfo
  {author} {\bibfnamefont {L.~T.}\ \bibnamefont {Buchman}}, \bibinfo {author}
  {\bibfnamefont {L.~E.}\ \bibnamefont {Kidder}}, \bibinfo {author}
  {\bibfnamefont {H.~P.}\ \bibnamefont {Pfeiffer}}, \ and\ \bibinfo {author}
  {\bibfnamefont {M.~A.}\ \bibnamefont {Scheel}},\ }\href {\doibase
  10.1103/PhysRevD.84.124052} {\bibfield  {journal} {\bibinfo  {journal} {Phys.
  Rev. D}\ }\textbf {\bibinfo {volume} {84}},\ \bibinfo {pages} {124052}
  (\bibinfo {year} {2011})},\ \Eprint {http://arxiv.org/abs/1106.1021}
  {arXiv:1106.1021 [gr-qc]} \BibitemShut {NoStop}%
\bibitem [{\citenamefont {Bhagwat}\ and\ \citenamefont
  {Pacilio}(2021)}]{Bhagwat:2021kfa}%
  \BibitemOpen
  \bibfield  {author} {\bibinfo {author} {\bibfnamefont {S.}~\bibnamefont
  {Bhagwat}}\ and\ \bibinfo {author} {\bibfnamefont {C.}~\bibnamefont
  {Pacilio}},\ }\href {\doibase 10.1103/PhysRevD.104.024030} {\bibfield
  {journal} {\bibinfo  {journal} {Phys. Rev. D}\ }\textbf {\bibinfo {volume}
  {104}},\ \bibinfo {pages} {024030} (\bibinfo {year} {2021})},\ \Eprint
  {http://arxiv.org/abs/2101.07817} {arXiv:2101.07817 [gr-qc]} \BibitemShut
  {NoStop}%
\bibitem [{\citenamefont {Barausse}\ and\ \citenamefont
  {Rezzolla}(2009)}]{Barausse:2009uz}%
  \BibitemOpen
  \bibfield  {author} {\bibinfo {author} {\bibfnamefont {E.}~\bibnamefont
  {Barausse}}\ and\ \bibinfo {author} {\bibfnamefont {L.}~\bibnamefont
  {Rezzolla}},\ }\href {\doibase 10.1088/0004-637X/704/1/L40} {\bibfield
  {journal} {\bibinfo  {journal} {Astrophys. J. Lett.}\ }\textbf {\bibinfo
  {volume} {704}},\ \bibinfo {pages} {L40} (\bibinfo {year} {2009})},\ \Eprint
  {http://arxiv.org/abs/0904.2577} {arXiv:0904.2577 [gr-qc]} \BibitemShut
  {NoStop}%
\bibitem [{\citenamefont {Sesana}(2016)}]{Sesana:2016ljz}%
  \BibitemOpen
  \bibfield  {author} {\bibinfo {author} {\bibfnamefont {A.}~\bibnamefont
  {Sesana}},\ }\href {\doibase 10.1103/PhysRevLett.116.231102} {\bibfield
  {journal} {\bibinfo  {journal} {Phys. Rev. Lett.}\ }\textbf {\bibinfo
  {volume} {116}},\ \bibinfo {pages} {231102} (\bibinfo {year} {2016})},\
  \Eprint {http://arxiv.org/abs/1602.06951} {arXiv:1602.06951 [gr-qc]}
  \BibitemShut {NoStop}%
\bibitem [{\citenamefont {Carson}\ and\ \citenamefont
  {Yagi}(2020)}]{Carson:2019kkh}%
  \BibitemOpen
  \bibfield  {author} {\bibinfo {author} {\bibfnamefont {Z.}~\bibnamefont
  {Carson}}\ and\ \bibinfo {author} {\bibfnamefont {K.}~\bibnamefont {Yagi}},\
  }\href {\doibase 10.1103/PhysRevD.101.044047} {\bibfield  {journal} {\bibinfo
   {journal} {Phys. Rev. D}\ }\textbf {\bibinfo {volume} {101}},\ \bibinfo
  {pages} {044047} (\bibinfo {year} {2020})},\ \Eprint
  {http://arxiv.org/abs/1911.05258} {arXiv:1911.05258 [gr-qc]} \BibitemShut
  {NoStop}%
\bibitem [{\citenamefont {Kawamura}\ \emph {et~al.}(2011)\citenamefont
  {Kawamura} \emph {et~al.}}]{Kawamura:2011zz}%
  \BibitemOpen
  \bibfield  {author} {\bibinfo {author} {\bibfnamefont {S.}~\bibnamefont
  {Kawamura}} \emph {et~al.},\ }\href {\doibase 10.1088/0264-9381/28/9/094011}
  {\bibfield  {journal} {\bibinfo  {journal} {Class. Quant. Grav.}\ }\textbf
  {\bibinfo {volume} {28}},\ \bibinfo {pages} {094011} (\bibinfo {year}
  {2011})}\BibitemShut {NoStop}%
\bibitem [{\citenamefont {Brito}\ \emph {et~al.}(2018)\citenamefont {Brito},
  \citenamefont {Buonanno},\ and\ \citenamefont {Raymond}}]{Brito:2018rfr}%
  \BibitemOpen
  \bibfield  {author} {\bibinfo {author} {\bibfnamefont {R.}~\bibnamefont
  {Brito}}, \bibinfo {author} {\bibfnamefont {A.}~\bibnamefont {Buonanno}}, \
  and\ \bibinfo {author} {\bibfnamefont {V.}~\bibnamefont {Raymond}},\ }\href
  {\doibase 10.1103/PhysRevD.98.084038} {\bibfield  {journal} {\bibinfo
  {journal} {Phys. Rev. D}\ }\textbf {\bibinfo {volume} {98}},\ \bibinfo
  {pages} {084038} (\bibinfo {year} {2018})},\ \Eprint
  {http://arxiv.org/abs/1805.00293} {arXiv:1805.00293 [gr-qc]} \BibitemShut
  {NoStop}%
\bibitem [{\citenamefont {Yang}\ \emph {et~al.}(2017)\citenamefont {Yang},
  \citenamefont {Yagi}, \citenamefont {Blackman}, \citenamefont {Lehner},
  \citenamefont {Paschalidis}, \citenamefont {Pretorius},\ and\ \citenamefont
  {Yunes}}]{Yang:2017zxs}%
  \BibitemOpen
  \bibfield  {author} {\bibinfo {author} {\bibfnamefont {H.}~\bibnamefont
  {Yang}}, \bibinfo {author} {\bibfnamefont {K.}~\bibnamefont {Yagi}}, \bibinfo
  {author} {\bibfnamefont {J.}~\bibnamefont {Blackman}}, \bibinfo {author}
  {\bibfnamefont {L.}~\bibnamefont {Lehner}}, \bibinfo {author} {\bibfnamefont
  {V.}~\bibnamefont {Paschalidis}}, \bibinfo {author} {\bibfnamefont
  {F.}~\bibnamefont {Pretorius}}, \ and\ \bibinfo {author} {\bibfnamefont
  {N.}~\bibnamefont {Yunes}},\ }\href {\doibase 10.1103/PhysRevLett.118.161101}
  {\bibfield  {journal} {\bibinfo  {journal} {Phys. Rev. Lett.}\ }\textbf
  {\bibinfo {volume} {118}},\ \bibinfo {pages} {161101} (\bibinfo {year}
  {2017})},\ \Eprint {http://arxiv.org/abs/1701.05808} {arXiv:1701.05808
  [gr-qc]} \BibitemShut {NoStop}%
\bibitem [{\citenamefont {Carullo}\ \emph {et~al.}(2018)\citenamefont {Carullo}
  \emph {et~al.}}]{Carullo:2018sfu}%
  \BibitemOpen
  \bibfield  {author} {\bibinfo {author} {\bibfnamefont {G.}~\bibnamefont
  {Carullo}} \emph {et~al.},\ }\href {\doibase 10.1103/PhysRevD.98.104020}
  {\bibfield  {journal} {\bibinfo  {journal} {Phys. Rev. D}\ }\textbf {\bibinfo
  {volume} {98}},\ \bibinfo {pages} {104020} (\bibinfo {year} {2018})},\
  \Eprint {http://arxiv.org/abs/1805.04760} {arXiv:1805.04760 [gr-qc]}
  \BibitemShut {NoStop}%
\end{thebibliography}%
\end{document}